\documentclass[twocolumn]{aastex63}
\usepackage{ulem}

\shorttitle{Comparison of FSF}
\shortauthors{Lonoce et al.}

\begin{document}

\title{A Comparison of Full Spectral Fitting Codes for Measuring the Stellar Initial Mass Function and Other Stellar Population Properties in Elliptical Galaxies \footnote{This paper includes data gathered with the 6.5 meter Magellan Telescopes located at Las Campanas Observatory, Chile.}}

\correspondingauthor{Ilaria Lonoce}
\email{ilonoce@uchicago.edu}

\author[0000-0001-8421-1005]{Ilaria Lonoce}
\affiliation{Department of Astronomy $\&$ Astrophysics, The University of Chicago, 5640 South Ellis Avenue, Chicago, IL 60637, USA}

\author[0000-0002-0160-7221]{Anja Feldmeier-Krause}
\affiliation{Department of Astrophysics, University of Vienna, T\"urkenschanzstrasse 17, 1180 Wien, Austria}
\affiliation{Max-Planck-Institut f\"ur Astronomie, K\"onigstuhl 17, 69117, Heidelberg, Germany}

\author[0000-0002-3361-2893]{Alexandra Masegian}
\affiliation{Department of Astronomy $\&$ Astrophysics, The University of Chicago, 5640 South Ellis Avenue, Chicago, IL 60637, USA}
\affiliation{Department of Astronomy, Columbia University, 538 West 120th Street, New York, NY 10027, USA}

\author[0000-0003-3431-9135]{Wendy L. Freedman}
\affiliation{Department of Astronomy $\&$ Astrophysics, The University of Chicago, 5640 South Ellis Avenue, Chicago, IL 60637, USA}



\begin{abstract}

We present a comparative test of four widely used full spectral fitting codes, with the aim of answering the question: how robust is the retrieval of the stellar initial mass function (IMF) and other stellar properties of galaxies? We used \textsc{ALF}, \textsc{PyStaff}, \textsc{Starlight}, and \textsc{pPXF} to fit a set of optical+near-infrared spectroscopic data from the Magellan telescope of the two brightest galaxies in the Fornax cluster, NGC1399 and NGC1404. By fitting the same data set with the same models, we can compare the radial trends (out to $\sim1$ R$_e$) of IMF slope, age, metallicity and $19$ elemental abundances when allowed with the four codes.
To further test the robustness of our analysis, we carried out parallel simulations by creating inputs with different star formation history (SFH) complexity. The results from simulations show that codes such as \textsc{ALF} and \textsc{PyStaff}, which both assume a simple stellar population (SSP) return greater precision and accuracy only when the underlying population is a pure SSP; however, in cases where the SFH is more complex, these codes return erroneous results. Although codes like \textsc{Starlight} and \textsc{pPXF}, which retrieve the best-fit SFH without prior assumptions, tend to produce results with greater scatter and bias, they are generally more reliable in identifying secondary components.  
Our analysis on the two targets shows that \textsc{ALF} and \textsc{PyStaff}, that assume an SSP, give results pointing to a single old age, a decreasing metallicity with radius and a flat super-Salpeter IMF. In contrast, \textsc{Starlight} and \textsc{pPXF} suggest the presence of a secondary component with different metallicity and IMF characteristics.

\end{abstract}

\keywords{Unified Astronomy Thesaurus concepts: Early-type galaxies (429), Initial mass function (796)}


\section{Introduction} 
\label{sec:intro}

Spectroscopic data of stellar systems encapsulate a wealth of information on their physical and chemical characteristics that can be used to derive their past assembly history, and intense efforts have been made to fully exploit them. Since the first spectral indices were defined to quantify the depth of observed spectral features in the '70s and '80s, followed by the widely used optical Lick indices (\citealt{worthey94}, \citealt{trager98}), many others have been added both in the near-UV (e.g.: \citealt{fanelli90}) and IR region (e.g.: \citealt{conroy12a}).  
Exploring larger wavelength ranges generally produces a better constraint on the main stellar population properties such as stellar age, metallicity and elemental abundances. To be noted, however, is that different spectral ranges are sensitive to different types of stars, (for instance, the UV is sensitive to hot stars,
whether main-sequence or evolved), possibly causing problems with degeneracies among stellar parameters (e.g.: \citealt{feldmeier-krause21}). In the last $20$ years, an expanded methodology to derive stellar parameters from spectra has been introduced (e.g.: \citealt{cappellari04}, \citealt{cidfernandes05}, \citealt{sarzi06}, \citealt{koleva09}, \citealt{chevallard16}, \citealt{Gomes17}, \citealt{wilkinson17}, \citealt{conroy18}, \citealt{2018MNRAS.479.2443V}, \citealt{Johnson21}), which is now widely accepted (e.g.: \citealt{koleva08, ge19, goncalves20}):  
full spectral fitting (hereafter, FSF). The advantage of FSF is that, in the comparison with models with different stellar parameters, not only particular regions of the spectra are taken into account but each (good) pixel of the spectrum contributes to the fit. 
Since spectra usually have large numbers of pixels, this allows the fit to include many free parameters, such as the presence of a second stellar component with different characteristics (e.g. the stellar age, \citealt{conroy18}). On the other hand, FSF has some drawbacks. For example, it is not possible to determine where the information for a given fitting parameter originates within the fitted spectrum, as all pixels contribute equally to the $\chi^2$ computation. Another limitation of FSF is that strong features, highly sensitive to stellar properties, have equal weight in the fit compared to spectral regions containing far less information. As a consequence, 
a hybrid approach between index and full spectral fitting also exists \citep{navarro19, meyer19} and consists of fitting only pixels within selected spectral features known to hold important stellar population information.

Among the stellar population properties, the stellar initial mass function (IMF) plays a leading role due to its importance in many aspects of cosmology and galaxy formation and evolution (see \citealt{hopkins18} for a review). 
Owing to the availability of very high-quality spectroscopic data, the IMF has been measured and characterized for an increasing number of local early-type galaxies, demonstrating its non-universality (e.g.: \citealt{vandokkum11, cappellari12, labarbera13, hopkins13, chabrier14}). In particular, great attention has been devoted to the radial trend of the IMF slope within a given galaxy, which is fundamental to unveiling the processes that lead to the formation of these stellar systems (e.g.: \citealt{navarro15, labarbera16, vandokkum17, alton17, sarzi18, labarbera19}). A general radial trend has been found where galaxies have bottom-heavy IMFs in their center and a more Milky-Way-like IMF in their outskirts. However, different trends have been also observed (e.g.: \citealt{2018MNRAS.479.2443V, feldmeier-krause21}), attesting to the complex nature of galaxy assembly history.   

Importantly, these works have also shown how the measurement of the IMF is difficult despite the availability of high signal-to-noise (S/N) ratio data. 
Indeed, the tiny variations of spectral features potentially due to different IMFs may be masked or themselves mask variations of other stellar population properties such as age, metallicity, and elemental abundances. If not carefully taken into account, all of these dependencies can easily lead to biased results on the retrieved IMF trends, as we demonstrated in \citet{lonoce21} and \citet{feldmeier-krause20}. 
Most importantly, results obtained from different data sets, analysis methods, and assumptions made in the models can exhibit different biases that prevent a comparison of the increasing number of important findings obtained by different studies within the same framework. 

In our previous papers, we have quantitatively shown the biases that occur when extracting information about the IMF using different methods and assumptions \citep{feldmeier-krause20, lonoce21}. For instance, we tested the role of elemental abundance in index fitting. We found that, in order to avoid severe biases on the IMF, once a set of spectral indices to retrieve the IMF slopes is chosen, all the other parameters, including all the elemental abundances that the indices are sensitive to, must be well-constrained; we compared index versus full spectral fitting demonstrating that when fitting all the necessary elemental abundances together, the IMF retrieval from index and FSF is consistent, with generally the FSF giving more precise results.  We also showed and quantified how the retrieved IMF differs when the information is extracted from different wavelength ranges, and verified that the inclusion, for example, of the NIR region $>8000$\AA\space improves the results (when also fitting elemental abundances). In this paper, our goal is to test the robustness of FSF analysis and highlight which of the many stellar population properties can or cannot be reliably retrieved. We compare and contrast the results from four different FSF codes, paying particular attention to the IMF. We also test the commonly made assumption of a simple stellar population and explore the case where multiple stellar components are involved. 

A few other works have addressed the problem of testing different FSF codes under different assumptions and on the same data set (e.g.: \citealt{Ge18, ge19, saracco23}),
finding in some cases different results.
In particular, \citet{Ge18} fitted mock spectra with \textsc{pPXF} and \textsc{Starlight} to test the retrieval of age, metallicity, dust extinction, and M/L with different S/N. They found that \textsc{pPXF} gives more accurate and precise results than \textsc{Starlight} for E(B-V) $\gtrapprox 0.2$; nevertheless, \citet{cidfernandes18} showed that with a better tuning of its initial setup, \textsc{Starlight} results can be efficaciously improved. \citet{ge19} focused on \textsc{pPXF}'s performance under different assumptions for the IMF, isochrones, and spectral libraries, and found biases on the order of $\sim0.1-0.2$ dex on metallicity and M/L values. \citet{saracco23} compared the results of fitting the same observed spectra with \textsc{Starlight} and \textsc{pPXF} and found an average systematic shift of $\sim0.5$ Gyr and $0.05$ dex for age and metallicity, respectively.
However, to our knowledge, no one has focused on the retrieval of the IMF slope.

We chose four public FSF codes for our comparative analysis: \textsc{ALF} (Absorption Line Fitter, \citealt{conroy12a, conroy18}), \textsc{PyStaff} \citep[Python Stellar Absorption Feature Fitting,][]{2018MNRAS.479.2443V},  \textsc{Starlight} \citep{cidfernandes05} and \textsc{pPXF} (Penalized PiXel-Fitting, \citealt{2017MNRAS.466..798C}). 
These are well-established and widely used codes to derive stellar population parameters and star formation histories (SFHs). Most importantly, they allow the retrieval of the IMF slope. Other codes are available in the literature, for example: Firefly \citep{wilkinson17}, Prospector \citep{Johnson21}, Fado \citep{Gomes17}, Paintbox \citep{barbosa21} but they either do not allow the determination of the IMF or are still in a testing phase and thus not yet widely used.

We have directly compared the results from the analysis of data for the same galaxies, in each case applying the same models (i.e. simple stellar populations by \citealt{conroy18}) under the same set of assumptions. We based our analysis on high-S/N ($>100$ \AA$^{-1}$) spectroscopic data of the two elliptical local galaxies NGC1399 and NGC1404, both members of the Fornax cluster. These two objects constitute a good sample for this analysis because their stellar population properties, including the IMF and elemental abundances, have already been studied by \citet[NGC1399]{2018MNRAS.479.2443V}  and \citet[NGC1404, with the same data used in this work]{feldmeier-krause21}, offering a parallel and partially external comparison with our results. Moreover, being a close pair, a comparison of their retrieved stellar population properties can also help to understand whether their formation and evolution have been affected by their proximity or their location in the Fornax cluster.  

The paper is organized as follows: in Section \ref{sec:data} we present our targets, the observations and the data reduction process; in Section \ref{sec:analysis} we detail the fitting set-up for the data and simulations, and the characteristics of the four FSF codes; in Section \ref{sec:results} we comment on our results and in Section \ref{sec:discussion} we discuss them in light of simulations and literature results ; in Section \ref{sec:conclusions} we conclude with a summary of this work and our final conclusions. The appendix includes: A) a description of our results in terms of elemental abundances, B) all the details on the simulations results, C) a discussion about residuals, and D) a focus on the obtained cross-correlations among retrieved parameters.


\section{Spectroscopic Data}
\label{sec:data}

Our targets, the two elliptical galaxies NGC1399 and NGC1404, belong to the Fornax cluster at a distance of $\sim19$~Mpc. NGC1399 is the brightest cluster galaxy with a total B band magnitude of $10.33$ mag and effective radius R$_e\sim5$ kpc. NGC1404, which lies only $\sim10$ arcmin from NGC1399, is slightly less bright (M$_B=10.90$ mag) and significantly smaller (R$_e\sim2.2$ kpc, \citealt{ho11}). Relevant information on the two sources is listed in Table \ref{tab:info}.
Because of their proximity in the sky, the two galaxies were observed simultaneously during the nights of 2019 December 3-4, with the Inamori-Magellan Areal Camera \& Spectrograph (IMACS, \citealt{dressler06}) on the Magellan-Baade 6.5m telescope at Las Campanas Observatory (Chile). The $15$ arcmin slit length with the choice of $2.5$ arcsec slit width allowed us to obtain spectroscopic data with spectral resolution R$\sim1000$ out to a few effective radii for both objects. The observed position angle was $\sim 152$\textdegree. We observed with two optical grating configurations for $1$ hour each, i.e. $600-8.6$ with grating angle (GA) $10.45$\textdegree\space and $9.70$\textdegree, which resulted in a continuous spectral range of $3500-7000$\AA\space without chip gaps. We also observed in the near-IR for $40$ min, using the $600-13.0$ grating with GA$=17.11$\textdegree\space to cover the Calcium Triplet region around $8000$\AA.    

\begin{table*}
 \caption{Main information on the galaxy pair NGC1399 and NGC1404: }
 \label{tab:info}
 \begin{tabular}{lcccccc}
 \hline
 \hline
 Galaxy  & RC3 Type  & Distance    & R$_e$    & Major axis P.A.  & M$_B$  & $\sigma_{center}$ \\
         &           & (Mpc)       & (arcsec)  & (degree)         & (mag)  &    (km s$^{-1}$)           \\
 \hline
 NGC 1399   & E1  & 18.9 & 54.6  & 78.3  &  10.33 & 375 \\
 NGC 1404   & E1  & 18.6 & 24.6  & 163.6 &  10.90 & 260  \\
 \hline
\end{tabular}
\vspace{0.15cm}
\\
 \textbf{Note.} Type, distance, effective radius, position angle and magnitude are taken from the Carnegie-Irvine Galaxy Survey database \citep{ho11}. Optical central velocity dispersion is from \citet{vanderbeke11}.
\end{table*}

\subsection{Data Reduction} 

The raw data for NGC1399 were reduced using \textsc{IRAF} \citep{tody93} and custom scripts. We removed cosmic rays and bad pixels using the \textsc{IRAF} tasks \textsc{COSMICRAYS} and \textsc{IMEDIT}. The bias was estimated from the overscan region of each exposure and subtracted. 

Calibration to air wavelengths was performed simultaneously with the correction of distortion along chip rows. Using the \textsc{IRAF} tasks \textsc{IDENTIFY}, \textsc{REIDENTIFY}, and \textsc{FITCOORDS}, we traced several emission lines on the He/Ne/Ar arc lamp exposures and matched them to their known wavelengths. We then applied the resulting calibration to our data using the \textsc{IRAF} task \textsc{TRANSFORM}. The data were flat-fielded and background subtraction was performed using the \textsc{IRAF} task \textsc{BACKGROUND}. Due to the long length of the instrument slit, we were able to estimate the background directly from the science exposures by taking all flux beyond 2R$_e$ from the center of the galaxy. The three science exposures, each $20$ min long, were then combined to produce one image per chip and GA.

We extracted the one-dimensional spectra from each image using a custom \textsc{IDL} script. To facilitate our investigation of radial trends, spectra were extracted in multiple radial bins: within R$_e/100$, R$_e/100$-R$_e/30$, R$_e/30$-R$_e/15$, R$_e/15$-R$_e/10$, R$_e/10$-R$_e/7$, R$_e/7$-R$_e/5$, R$_e/5$-R$_e/3$, R$_e/3$-$1$R$_e$ (NGC1399) and within R$_e/100$, R$_e/100$-R$_e/30$, R$_e/30$-R$_e/15$, R$_e/15$-R$_e/10$, R$_e/10$-R$_e/5$, R$_e/5$-R$_e/3$, R$_e/3$-R$_e/2$, R$_e/2$-$3/4$R$_e$, $3/4$R$_e$-$8/5$R$_e$ (NGC1404). We applied flux calibrations derived from observations of the standard star HD40285 using the \textsc{IRAF} task \textsc{CALIBRATE}. 

\begin{figure*}[ht!]
\begin{centering}
\includegraphics[width=18.5cm]{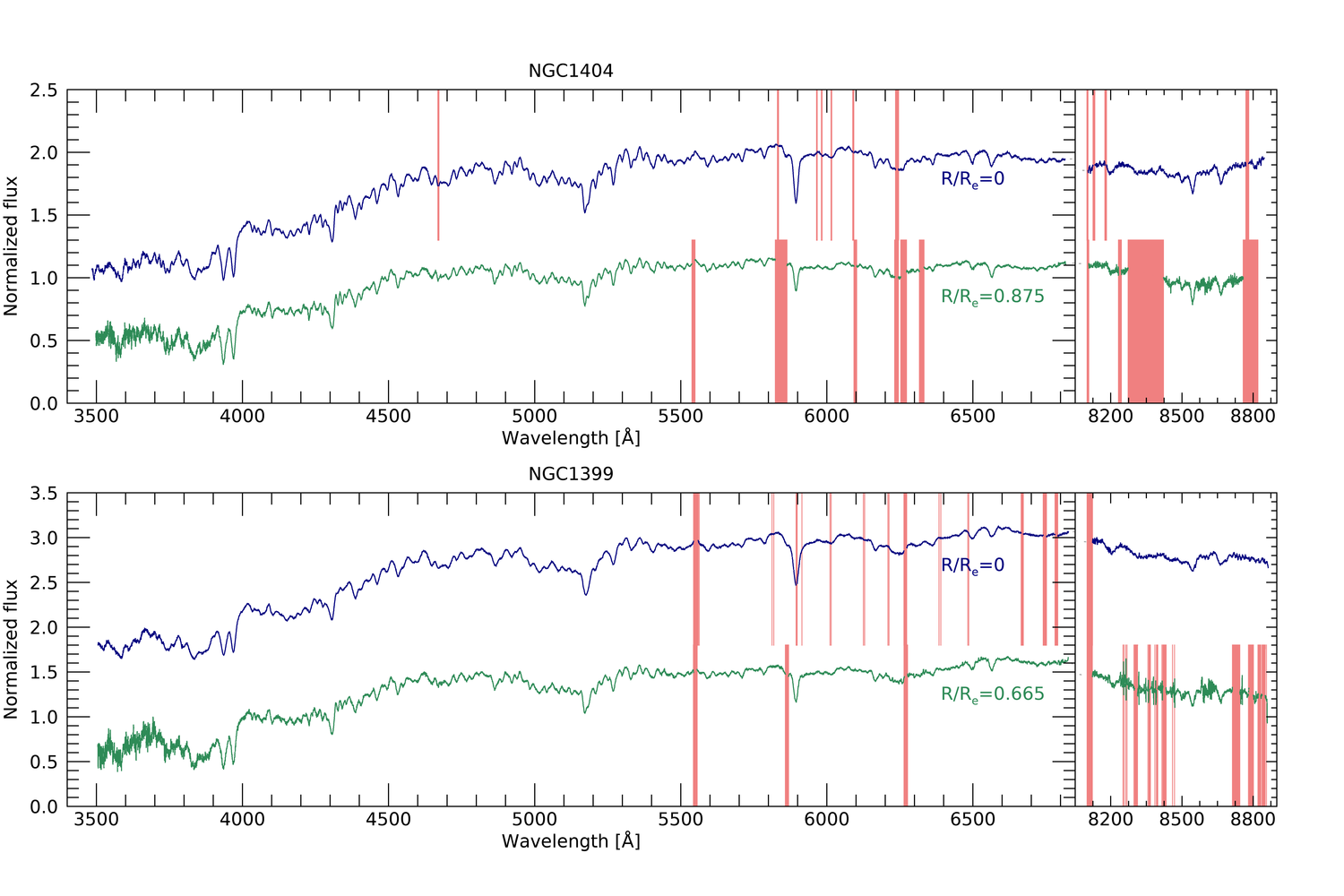}
\caption{\small{Examples of reduced spectra for NGC1404 (upper panel) and NGC1399 (lower panel). Blue spectra are the innermost radial bin, while green ones are the outermost. Bad pixels have been masked as indicated by pink vertical lines. Wavelengths are at rest-frame, optical region on the left, NIR region on the right.}}
\label{fig:spectra}
\end{centering}
\end{figure*}

As certain regions of interest in our spectra were affected by telluric absorption lines, we used ESO’s \textsc{MOLECFIT} \citep{2015A&A...576A..78K,2015A&A...576A..77S} to correct for this absorption. Despite our best efforts, we were not always able to remove the absorption completely. As can be seen in Figure \ref{fig:spectra}, there are residuals in the infrared portion of the spectrum, especially in the outer radial bins. These residuals are accounted for in the per-pixel noise spectrum, which we provide as input to each FSF code.

In the near-infrared (NIR) region, our spectroscopic data were affected by fringing. We refer to \citet{lonoce21} for the full description of how we removed the fringes.

To measure and correct the radial velocity gradient of the galaxy, we used the \textsc{IDL} version of \textsc{pPXF}. Specifically, \textsc{pPXF} produced a radial velocity curve that we applied as a correction to shift all of the spectra to the rest frame. We then combined the spectra above and below the center of the galaxy to produce a single spectrum for each radial bin. 

\begin{figure*}[ht!]
\begin{centering}
\includegraphics[width=8.8cm]{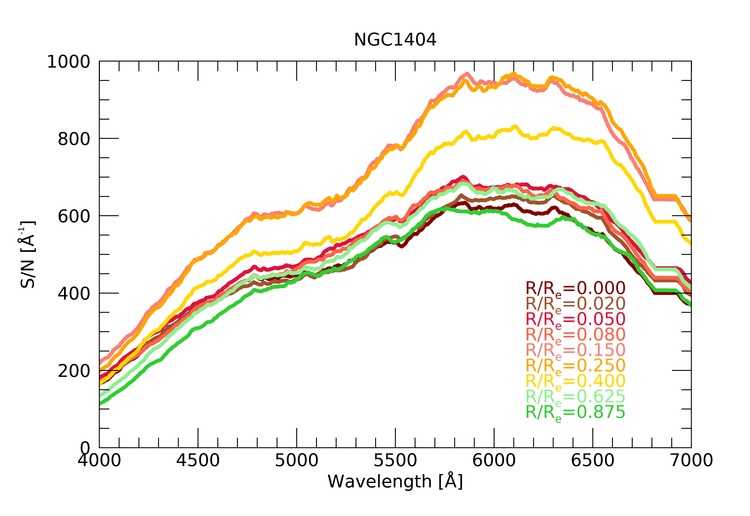}
\includegraphics[width=8.8cm]{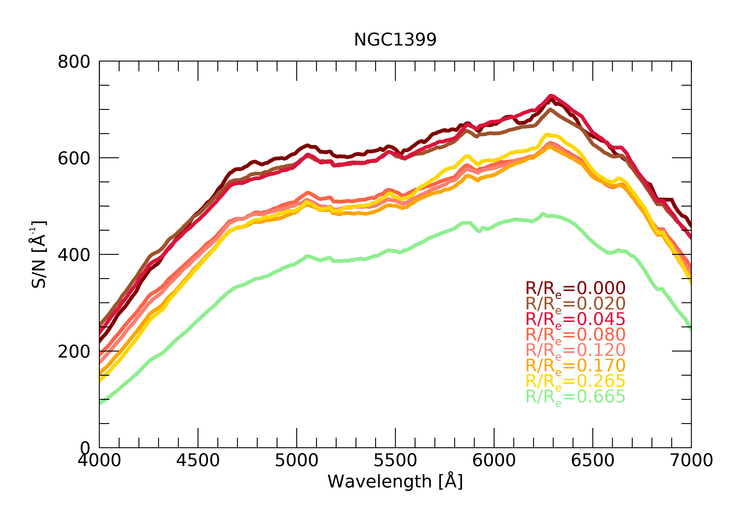}
\caption{\small{Signal-to-noise trends in the optical regions measured on the final spectra of each radial bin. NGC1404 on the left, NGC1399 on the right. In the NIR region, the trends span from $400$ to $100$ \AA$^{-1}$} from the inner to the outer radii.}
\label{fig:sn}
\end{centering}
\end{figure*}

The S/N curves for the final spectra are shown in Figure \ref{fig:sn}. We note that these curves are intended to present mean trends at different radii, so they have been smoothed significantly. When deriving the error for fitted parameters later in our analysis, the per-pixel noise is taken into account rather than these smoothed spectra. We note that the obtained S/N curves are also all above $100$ \AA$^{-1}$ in the NIR region (not shown). 

In the case of NGC1404, the spectroscopic data were presented in \cite{feldmeier-krause21}. The data reduction, described in more detail in that paper, was performed in a similar way to NGC1399. Examples of the reduced spectra of both NGC1404 and NGC1399 are shown in Figure \ref{fig:spectra}.

\section{Analysis}
\label{sec:analysis}

We performed FSF of the spectroscopic data of NGC1399 and NGC1404 (described in Section \ref{sec:data}) with four different codes publicly available in the literature: \textsc{ALF}, \textsc{PyStaff}, \textsc{Starlight} and \textsc{pPXF} (presented below). The choice of these codes was based on their ability to retrieve not only the age and metallicity of the stellar population but also its IMF slope. To our knowledge, only these four public codes allow this. Fortunately, we can also directly compare two parametric FSF codes (i.e. \textsc{ALF} and \textsc{PyStaff}) and two non-parametric FSF ones (i.e. \textsc{Starlight} and \textsc{pPXF}). The difference is that parametric FSF codes adopt an assumption about the star formation history, in this case, a simple stellar population (SSP), while non-parametric FSF codes fit a linear combination of SSPs and retrieve the minimum $\chi^2$ solution, which takes the form of a composite stellar population with a characteristic star formation history. In this latter case, the stellar parameters obtained are in the form of light/mass-weighted values.
As hinted in the introduction (Section \ref{sec:intro}), with this test we want to answer the question: what are the fitted stellar population parameters that can be robustly retrieved with FSF regardless of the code choice? \\

To produce comparable results from the four fitting runs, we adopted the same models and similar settings, as described below and further detailed in Tables \ref{tab:bounds}, \ref{tab:emline} and \ref{tab:section}:
\begin{itemize}
    \item Conroy et al. 2018 models \citep{conroy18}: covering the $3500-9000$\AA\space spectral range with a spectral resolution of $100$ km s$^{-1}$.
    
    \item Response functions: we applied the Conroy et al. 2018 response functions \citep{conroy18} to obtain non-solar values of the following $19$ elemental abundances: Fe, O, C, Ca, Na, N, Mg, Ti, Si, K, V, Cr, Mn, Co, Ni, Cu, Sr, Ba and Eu. The response functions cover the same wavelength region as the models and the same spectral resolution (i.e. $100$ km s$^{-1}$) and are provided for a Kroupa IMF \citep{kroupa}. Values of elemental abundances provided by the response functions are typically in the range between $-0.3$ to $+0.3$ dex, with the exception of C ($-0.15$ to $+0.15$) and Na ($-0.3$ to $+1.0$), but they have been extended to higher and/or lower values whenever the retrieved parameter has been found at its range limits.  
    Regarding interpolation and extrapolation to values of elemental abundances not provided by the response functions, the fitting codes have different approaches: the code \textsc{ALF} performs linear interpolation/extrapolation, while the code \textsc{PyStaff} applies a Taylor expansion near the reference value. Extrapolated response functions can thus be quite different for the two codes. A quantitative analysis of the differences that occur when applying a linear vs Taylor extrapolation can be found in \cite{feldmeier-krause21}.
    See Table \ref{tab:bounds} for the elemental abundance limits used in each fit.
    Not all FSF codes allow for the retrieval of elemental abundances, so we applied response functions only to \textsc{ALF} and \textsc{PyStaff} fits.  
    In Appendix \ref{app:elem} we discuss how the results for age, metallicity and IMF would change when fitting with a set of templates built with non-solar elemental abundances.

    \item Age and metallicity: the age values provided by the adopted models go up to $13.5$ Gyr, and [Z/H] ranges from $-1.50$ to $+0.2$ dex. However, the actual boundaries used in each fit are listed in Table \ref{tab:bounds}. Some cuts in the templates were necessary since, for example, \textsc{Starlight} allows only a maximum number of $300$ base models. In this case, we penalized the very young ages $<5$ Gyr.
    
    \item IMF slope: Conroy et al. 2018 models follow the IMF parametrization $dN/dm \propto m ^{-x_i}$, where X1 and X2 describe the two IMF slopes in the mass ranges $0.08-0.5$M$_{\odot}$ and $0.5-1.0$M$_{\odot}$, respectively. Above $1$M$_{\odot}$ the Salpeter slope value holds \citep{salpeter}. Due to the mutual degeneracy of X1 and X2 during FSF fits observed in \citet{lonoce21}, we decided to fix X1=X2, thus focusing on the results of a single IMF slope within $0.08-1.0$M$_{\odot}$.

    \item Wavelength range: we fitted the spectra of both galaxies in the rest-frame wavelength range $4040-6640$\AA\space and $8143-8600$\AA. 
    To account for systematic uncertainties in the choice of the wavelength range in the fit, we run a full fit of each spectrum twice, with slightly different wavelength ranges. The ranges are detailed in Table \ref{tab:section}. Note that the first range includes the Balmer transitions H$\delta$ and H$\alpha$, while the second range does not. The inclusion or omission of these features influences the gas emission fit, which correlates with the best-fit age, which correlates with the best-fit metallicity \citep{feldmeier-krause21}. To account for this uncertainty, we used our results from the two different ranges to compute the standard deviation as systematic uncertainty.
        
    \item Gas component: three out of the four FSF codes allowed us to include in the fit a non-stellar component, i.e. a gas component which shows emission lines. The optical emissions we took into account are listed in Table \ref{tab:emline}. For \textsc{Starlight}, the only code that does not fit emission lines, we masked the regions likely to be affected by gas emission. 

    \item Bad pixel mask: for each spectrum, we created a mask that flagged uncertain pixels (such as outliers, gaps, and regions directly adjacent to gaps). These masks were provided as input to all four FSF codes. For \textsc{Starlight}, this mask was combined with the gas emission mask described above.
    
    \end{itemize}

\subsection{\textsc{PyStaff}}
\label{sec:pystaff}

\textsc{PyStaff} \citep[Python Stellar Absorption Feature Fitting,][]{2018MNRAS.479.2443V} is a \textsc{python} code available on \textsc{GitHub} \footnote{\textit{https://github.com/samvaughan/PyStaff}}. It utilizes several features of \textsc{pPXF} \citep[Penalized PiXel-Fitting,][]{2017MNRAS.466..798C} to extract the stellar and gas kinematics. The posterior is sampled using \textsc{emcee} \citep{2013PASP..125..306F}. 

The spectrum is divided into four wavelength sections, and in each section a multiplicative Legendre polynomial is fit to account for small differences in continuum shape between model and data. Such differences may be caused by the instrument or flux calibration problems. Dividing the spectrum into four regions makes finding the polynomials easier and the code faster. 

We set up the code to simultaneously fit the following SSP parameters: age, metallicity, 19 elemental abundances, and one low-mass IMF slope from 0.08-1.0\,M$_{\odot}$. In addition, the code derives the stellar radial 
velocity $V$, velocity dispersion $\sigma$, mean gas velocity, gas velocity dispersion, and a velocity offset term for the fourth wavelength section, as there is a gap of more than 1500\,\AA\space to the third section. In addition, we fit the Balmer emission line flux, but keep the flux ratios between the different transitions fixed (see Table \ref{tab:emline}). The same is done for the [N I], [N II], and [O III] emission line doublets; we use the theoretical flux ratios of the doublets, and thus have only one fitting parameter per species. Further details can be found in Tables \ref{tab:bounds}-\ref{tab:section}.

\begin{table*}
\begin{centering}
 \caption{Fitting bounds }
 \label{tab:bounds}
 \begin{tabular}{lcccc}
 \hline
 \hline
& \textsc{PyStaff}  & \textsc{ALF}  & \textsc{Starlight}  & \textsc{pPXF}  \\ 
 \hline
 age [Gyr]         & 1.0--14.0 & 0.5--14.0  & 5.0--13.5 & 1.0--13.5 \\
 \text{[Z/H]} [dex]& -1.5--0.3 & -1.8--0.3  & -1.0--0.2 & -1.5--0.2 \\
 IMF slope         & 0.5--3.5  & 0.5--3.9   & 0.7--3.5  & 0.5--3.5 \\
\hline 
\text{[Fe/H]} [dex] & -0.45--0.45 & -0.5--0.6 &-&-  \\
\text{[Na/H]} [dex] & -0.45--1.0  & -0.5--1.0 &-&- \\
\text{[C/H]} [dex] & -0.2--0.3    & -0.5--0.6 &-&- \\
\text{[Ca/H]} [dex] & -0.45--0.45 & -0.5--0.6 &-&- \\
\text{[N/H]} [dex] & -0.45--0.45  & -0.5--1.0 &-&- \\
\text{[Mg/H]} [dex] & -0.45--0.45 & -0.5--0.6 &-&- \\
\text{[Si/H]} [dex] & -0.45--0.45 & -0.5--0.6 &-&- \\
\text{[Ti/H]} [dex] & -0.45--0.45 & -0.5--0.6 &-&- \\
\text{[Ba/H]} [dex] & -0.45--0.45 & -0.5--0.6 &-&- \\
\text{[O/Fe]} [dex] & 0.00--0.45  & -0.5--0.6 &-&- \\
\text{[Cr/H]} [dex] & 0.00--0.45  & -0.5--0.6 &-&- \\
\text{[Mn/H]} [dex] & 0.00--0.45  & -0.5--0.6 &-&- \\
\text{[Ni/H]} [dex] & 0.00--0.45  & -0.5--0.6 &-&- \\
\text{[Co/H]} [dex] & 0.00--0.45  & -0.5--0.6 &-&- \\
\text{[Eu/H]} [dex] & 0.00--0.45  & -0.5--0.6 &-&- \\
\text{[Sr/H]} [dex] & 0.00--0.45  & -0.5--0.6 &-&- \\
\text{[K/H]} [dex] & 0.00--0.45   & -0.5--0.6 &-&- \\
\text{[V/H]} [dex] & 0.00--0.45   & -0.5--0.6 &-&- \\
\text{[Cu/H]} [dex] & 0.00--0.45  & -0.5--0.6 &-&- \\
\hline
$V$ [km$\cdot$s$^{-1}$] & -100--300                 & -100--300 & -10--100 & -2000--2000\\
$\sigma$ [km$\cdot$s$^{-1}$] & 70--400              & 70--400  & 100--400 & $\sim$0--1000\\
$V_{gas}$ [km$\cdot$s$^{-1}$] & $V \pm$25           & -100--300 & - & =V\\
$\sigma_{gas}$ [km$\cdot$s$^{-1}$] & $\sigma \pm$25 & 10--400   & - & =$\sigma$ \\
\hline
\end{tabular}
\end{centering}
\end{table*}

\begin{table*}
\begin{centering}
 \caption{Emission line treatment}
  \label{tab:emline}
 \begin{tabular}{ll}
 \hline
 \hline
Emissions & Relative strengths formula  \\ 
\hline
Balmer lines  &H$\delta$=0.259 H$\beta$, \\
              &H$\gamma$=0.468 H$\beta$, \\ 
              &H$\alpha$=2.86 H$\beta$ (\citealt{1989agna.book.....O, 1997MNRAS.291..403R}) \\
\text{[O III]} $4960$\AA, $5008$\AA\  &\text{[O III]} $4960$\AA=0.33$\cdot$\text{[O III]} $5008$\AA \\
\text{[N I]} $5203$\AA\  &\text{[N I]} $5199$\AA=0.658$\cdot$\text{[N I]} $5202$\AA \\
\text{[N II]} $6550$\AA, $6585$\AA\ &\text{[N II]} $6550$\AA=0.33$\cdot$\text{[N I]} $6585$\AA \\
\hline
\textsc{PyStaff}    & Included \\
\textsc{ALF}        & Included \\
\textsc{Starlight}  & Masked   \\
\textsc{pPXF}       & Included \\
\hline
\end{tabular}
\end{centering}
\end{table*}

\begin{table*}
\begin{centering}
 \caption{Fitting setup }
 \label{tab:section}
 \begin{tabular}{lcccc}
 \hline
 \hline
& \textsc{PyStaff}  & \textsc{ALF}  & \textsc{Starlight} & \textsc{pPXF} \\ 
 \hline
Number of $\lambda$ sections & 4 & 2 & 1 & 1\\
Wavelength range 1 \text{[\AA]}& 4040--4975 & 4040--6640 & 4040--6640 & 4040--6640 \\
                               & 4975--5843 & 8143--8600 & 8143--8600 & 8143--8600\\
                               & 5843--6640 & & &\\
                               & 8143--8600 & & &\\
Wavelength range 2 \text{[\AA]}& 4251--4975 & 4251--6487 & 4251--6487 & 4251--6487\\
                               & 4975--5843 & 8184--8643 & 8184--8643 & 8184--8643\\
                               & 5843--6487 & & &\\
                               & 8184--8643 & & &\\

Polynomial fit & Multiplicative
               & Multiplicative 
               & No correction
               & Additive--Multiplicative\\
Order &  
         ($\lambda$$_{upper}$--$\lambda$$_{lower}$)/100& ($\lambda$$_{upper}$--$\lambda$$_{lower}$)/100  & - & 4--10\\
Number of walkers & 100 & 1,024 &300 & -\\
Number of steps & 10,000 & 100 &7 & -\\
Burn-in & 5,000 & 10,000 & threshold =1.2 & -\\    
Spectral resampling  \text{[\AA]} & 1.25& - & - & -\\
Resolution change & - & - & to constant $\sigma_{inst}$ & to $\Delta\lambda$=6.85\AA\space at $\lambda>7638$\AA\\
\hline
\end{tabular}
\end{centering}
\end{table*}

\subsection{\textsc{ALF}}
\label{sec:alf}

\textsc{ALF}, the Absorption Line Fitting code by \citet{conroy18} (publicly available on \textsc{GitHub} \footnote{\textit{https://github.com/cconroy20/alf}}), has been designed to constrain the stellar population properties of massive and old galaxies, with a special focus on the IMF, which can be retrieved with different parametrizations. \textsc{ALF} works in continuum-normalized space, meaning that the shape of the continuum is not used to derive information. Similarly to \textsc{PyStaff}, the fitting process is performed by means of the ensemble MCMC sampler \textsc{emcee} \citep{2013PASP..125..306F}. 
A high-order polynomial fit is applied to the ratio between the continuum and model in each wavelength section, with the order $n$ defined by $n=(\lambda_{upper}-\lambda_{lower})/100$.

Used in the ``full mode", \textsc{ALF} simultaneously retrieves the following parameters: stellar radial 
velocity, velocity dispersion, stellar age, age of a second younger component, the mass fraction of the younger component, [Z/H], $19$ elemental abundances, logarithmic slopes of the IMF in the low-mass ranges $0.1-0.5$ M$_{\odot}$ and $0.5-1.0$ M$_{\odot}$, 
several emission line strengths
, gas kinematics including gas velocity dispersion and radial velocity offset from the stellar component, 
a jitter multiplicative term to enlarge the errors, and two additional terms to correct the quoted errors in regions where sky and telluric residuals can be present. 
Although \textsc{ALF} permits fitting of a secondary stellar component with a different (younger) age (but with all other parameters the same), in order to compare the results with the other codes (in particular with \textsc{PyStaff}), we only allowed for a single age component (SSP) and a single IMF slope, as discussed previously.

\subsection{\textsc{Starlight}}
\label{sec:SL}

\textsc{Starlight} \citep{cidfernandes05}, a non-parametric Fortran 77 public code available at \textit{www.starlight.ufsc.br}, fits an observed spectrum by creating the minimum $\chi^2$ model built as a composition of SSP components extracted from a set of base spectra (in our case, Conroy et al. 2018 models). 

\textsc{Starlight} retrieves radial velocity and velocity dispersion, dust extinction (with a large choice of extinction laws, we adopted \citealt{calzetti00}), galaxy masses, luminosity/mass ratios, age, metallicity, and [$\alpha$/Fe]. It also retrieves the stellar parameters like age and metallicity under the assumption of a SSP, where the best superposition of models is traded for the best single population one.
Since we were interested in retrieving the IMF slope and not the $\alpha-$element value, we substituted this new parameter by putting IMF slope values in the input file column describing the base templates. Differently from \textsc{PyStaff} and \textsc{ALF}, with \textsc{Starlight} we could not retrieve any information on the chemical content of our galaxies besides their total metallicity.

Unlike the other three codes, \textsc{Starlight} does not fit for a gas component. Instead, we excluded regions that could be affected by gas emission lines by adding them to the bad pixel mask for each spectrum.

In order to use \textsc{Starlight} correctly, an input spectrum with a constant instrumental velocity dispersion along the wavelength range is required. We accordingly transformed our observed spectra from a constant instrumental full-width-half-maximum ($\sim6$\AA) to a constant velocity dispersion of $228.8$ km s$^{-1}$ (NGC1404) and $207.4$ km s$^{-1}$ (NGC1399).  

Statistical errors have been estimated by repeating the fit on $100$ Monte-Carlo noise realizations of each spectrum. For each pixel of the spectrum, a random noise value, extracted by a Gaussian distribution centered at zero and with a width of the value of the noise spectrum, has been added. The standard deviation calculated from the distribution of values retrieved for each parameter gave the statistical error on the light-weighted age, metallicity and IMF slope. 
Systematic errors were computed by repeating the fit on a shorter wavelength range (see Table \ref{tab:section}, wavelength range 2). The systematic errors were obtained by calculating the standard deviation of the weighted-means of the two fits over the different wavelength ranges. The final errors, shown in Figures \ref{fig:results14} and \ref{fig:results13}, were obtained by adding the statistical and systematic uncertainties in quadrature.

\subsection{\textsc{pPXF}}
\label{sec:ppxf}

Similarly to \textsc{Starlight}, the Penalized PiXel-Fitting code (\textsc{pPXF}, \citealt{cappellari04, 2017MNRAS.466..798C}), is a non-parametric FSF code that combines stellar templates and gives the best composite solution without an a-priori assumption on the SFH. The \textsc{Python} or \textsc{IDL} versions of \textsc{pPXF} are available at \textit{https://www-astro.physics.ox.ac.uk/$\sim$cappellari/software/}. \textsc{pPXF} uses a maximum penalized likelihood method in the pixel space to fit the observed spectra and extracts both the galaxy stellar kinematics (radial velocity, velocity dispersion and four Gauss–Hermite coefficients) as well as stellar parameters as included in the templates input library. To save time, we did not include non-solar elemental abundances values, and focused on the retrieval of age, metallicity and IMF slope. 

Differently from \textsc{Starlight}, \textsc{pPXF} allows the fitting of a gas component with independent kinematics. Moreover, for the stellar parameters, it performs a regularization of the solutions (i.e. smoothing the template weights during the fit) which helps in constraining the galaxy SFH. 

To deal with model templates, \textsc{pPXF} requires a constant spectral resolution $\Delta\lambda$ along the wavelength range. As stated before, \citet{conroy18} models have a constant velocity dispersion of $100$ km s$^{-1}$ over $\lambda$. We thus converted the constant velocity dispersion of templates to a constant $\Delta\lambda=6$\AA, the same as our observed data. However, at $\lambda>7638$\AA, we had to increase the $\Delta\lambda$ value to $6.85$\AA. To be consistent, we also downgraded the red region ($\lambda>7638$\AA) of each observed spectrum before fitting with \textsc{pPXF}. We carefully tested the downgrading process and verified that extracting the kinematics by fitting the same spectra with EMILES models (\citealt{vazdekis16}, built with a constant $\Delta\lambda=2.5$\AA) resulted in perfectly consistent kinematics values.  

As suggested in the \textsc{pPXF} documentation, to properly fit our spectra we firstly ran \textsc{pPXF} retrieving only the kinematics and applying a fourth-degree additive polynomial to adjust
the continuum shape of the templates to our observed spectrum. After the kinematics were extracted, we fixed the retrieved values (for both stellar and gas components) and performed another fit focused on the retrieval of stellar parameters and gas emissions. In this second fit, we used a tenth-degree multiplicative polynomial. 
We repeated the fit until the optimal regularization was found. We did not include reddening. 

To account for statistical errors, for each galaxy and for each radial bin, we performed $100$ Monte-Carlo noise realizations of the same spectrum and repeated the fit with the same set-up and regularization. The standard deviation of each distribution of values was considered our statistical noise on the extracted weighted means of the initial fit.
As for the other codes, systematic errors were estimated by repeating the fit over a shorter wavelength range (see Table \ref{tab:section}, wavelength range 2). The final errors were computed by adding the two sources of error in quadrature, as shown in Figures \ref{fig:results14} and \ref{fig:results13}.

\subsection{Simulations}
\label{sec:sim}

Before presenting our results from fitting the real data, here we briefly discuss how we set up simulations to aid in the interpretation of our main results (see Section \ref{sec:discussion}).
We prepared four sets of simulations based on mock spectra with different SFH complexities. The base templates were taken from the \citet{conroy18} models. Information on the stellar parameters of the input spectra is shown in Table \ref{tab:sim}. We started with a SSP with old age and super-solar metallicity, typical of nearby elliptical galaxies. We tested both a bottom-heavy (BH) and a top-heavy (TH) IMF slope since we know from \citet{lonoce21} that the retrieval of stellar properties can be different depending on the IMF slope. We also applied response functions to all our simulated data to have non-solar values of elemental abundances, using as reference the mean values retrieved with \textsc{PyStaff} from real data. In the second set, we increased the complexity of the SFH by adding a secondary ($20$\% of the primary mass) stellar component with a younger age, sub-solar metallicity and equal IMF. We call this set CSP, for composite stellar population. Again we tested both BH and TH IMF slopes. In the last two sets (CSP2 and CSP3) we also mixed the IMF slope values, and changed the fractions of mass of each component, as detailed in Table \ref{tab:sim}.

For each input, we took the desired basic template, downgraded its spectral resolution to match the real data, i.e. $6$\AA, and broadened the spectrum to a constant velocity dispersion of $250$ km s$^{-1}$, typical of elliptical galaxies. We then applied response functions that had been similarly downgraded to the same instrumental resolution and velocity dispersion.
We cut the spectrum to the wavelength range $4040-8600$\AA\space and created a wavelength gap between the optical and NIR region ($6815-8182$\AA), as in our real data. We finally applied a noise spectrum built from our observed noise spectra to create $10$ realizations of the same mock spectrum.

\begin{table*}
\begin{centering}
 \caption{Simulations: mock input spectra properties. All mock spectra have non-solar elemental abundances with values taken from \textsc{PyStaff} results.}
 \label{tab:sim}
 \begin{tabular}{lcccccccc}
 \hline
 \hline
         & \% 1 & \% 2 & Age 1 & Age 2 & [Z/H] 1 & [Z/H] 2 & IMF slope 1 & IMF slope 2 \\
         &      &      & [Gyr] & [Gyr] & dex     & dex     &             &             \\
 \hline
 1. SSP  &  -   &   -  & 13.5  &  -    & 0.20    &  -      & 3.1         & -  \\
         &  -   &   -  & 13.5  &  -    & 0.20    &  -      & 1.5         & -  \\
 2. CSP  &  80  &   20 & 13.5  &  7.0  & 0.20    & -0.50   & 3.1         & -  \\
         &  80  &   20 & 13.5  &  7.0  & 0.20    & -0.50   & 1.5         & -  \\
 3. CSP2 &  80  &   20 & 13.5  &  7.0  & 0.20    & -0.50   & 3.1         & 1.5 \\
 4. CSP3 &  50  &   50 & 13.5  & 11.0  & 0.20    & -1.00   & 1.1         & 3.5 \\
 \hline
\end{tabular}
\end{centering}
\end{table*}

We tested the quality of the parameter retrieval for both kinematics and stellar parameters with the four codes and the same set-up as used for the real data as described at the beginning of this section. When fitting with \textsc{ALF} for the cases of double stellar components, we allowed for a secondary component which could only differ in age, and whose age was confined to be lower than that of the main component.

All plots showing the results obtained from simulation tests are shown and described in Appendix \ref{app:sim}. A brief summary of the main results and a discussion of their implications in our analysis are given in Section \ref{sec:discussion}.


\section{Results}
\label{sec:results}

\begin{figure*}[h]
\begin{centering}
\includegraphics[width=14.0cm]{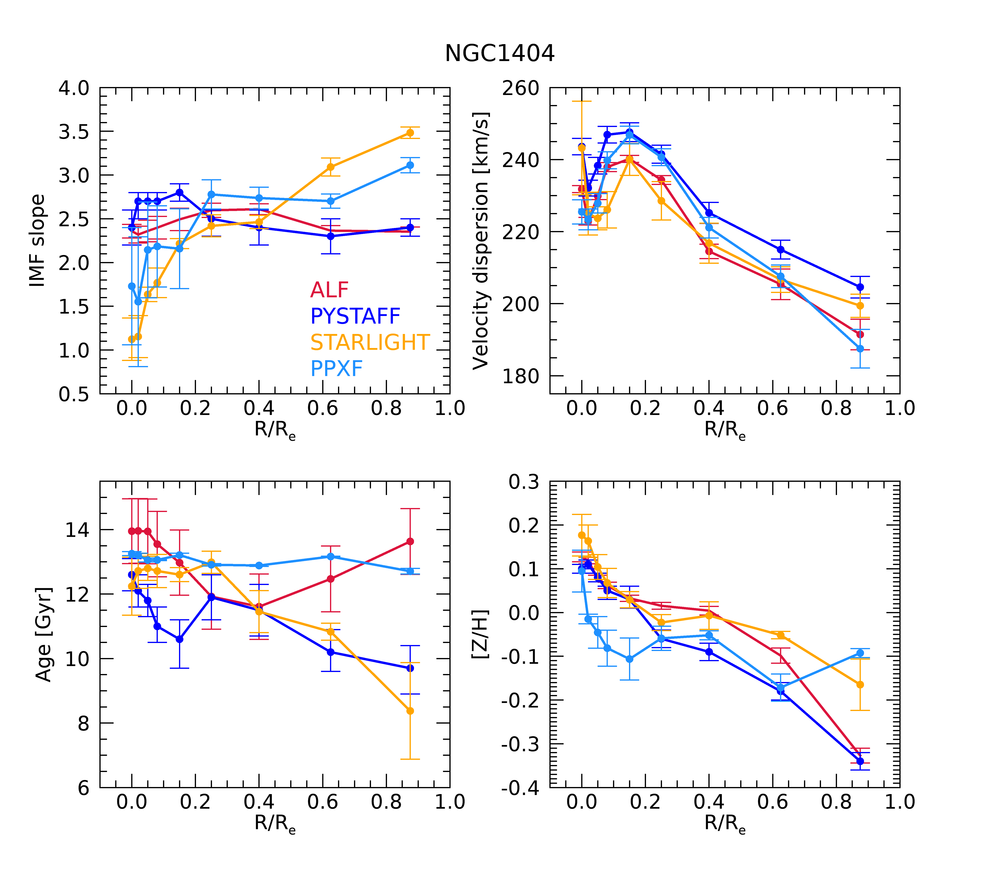}
\caption{\small{NGC1404 results: IMF slope, stellar velocity dispersion, age, and metallicity radial trends as retrieved from the four codes: \textsc{ALF} (red), \textsc{PyStaff} (blue), \textsc{Starlight} (orange) and \textsc{pPXF} (light blue). Points show the weighted means (in particular light-weighted means for \textsc{Starlight} and \textsc{pPXF}) and the error bars include both statistical uncertainties (obtained from MCMC for \textsc{ALF} and \textsc{PyStaff}, and from 100 Monte-Carlo simulations for \textsc{Starlight} and \textsc{pPXF}) and systematic uncertainties (see text, obtained by fitting over a smaller wavelength range). }}
\label{fig:results14}
\end{centering}
\end{figure*}

\begin{figure*}[h]
\begin{centering}
\includegraphics[width=14.0cm]{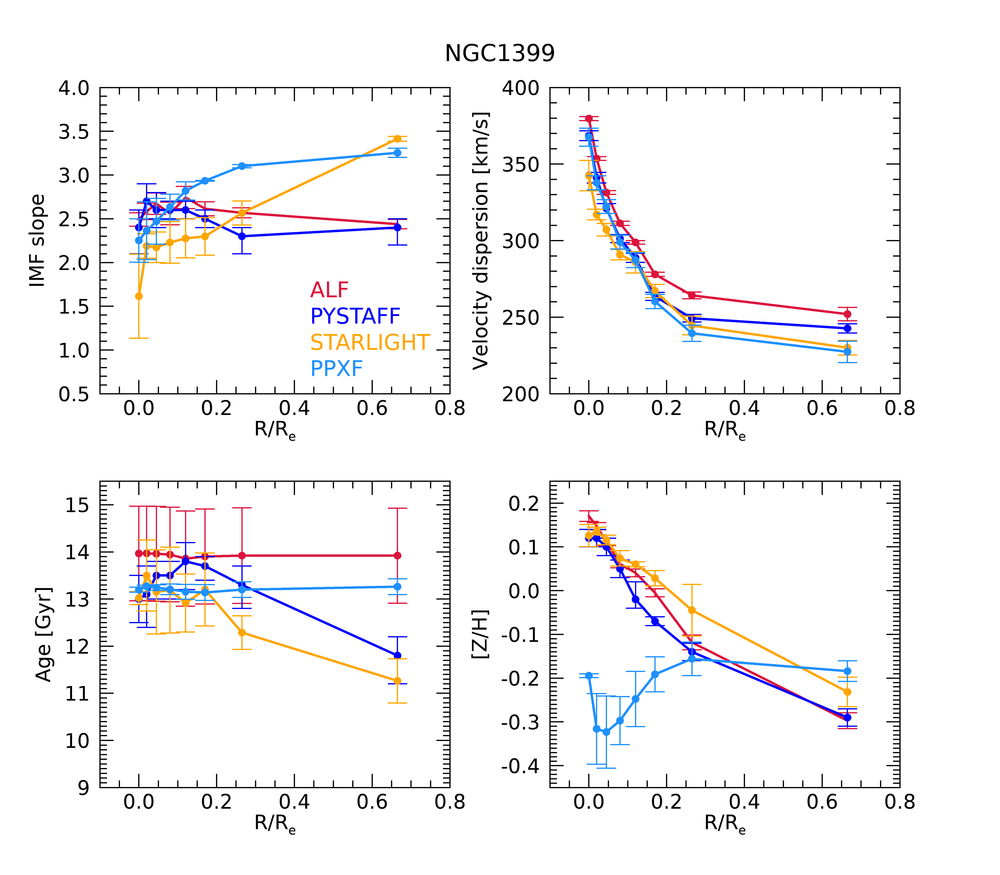}
\caption{\small{NGC1399 results. Same description as for Figure \ref{fig:results14}. }}
\label{fig:results13}
\end{centering}
\end{figure*}

In Figures \ref{fig:results14} and \ref{fig:results13} we show the radial trends of the IMF slope, velocity dispersion, age and metallicity as retrieved from the four codes for NGC1404 and NGC1399, respectively. For \textsc{ALF} and \textsc{PyStaff} (fitting the best SSP) we show the mean values and their uncertainties as obtained from the converged MCMCs, while for \textsc{Starlight} and \textsc{pPXF} (fitting superposition of SSPs) we show the light-weighted means. 

For both galaxies, we found good agreement for the velocity dispersion as retrieved by all codes: NGC1399 shows a sharp decreasing trend, from high values $\sim380$ km s$^{-1}$ in the center toward $\sim240$ km s$^{-1}$ in the outskirts, while NGC1404 presents a milder decreasing trend, from $\sim250$ km s$^{-1}$ to $\sim200$ km s$^{-1}$, with a peculiar bump around R/R$_e\sim0.2$. Radial velocity trends (not shown) showed disagreement among codes, exhibiting  similar trends with radius but with offsets up to $\sim100$ km s$^{-1}$. However, these differences are an expected consequence of the necessary re-sampling/downgrading transformations applied to each code as described in Section \ref{sec:analysis}; we note that such differences cannot affect the retrieval of stellar population properties. \\

For NGC1404, the age is flat with radius on average and consistent with an old population, but there is a large spread among codes, mostly in the outer regions. For NGC1399, with the exception of the outermost radial bin, all codes agree on an old age around $13.5$ Gyr. \\
With the exception of \textsc{pPXF} (light blue),  for both galaxies, the metallicity trend shows good agreement among codes and indicates a negative gradient from $\sim+0.2$ dex to $\sim-0.3$ dex. \textsc{pPXF}'s metallicity values deviate significantly from these trends, in particular for NGC1399. This evident tension between the \textsc{pPXF}'s metallicity trends with respect to the other codes' trends is addressed in Section \ref{sec:discussion} in light of  results from our simulations.\\

Regarding the IMF radial trends, \textsc{ALF} and \textsc{PyStaff} give consistent values for both galaxies, showing a flat trend around $2.5$ for all radial bins. The other two codes suggest a positive radial gradient of the IMF for both galaxies, from TH values in the central regions toward BH IMF slopes in the outer regions (though the large error bars of some radial bins are generally the reason). \\
Elemental abundances, retrieved only from \textsc{ALF} and \textsc{PyStaff}, show general agreement with a few exceptions. The comparison is shown and discussed in Appendix \ref{app:elem}.

\vspace{0.5cm}

\begin{figure*}[h]
\begin{centering}
\includegraphics[width=8.9cm]{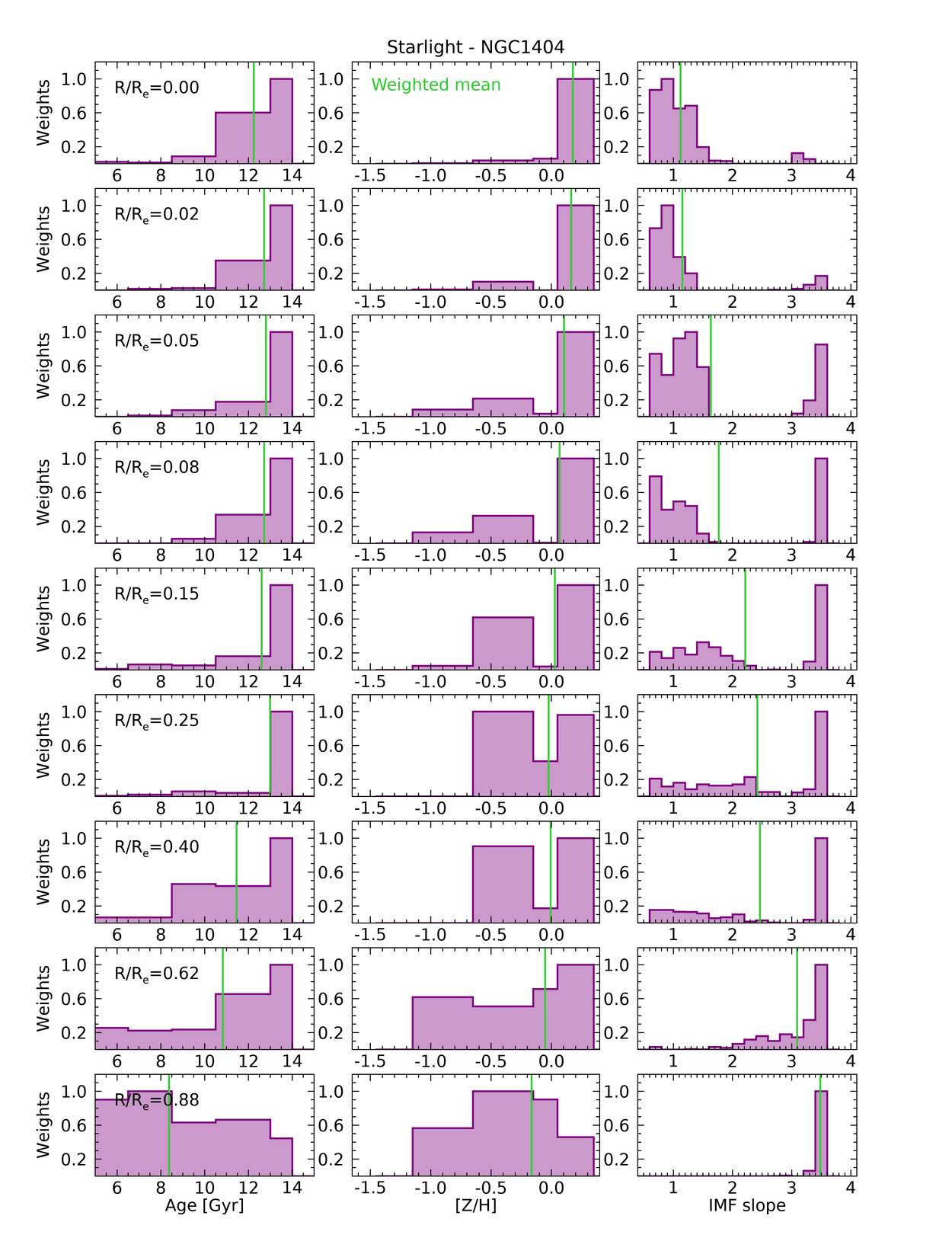}
\includegraphics[width=8.9cm]{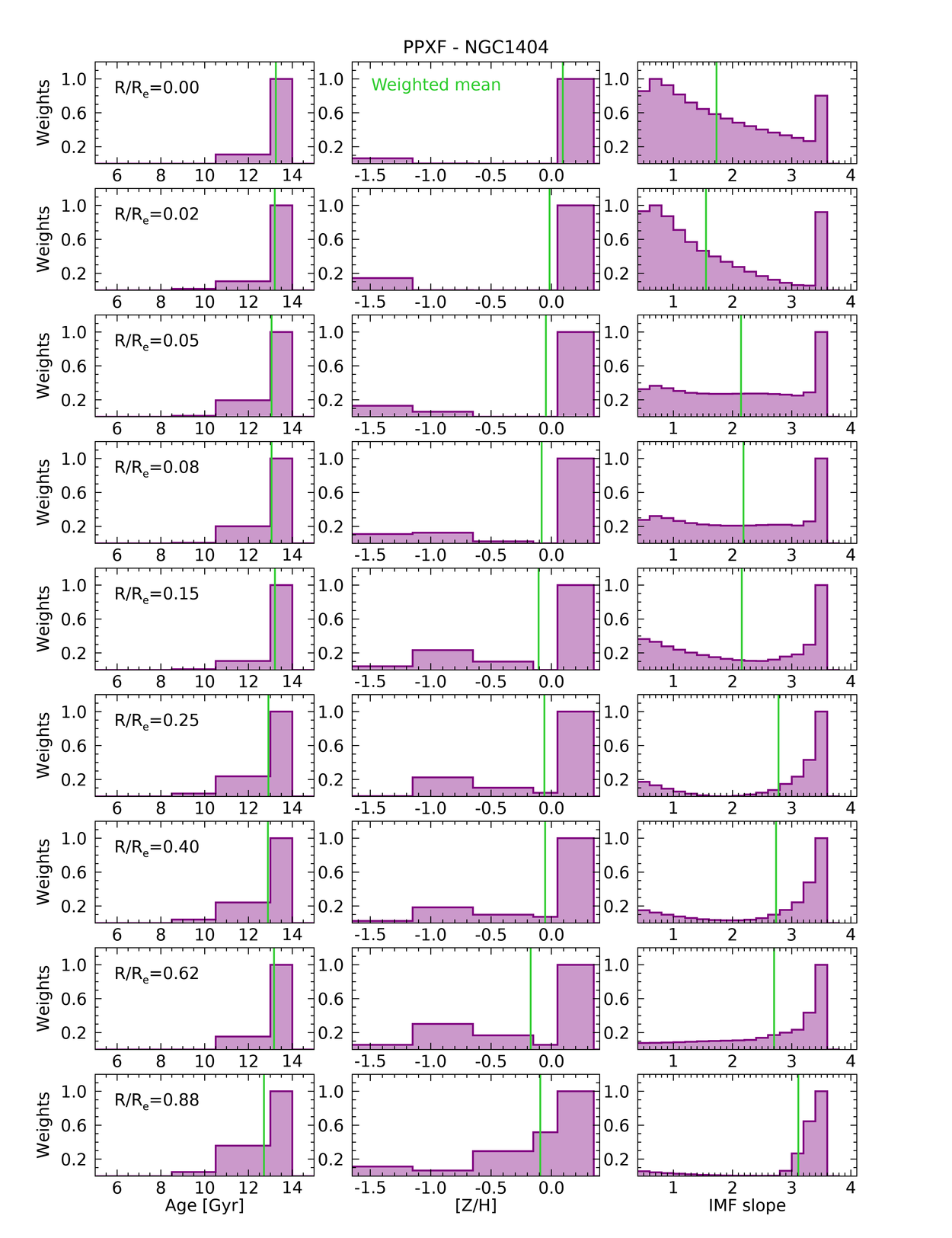}
\caption{\small{NGC1404 results. \textit{Left panel:} distribution of weights summed in each parameter bin for age (left), metallicity (middle) and IMF slope (right) as obtained by fitting with \textsc{Starlight} our 9 radial bin spectra (from top, innermost bin, to bottom, outermost bin). Green vertical lines show the weighted means (as they appear in Figures \ref{fig:results14} and \ref{fig:results13}). \textit{Right panel:} Same as left panel, but in this case the results are obtained by fitting with \textsc{pPXF}. }}
\label{fig:res_pdf14}
\end{centering}
\end{figure*}

\begin{figure*}[h]
\begin{centering}
\includegraphics[width=8.9cm]{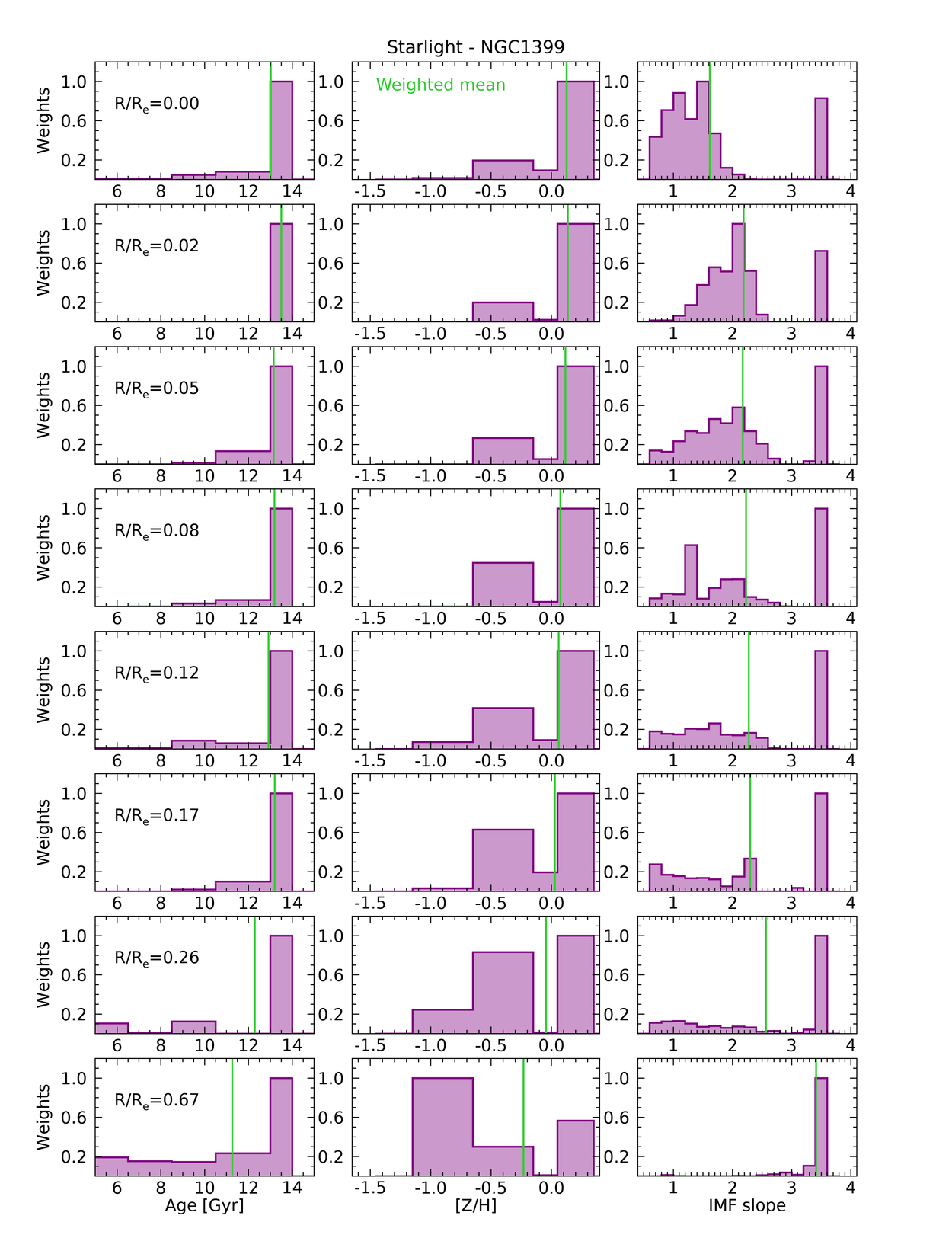}
\includegraphics[width=8.9cm]{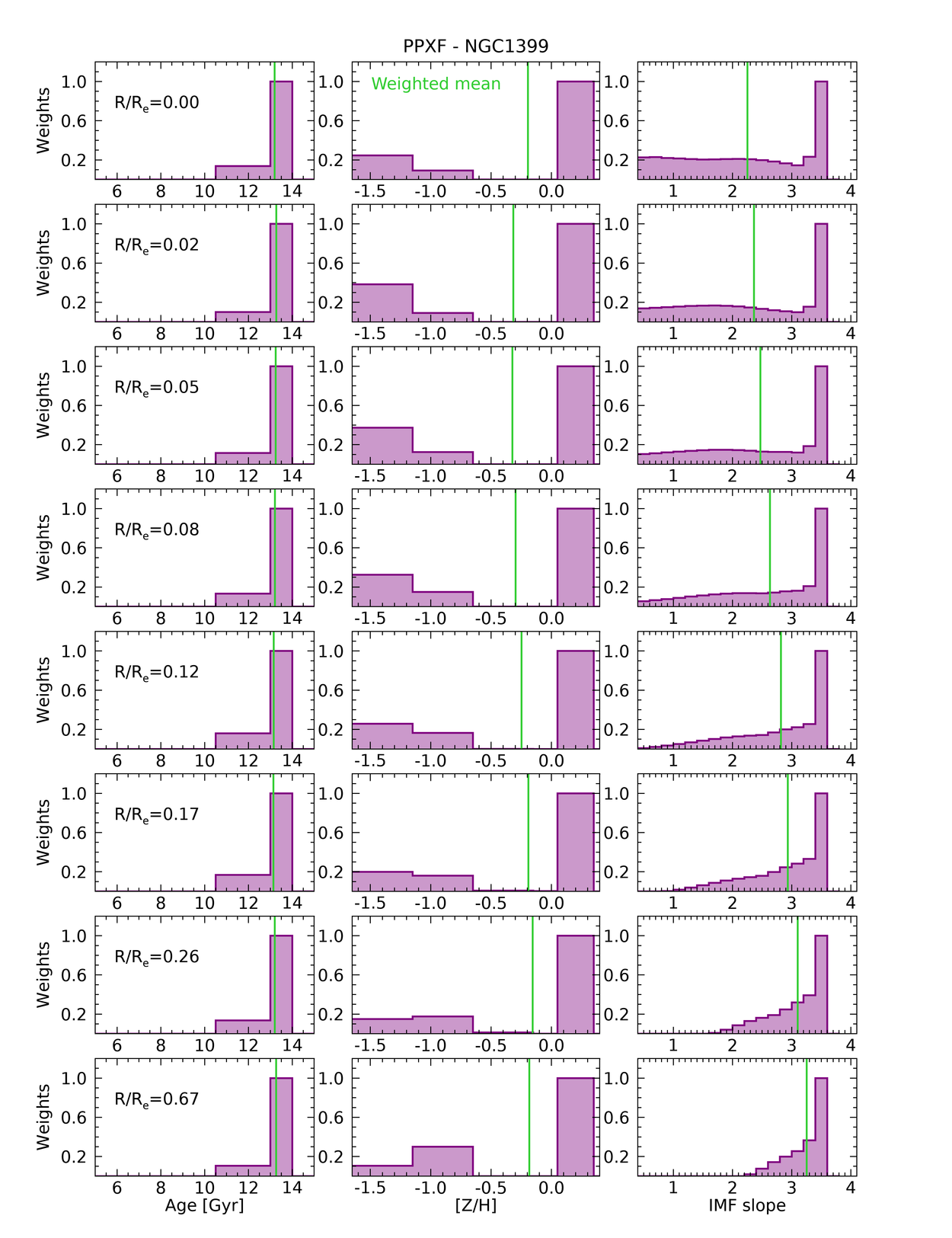}
\caption{\small{NGC1399 results. Same as in Figure \ref{fig:res_pdf14}, but for NGC1399. }}
\label{fig:res_pdf13}
\end{centering}
\end{figure*}

To better justify the discrepancy we found between the results of \textsc{ALF}-\textsc{PyStaff} and \textsc{Starlight}-\textsc{pPXF}, it is important to recall the quantities that we are initially comparing: the mean values of the posterior distribution assuming a SSP on one side, and light-weighted quantities computed over a distribution of solutions on the other. 
Interestingly, for \textsc{Starlight} and \textsc{pPXF}, we can inspect the shape of the distributions of weights adopted for each parameter (age, metallicity and IMF) in each final best-fit. In Figures \ref{fig:res_pdf14} and \ref{fig:res_pdf13} we show such distributions for each radial bin of NGC1404 and NGC1399 respectively. 
From inspecting these plots it can be seen that the light-weighted means (represented by the vertical green lines) are in most cases not representative of the weight distributions: metallicity and IMF (age only moderately) are clearly showing double components, whose proper mean value and intensity changes with radius. The light-weighted mean values, by definition, lie in between the two distributions and can mimic radial gradients. We can thus observe that the discrepancies between \textsc{Starlight}-\textsc{pPXF} and \textsc{ALF}-\textsc{PyStaff} can be explained by the presence of multiple components in the stellar populations of the analyzed galaxies, which the latter cannot detect due to the assumption that the star formation history is fixed to be an SSP. 


\begin{figure*}[h]
\begin{centering}
\includegraphics[width=8.9cm]{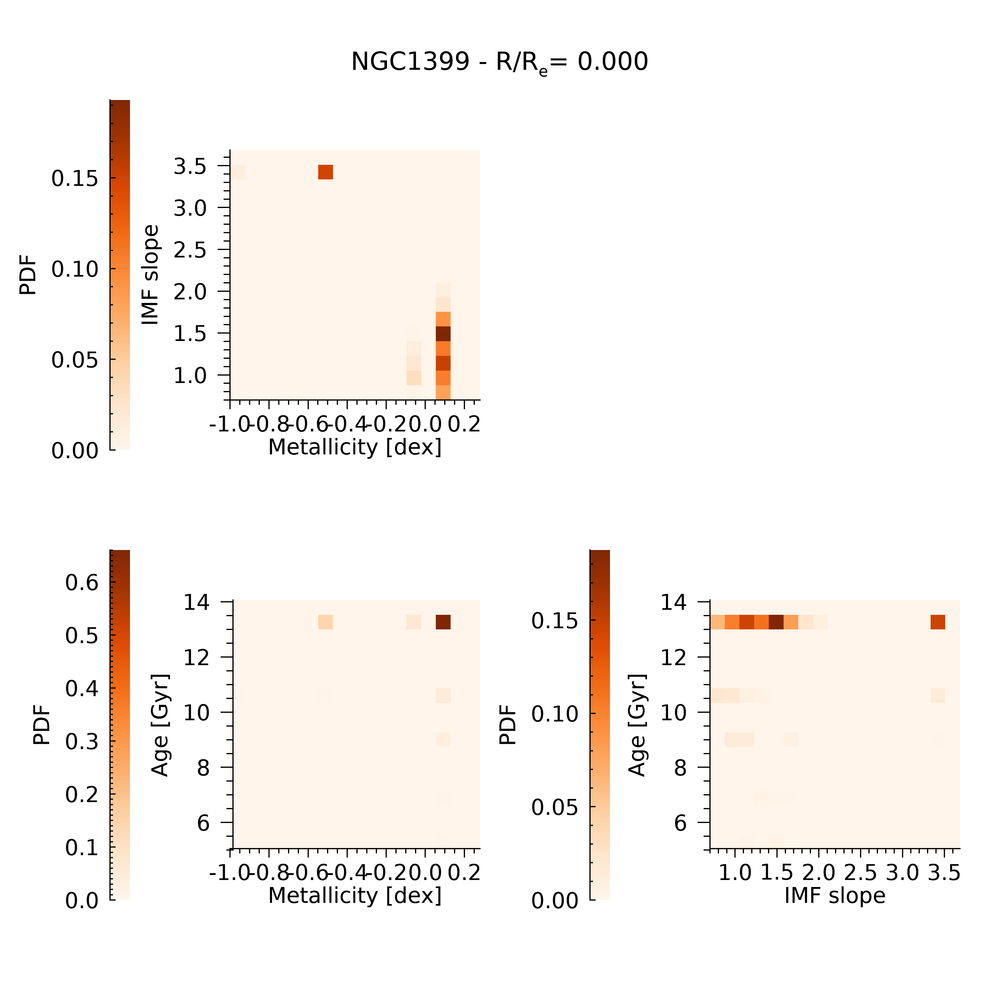}
\includegraphics[width=8.9cm]{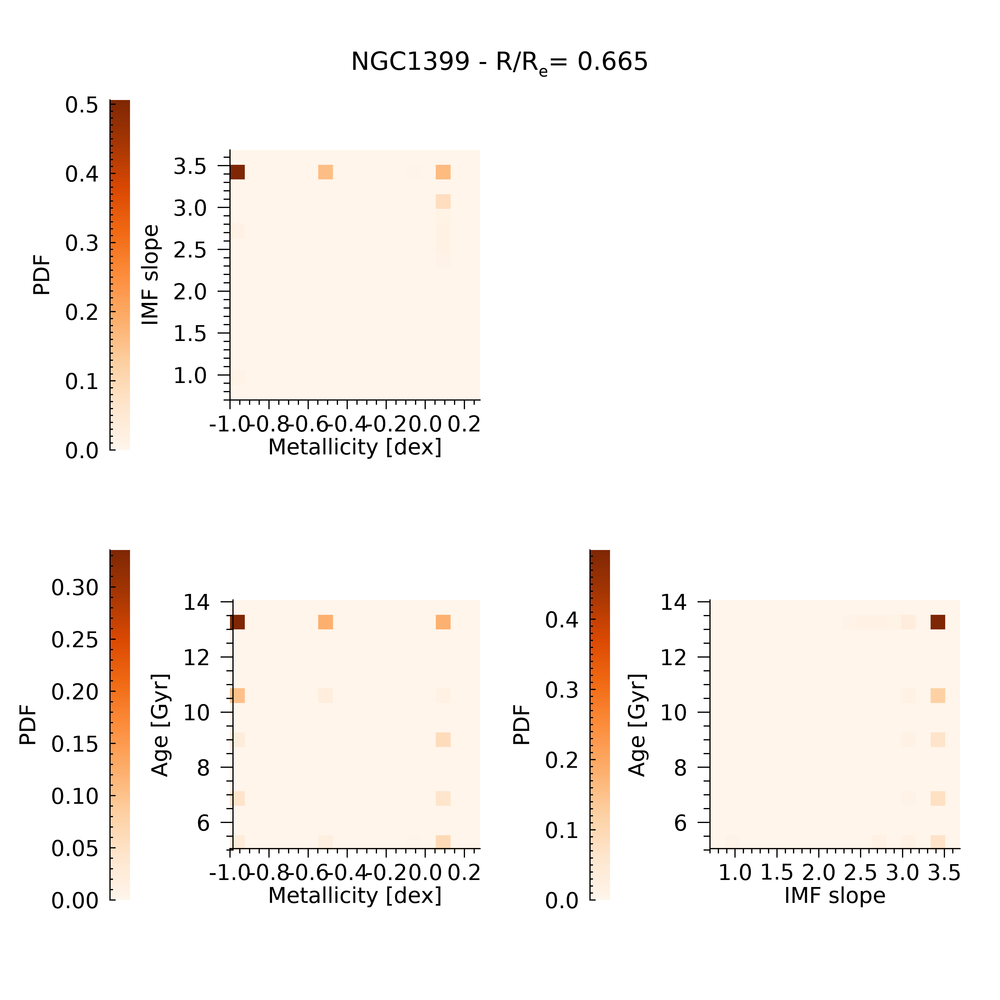}
\includegraphics[width=8.9cm]{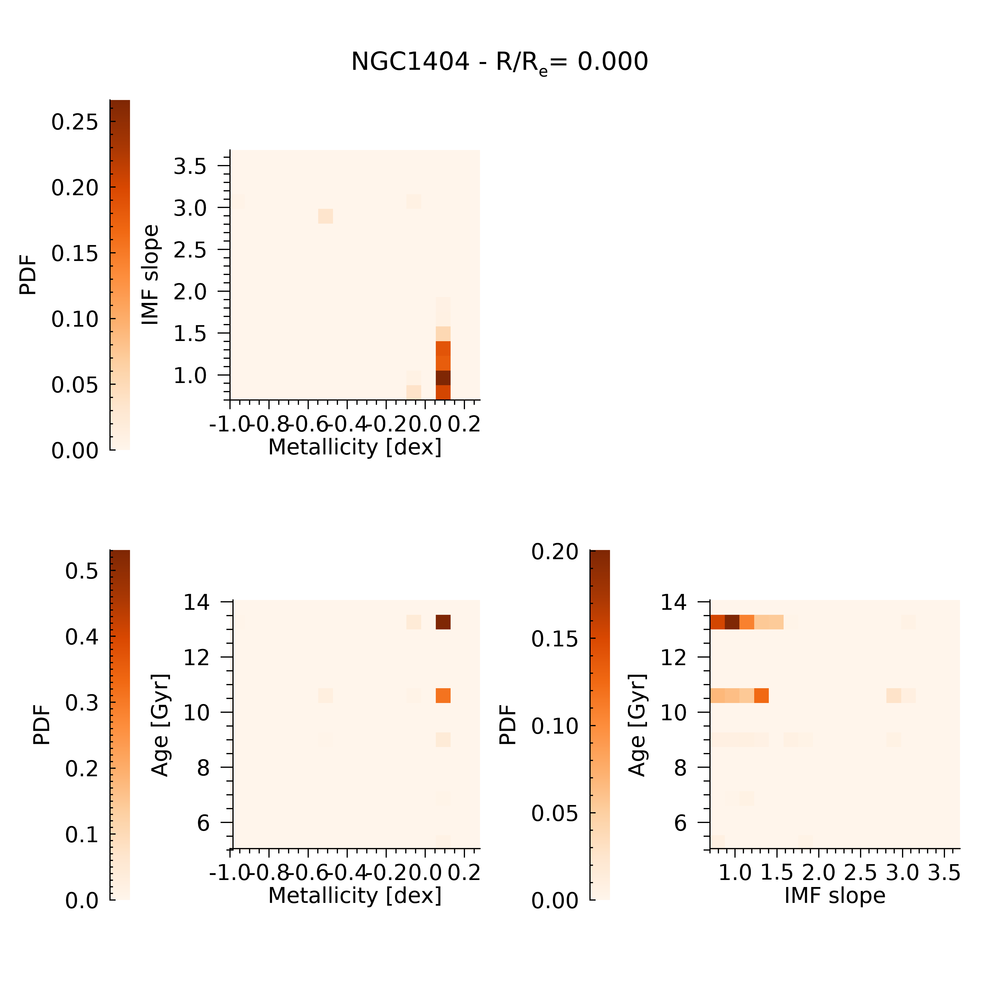}
\includegraphics[width=8.9cm]{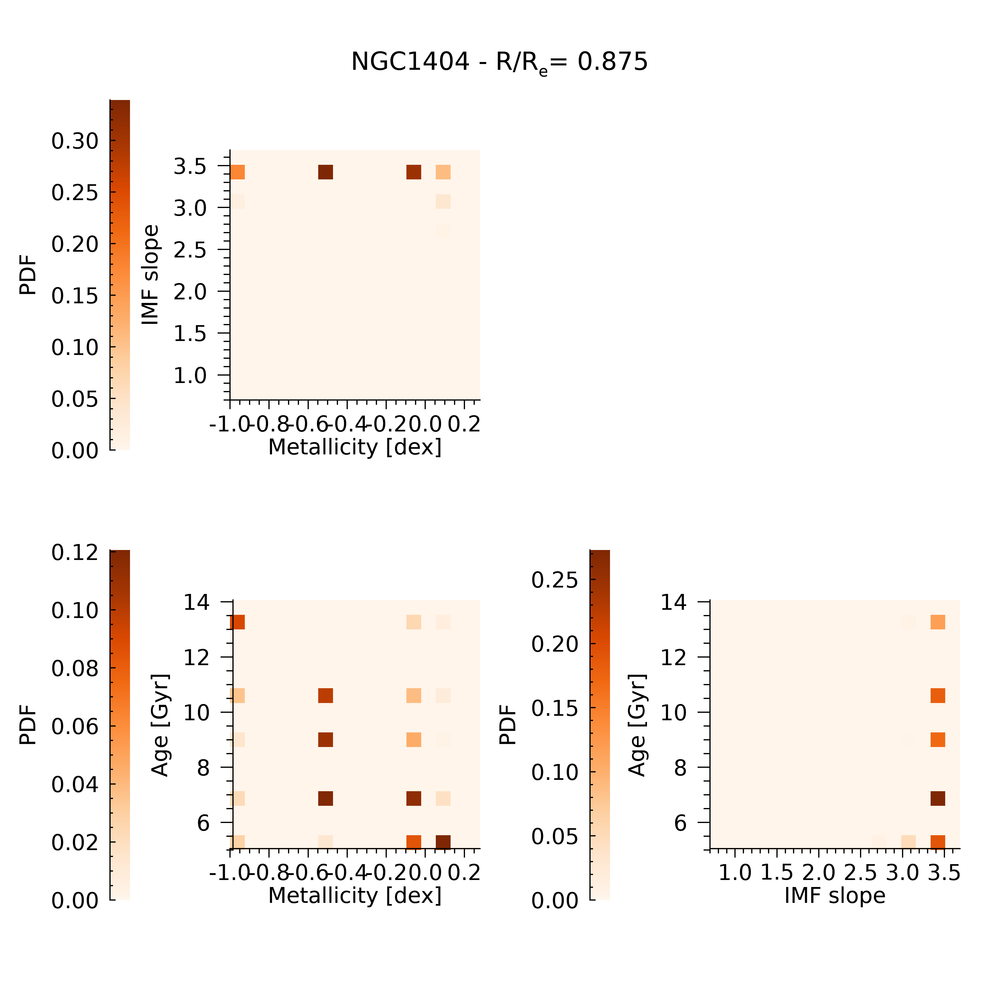}
\caption{\small{\textsc{Starlight} results for NGC1399 (upper panels) and NGC1404 (lower panels) for their central (left panels) and outermost (right panels) radial bins. Each panel shows three plots for the three planes IMF-Metallicity, Age-Metallicity and Age-IMF. Color intensity is proportional to the weights distribution (PDF) as indicated by the side bars.}}
\label{fig:res_contourSL}
\end{centering}
\end{figure*}

\begin{figure*}[h]
\begin{centering}
\includegraphics[width=8.9cm]{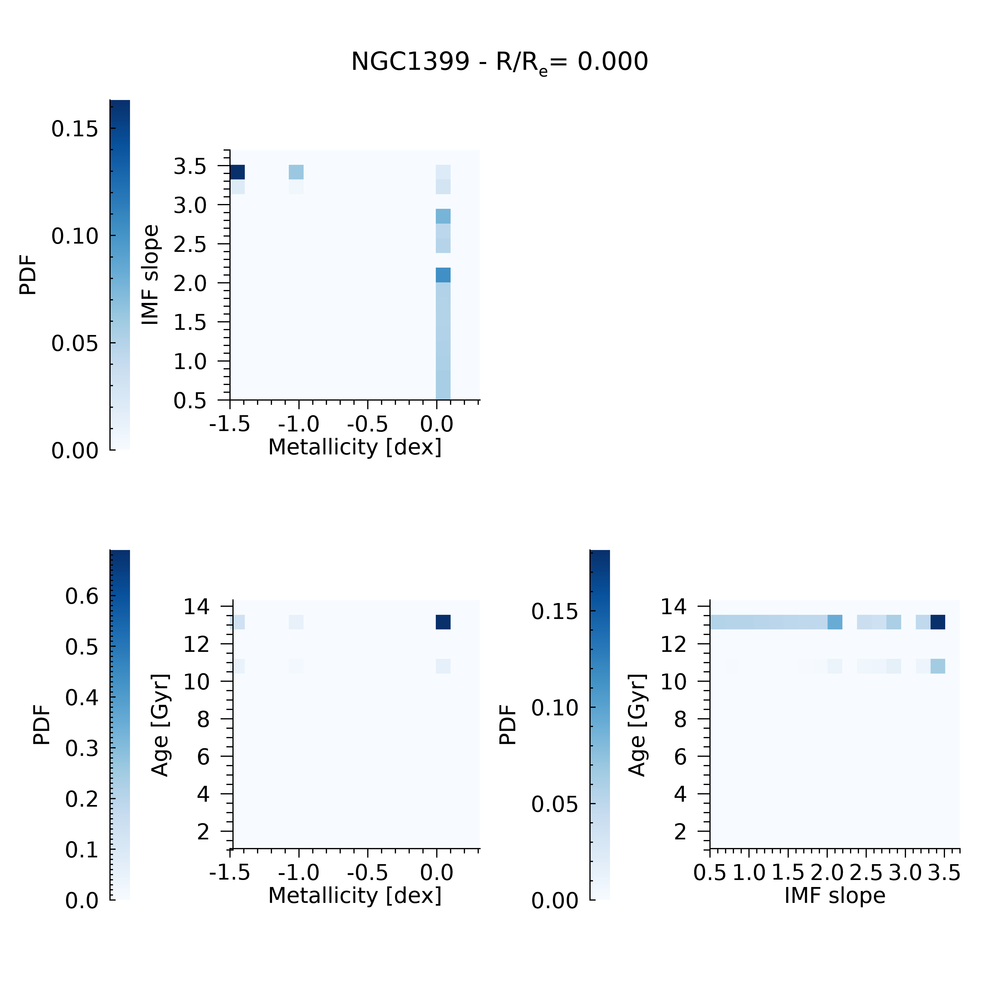}
\includegraphics[width=8.9cm]{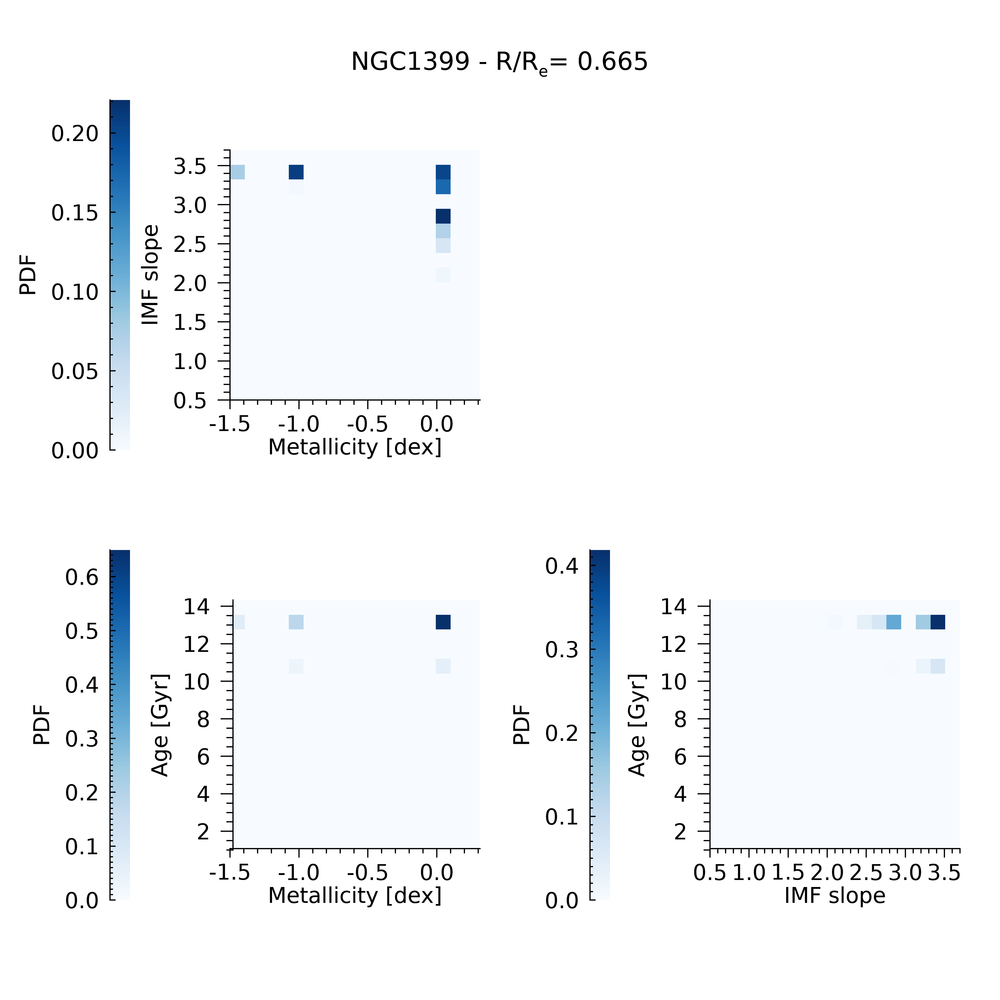}
\includegraphics[width=8.9cm]{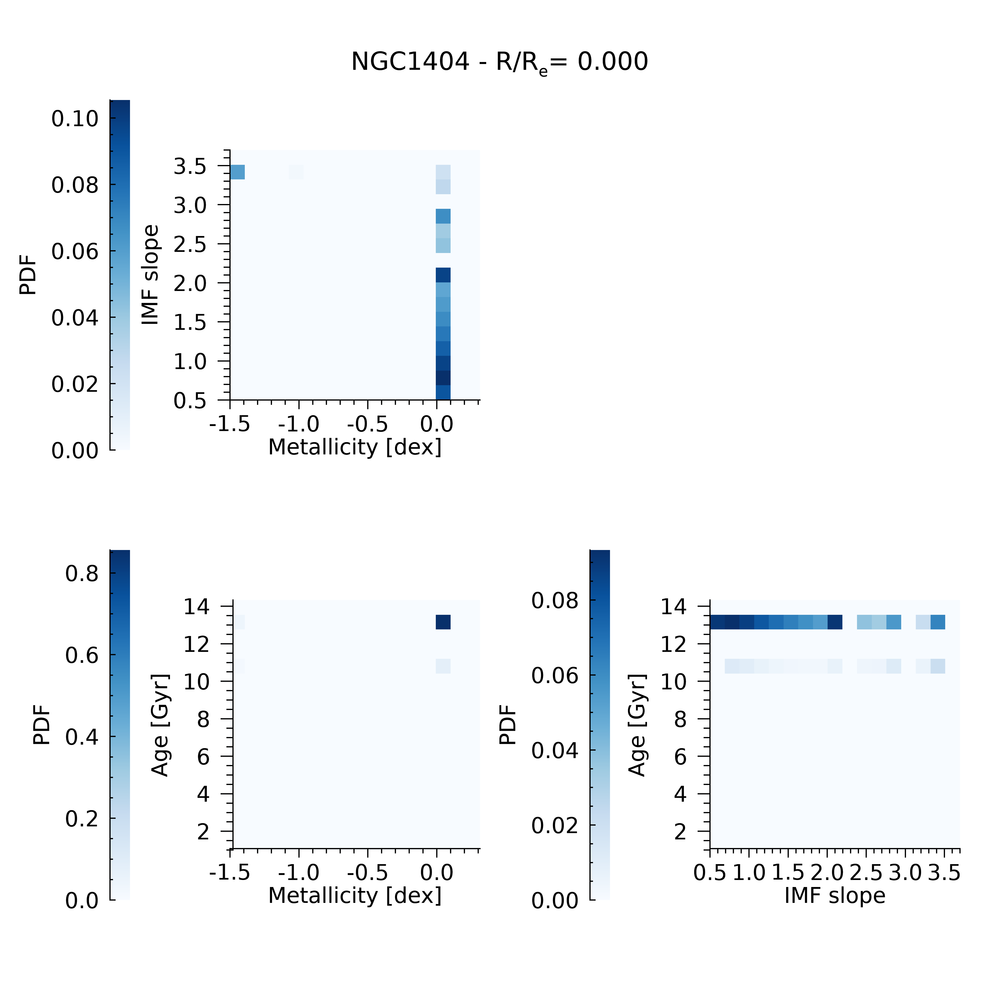}
\includegraphics[width=8.9cm]{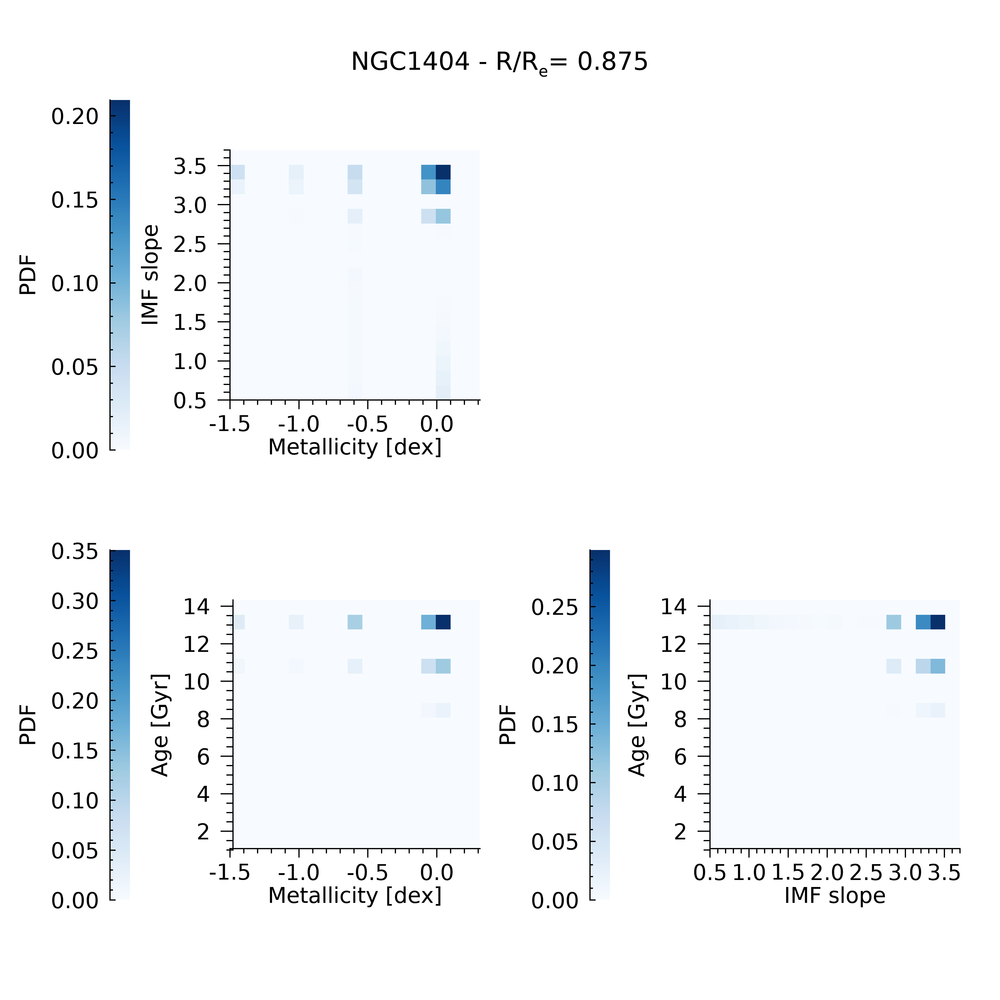}
\caption{\small{Same as Figure \ref{fig:res_contourSL} but for \textsc{pPXF} results. }}
\label{fig:res_contourPpxf}
\end{centering}
\end{figure*}

In Figures \ref{fig:res_contourSL} and \ref{fig:res_contourPpxf} we also show some examples of contour plots for the results obtained with \textsc{Starlight} and \textsc{pPXF} to better view the properties of each stellar component. In particular, for each galaxy we show the central and outermost radial bins, since they represent the two extreme cases. In between, the contour plots are just a smooth transition from one to the other, as can be seen from Figures \ref{fig:res_pdf14} and \ref{fig:res_pdf13}. From these contour plots, it can be noticed that, with only very few exceptions, we detect two distinct populations whose stellar properties are well defined.


\begin{figure*}[ht!]
\begin{centering}
\includegraphics[width=14.5cm]{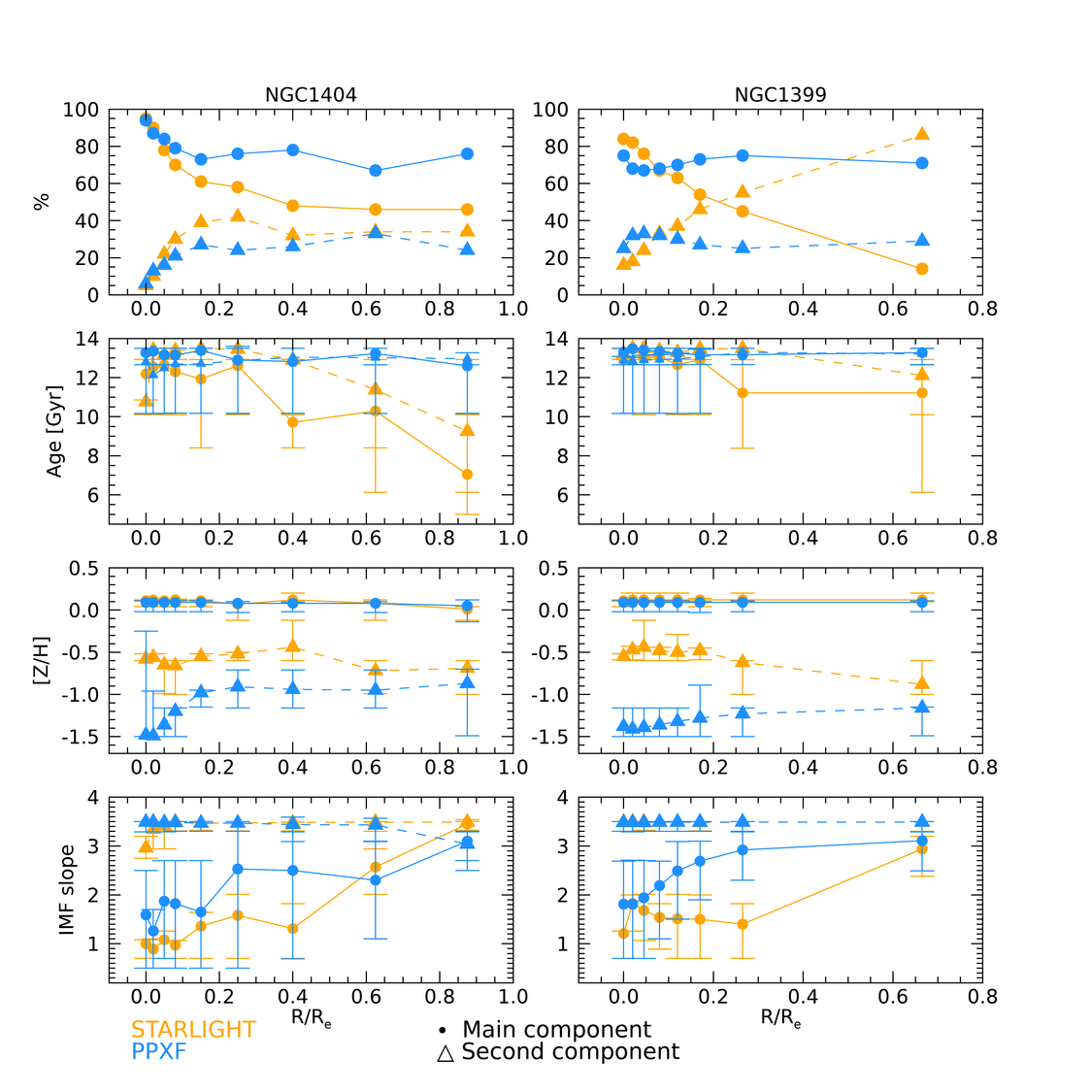}
\caption{\small{Detailed radial trends of mass fraction, age, metallicity and IMF slope for both stellar components as detected with \textsc{Starlight} (orange) and \textsc{pPXF} (light blue), for NGC1404 (left panels) and NGC1399 (right panels). In particular, as indicated in the upper panels, dots and solid lines refer to the main component, while triangles and dashed lines to the secondary component. }}
\label{fig:components}
\end{centering}
\end{figure*}

We quantified the fractions of the main and secondary component, together with their proper values of age, metallicity, and IMF slope, by isolating them in the three-dimensional best-fit weights matrix offered by both codes, with the help of contour plots as shown in Figures \ref{fig:res_contourSL} and \ref{fig:res_contourPpxf}. In particular, we isolated each stellar component and its parameters by inspecting the presence of delimited regions with higher concentrations of weights and calculated the total fraction belonging to each of them. The results are shown in Figure \ref{fig:components}. \textsc{Starlight} results are shown in orange and \textsc{pPXF} results in light blue. Dots and solid lines refer to the main component, while triangles and dashed lines refer to the secondary component. In general, we find good agreement between results from the two codes. From the upper panels we learn that the main component is dominant only in the first half of both galaxies; from the parameter distributions we see that the ages are all old and flat as a function of radius, with only \textsc{Starlight} pointing to a negative gradient for the main component of NGC1404, decreasing from around $12$ Gyr to around $7$  Gyr. For both galaxies the main component has a flat and super-solar metallicity, at the edge of the models' range, while the secondary component, again consistent with a flat trend, has values below $-0.5$ dex. The metal-rich and dominant component is associated with a top-heavy IMF slope, which tends to increase with radius. The sub-solar metallicity component instead shows a flat bottom-heavy IMF at the higher limit of the model's range.

In the next section, we will discuss these results in the context of the simulations results (see Section \ref{sec:sim}).

\section{Discussion}
\label{sec:discussion}

In Section \ref{sec:results} we have presented our results obtained by fitting the same data-set with the same stellar population synthesis models and four different FSF codes which are generally used to derive stellar population properties of passive galaxies, including information on their IMF.
Our results suggest that the two codes that fit an observed spectrum assuming a SSP (i.e. \textsc{ALF} and \textsc{PyStaff}) agree well with each other, but are in tension with the other two codes that do not assume a fixed SFH but allow the best-fit to be composed of a distribution of templates (i.e. \textsc{Starlight} and \textsc{pPXF}).
Indeed, we fitted the radial spectra for our targets NGC1404 and NGC1399 and found that for the former two codes, the results assuming a SSP indicate a flat IMF trend, while the latter codes both suggest the presence of two distinct stellar components. 

We exclude the possibility that the discrepancy found between the \textsc{ALF}-\textsc{PyStaff} and \textsc{Starlight}-\textsc{pPXF} results is completely due to the inclusion (as in the first two codes) or not (as in the second two codes) of non-solar elemental abundances in the fitting templates. Indeed, repeating the fit with \textsc{Starlight} and \textsc{pPXF} and fitting templates with \textit{fixed} non-solar elemental abundances produced consistent results for age and metallicity, and partially for the IMF, as is shown and described in more detail in Appendix \ref{app:elem}.

The question that naturally arises is: how robust is the retrieval for each code? And are \textsc{ALF} and \textsc{PyStaff} able to indicate the presence of double components (by showing large error bars, for example)? Are \textsc{Starlight} and \textsc{pPXF} able to give reliable information on the IMF of each detected stellar component?

A possible means of addressing these questions could be to look at the residuals of each fit for each code and examine which code gives the closest best-fit to the original spectrum. However, we warn that this comparison could be misleading, for the following two reasons: i) each code actually uses a different input spectrum, as described in Section \ref{sec:analysis}, with a different spectral resolution and/or sampling, thus a direct pixel-to-pixel comparison of spectral features is ruled out; ii) each code considers the continuum in a different way, i.e.: \textsc{ALF}, \textsc{PyStaff} and \textsc{pPXF} apply a polynomial function to better trace the continuum, producing best-fit spectra more similar to the original input one, while \textsc{Starlight} uses only the depth of the absorption lines and ignores the continuum shape. 
As an exercise, however, we checked the residual plots of both real data and simulation fits, and discuss these results in Appendix \ref{app:residuals}.

To better help in answering these questions in a more quantitative and reliable way, as mentioned above, we ran a set of simulations aimed at testing the retrieval of the same stellar parameters with the four codes in cases of increasing complexity of the input SFH. We set up simulations as described in Section \ref{sec:sim}. The SFH complexity of each input is shown in Table \ref{tab:sim}: we started with a SSP for an old and metal-rich population with non-solar elemental abundances, then added a secondary component which differs only in age and metallicity (CSP), then finally tested two cases of double components with mixed IMF slopes (CS2 and CSP3) but the same non-solar elemental abundances. In particular, the last set of simulations was tuned to resemble the results we obtained from NGC1404 and NGC1399, i.e. two components with both ages $>11$ Gyr, one more metal-rich and TH and the other with lower metallicity and BH. We fit the mock spectra in the same way as the real data, i.e. with the same code setup (Section \ref{sec:analysis}), with non-solar elemental abundances templates for \textsc{ALF} and \textsc{PyStaff}, and with only solar elemental abundances for \textsc{Starlight} and \textsc{pPXF}. 
All results are shown and discussed in Appendix \ref{app:sim}. 

Here we summarize our main findings that answer the above questions: 

\begin{enumerate}
    \item In the case of SSP input, although all four codes successfully retrieve the age and metallicity values, the IMF is better retrieved, in terms of both precision and accuracy, by \textsc{PyStaff}, and \textsc{ALF}, while \textsc{Starlight} and \textsc{pPXF} give more broadened and biased distributions (Figures \ref{fig:sim_sspBH}, \ref{fig:sim_sspTH}). In particular, \textsc{pPXF} exhibits double-peaked distributions for the metallicity and IMF slope values, a sign of a possible degeneracy.
    
    \item When including a secondary component (CSP), \textsc{ALF} and \textsc{PyStaff} start giving biased results for age and metallicity, and for the IMF when the input is TH. In particular, the retrieved values lie between the true values of the two components and off from their light-weighted means (light-green dotted vertical lines in Figures \ref{fig:sim_elemSFH}-\ref{fig:sim_elemSFH3}), and return small uncertainties (see point 4). 
    On the other hand, although with large uncertainties, \textsc{Starlight} is  able to retrieve all input parameters with more accuracy (Figures \ref{fig:sim_sfhBH}, \ref{fig:sim_sfhTH}). \textsc{pPXF} reveals the presence of double components but with biased peak values. In particular for the IMF slope, the values are pushed to both a very top-heavy and bottom-heavy IMF, similar to the case of SSP; for the metallicity, we notice lower than expected values.
    
    \item When the two stellar components also have different IMF slopes (CSP2 and CSP3), results from \textsc{ALF} and \textsc{PyStaff} are even more biased. Not only the IMF slope itself but also the age and metallicity are severely affected. Elements are in general well-retrieved, despite differences exceeding $1\sigma$ in some cases for, e.g., Na, O, C, Si, Mg, Ca, Ni and V. However, we made the unrealistic assumption that both populations have the same elemental abundances.
    \textsc{Starlight} is always able to detect the correct input set of parameters, although the light fraction attributed to each component is not always accurate (Figures \ref{fig:sim_sfh2}, \ref{fig:sim_sfh3}). As before, \textsc{pPXF} detects the presence of double components but returns biased distributions, particularly for the metallicity. It correctly retrieves the CSP3 input values.
    
    \item In general, \textsc{PyStaff} (and \textsc{ALF} in some cases) at each level of input complexity gives very precise results with small error bars on biased values, thus not offering any hint as to the presence of other components. However, the narrow distributions of \textsc{PyStaff} are partially due to the data re-sampling to $1.25$ \AA\space pixel$^{-1}$, which reduces the noise. Indeed, with higher signal-to-noise spectra  the fitting procedure is forced to prefer the best-fit model with less margin of error, with the consequence of offering sharper output parameter distributions. On the other hand, \textsc{Starlight} and \textsc{pPXF}, despite their ability to detect the secondary component, offer less precise distributions. 
\end{enumerate}

Knowing that \textsc{ALF} and \textsc{PyStaff} in the presence of a double stellar population retrieve values in between the two true values, and that \textsc{Starlight} and \textsc{pPXF} are similarly able to distinguish the two populations despite larger uncertainties, we are now able to interpret our results on NGC1404 and NGC1399 (plots shown in Section \ref{sec:results}).

First, we can now explain the systematic low metallicity values found with \textsc{pPXF} as shown in the lower-left panel of Figures \ref{fig:results13} (and  \ref{fig:results14}). By looking at the simulations results for \textsc{pPXF} (see Figure \ref{fig:sim_PPall}), we can see that this code in all cases retrieves biased distributions for the metallicity values. In particular, and in agreement with our findings, the metallicity distributions always present a main component centered on the correct input super-solar value, together with secondary peaks centered on sub-solar values which are generally significantly biased toward too-low values and erroneously present also in the cases of super-solar SSPs. We address this bias to the lack of fitting non-solar values of the elemental abundances with \textsc{pPXF}. Interestingly, \textsc{Starlight} also could not account for non-solar chemical ratios, but gave much less biased results (although with poor precision).
With regard to the presence of secondary components in our observed galaxies, we must proceed with caution. Both \textsc{Starlight} and \textsc{pPXF} are able to detect the presence of a secondary component and characterize its properties. By inspecting CSP3 results (Figure \ref{fig:sim_sfh3}), whose mock inputs were built with stellar parameters similar to our targets, we learn that both codes should be able to retrieve their age, metallicity and IMF slope with no evident biases. In this sense we have a chance to believe that our observed data are actually revealing a double stellar component. However, as we have already noticed for the case of \textsc{pPXF}, not considering the elemental abundances in the fit as free parameters can lead to hidden degeneracies, mostly with metallicity and IMF. The ideal way to perform such analysis would be having a non-parametric FSF code that would be able to fit also for non-solar elemental abundances in a reasonable amount of time. 
In conclusion, with the present analysis, we are inclined to rely only on \textsc{Starlight} results, which, although with large uncertainties, have not shown any significant bias in our simulations set. \textsc{Starlight} results suggest the presence of a secondary component, so we conclude that it is likely that both galaxies do have a double stellar population.\footnote{We note more precisely that it is likely that both galaxies have more than a single population; however, we have not tested for multiple populations.} 

In more detail, both galaxies similarly show the presence of (at least) two stellar populations
of similar old age and differentiated by metallicity and IMF slope, with the super-solar metallicity component associated with a TH IMF and the sub-solar one with a BH IMF. Results from \textsc{ALF} and \textsc{PyStaff}, which both consistently show a constant IMF, actually indicate a constant value in between the two distributions: a trade-off between the decreasing fraction of the main component with radius and the positive gradient of its IMF (see also Figure \ref{fig:components}). The metallicity trends of both components, separated by $\sim1.0$ dex, are flat, so the apparent negative gradient suggested by \textsc{ALF} and \textsc{PyStaff} is probably produced by the decreasing mass fraction with radius of the main component.   

In accordance with \citet{iodice16}, who found that the Fornax cluster is probably more dynamically evolved than other local clusters and that most of the interactions among the galaxies in its center happened in an early formation epoch, we observed that both our galaxies exhibit old ages in both of their stellar components. We thus confirm that their secondary component, if formed by merging or interaction, was accreted during that critical phase of cluster formation, $>12$ Gyr ago. 

Another confirmation of this is that both galaxies, regardless of their position in the cluster (recalling that NGC1399 is the central brightest cluster galaxy), or their mass (M$_{\mathrm{NGC}1404}=1.3 \cdot 10^{11}$ M$_\odot$ from \citealt{iodice19}, M$_{\mathrm{NGC}1399}=5.8 \cdot 10^{11}$ M$_\odot$ from \citealt{spavone17}) or velocity dispersion profile (see Figures \ref{fig:results14} and \ref{fig:results13}, upper right panels), show a similar SFH. Indeed, as detailed in Figure \ref{fig:components}, the two components of both objects are very similar in all their characteristics, suggesting that their formation occurred from similar processes and at similar times.

According to the two-phase formation scenario (\citealt{naab09}, \citealt{oser10}), massive ellipticals form their dominant stellar component (\textit{in situ}) in an initial burst of star formation, followed by subsequent merging events that make up the accreted component, which generally resides in the outskirts of galaxies. \citet{spavone17} analyzed the photometric profile of NGC1399 and found it consistent with the presence of three stellar components. In particular, the transition from the inner component to the secondary component is found right before our last radial bin. This limit is where our secondary component becomes important according to our \textsc{Starlight} results (Figure \ref{fig:components}, right panels); in addition, with our current work we managed to better describe the presence and properties of the second component all along the spanned radial profile.

Despite the large number of works devoted to the study of the Fornax cluster and its assembly history, only a few of them were focused on the stellar population analysis and in particular on the radial profile of the IMF. For NGC1404, we compare our age and metallicity results with the work of \citet{iodice19}, finding consistency between their values extracted within and beyond $0.5$ R$_e$. Their analysis was based on fitting the values of five spectral indices in the region around $5000$\AA\space without the possibility of detecting a secondary component. To our knowledge, the only IMF measurement is from \citet{feldmeier-krause21}, but that was obtained from our same data set and with a similar fitting procedure, i.e. using \textsc{PyStaff} with \citet{conroy18} models but including only $9$ elemental abundances. For NGC1399, the work by \citet{vaughan18} offers a good comparison since they could derive its radial profile for a large number of stellar properties, including the IMF. In particular, they used \textsc{PyStaff} to fit the spectral range from $4800$ to $9000$\AA\space (with only a gap between $7500$ and $7700$\AA) for $18$ radial bins out to $\sim44$\arcsec\space (for reference, our radial range extends to $38$\arcsec). From their Figure $6$, we can compare our results for the velocity dispersion, age, metallicity, Na, Fe, Mg and the mismatch parameter $\alpha$, which indicates the need for a bottom-heavier IMF with respect to the Milky Way. The trends are similar to our \textsc{PyStaff} results, but the gradients do not always span exactly the same range, though they are centered on the same values. This may be due to the fact that \citet{vaughan18} data are measured in annular bins, thus collecting the information all around the galaxy and showing very small error bars. In particular, with both \textsc{PyStaff} and \textsc{ALF} we found the same flat IMF trend within the common radial range, with slightly super-Salpeter values (IMF slope $\sim2.35$), as shown in the upper left panel of Figure \ref{fig:results14}. 
We stress again that although \textsc{PyStaff} retrieves quantities with small uncertainties, it assumes a SSP and is not able to detect multiple components. Thus all of the conclusions obtained regarding the IMF slope trends are impacted by this critical assumption.


\section{Summary and Conclusions}
\label{sec:conclusions}

The purpose of the present work was to directly compare, with an emphasis on the IMF, the results from applying four different FSF codes to the same data set and with the same adopted stellar population synthesis models.
The FSF codes chosen were all public codes that allow the retrieval of the IMF slope in addition to age and stellar metallicity: \textsc{ALF}, \textsc{PyStaff}, \textsc{Starlight} and \textsc{pPXF}. 
For this task, we analyzed the radial profile out to around $1$R$_e$ of the two brightest elliptical galaxies, NGC1399 and NGC1404, in the Fornax cluster. We exploited high S/N spectroscopic data in the optical and NIR range taken with IMACS on the $6.5$-m Magellan Baade telescope. We fit not only the observed data to retrieve multiple stellar parameters (age, metallicity, IMF slope, and elemental abundances when allowed), but also a set of mock spectra built with increasing SFH complexity, in order to test the robustness of each code in retrieving accurate results.

In both our data and simulations, we found a clear difference between the results obtained with \textsc{ALF}-\textsc{PyStaff} and those obtained with \textsc{Starlight}-\textsc{pPXF}. 
This was due to the assumption for the model SFH that the two pairs of codes make in order to reach the best-fit solution: a fixed SSP for the former, and no \textit{a-priori} SFH for the latter. In particular, \textsc{Starlight} and \textsc{pPXF} offer a best fit that is a weighted composition of the basic templates, which can have any arbitrary SFH. This difference causes disagreement in the final results and differences in the power of retrieving the stellar parameters. Indeed, from simulations we found that when the input is a SSP, \textsc{ALF} and \textsc{PyStaff} both work very well in retrieving precise and accurate values of all parameters, including the IMF, while the other two codes suffer biases in the IMF slopes. In addition, when the complexity of the input SFH increases, \textsc{ALF} and \textsc{PyStaff} lose the ability to retrieve accurate results not only for the IMF but also for age and metallicity. Conversely, \textsc{Starlight} and \textsc{pPXF} generally offer accurate results despite a lower precision, and are also able to detect the presence of a secondary component with a different IMF slope. Comparing \textsc{Starlight} and \textsc{pPXF}, neither of which account for elemental abundances, we found that \textsc{pPXF} produces more biased results, in particular for SSPs.

In light of these findings, we conclude that despite their differences in velocity dispersion profile, mass and position in the cluster, NGC1404 and NGC1399 share a similar SFH. For both galaxies, we detected two stellar components with similar old ages ($>11$ Gyr): a main metal-rich ([Z/H]$\sim0.1$ dex, flat with radius) component with IMF slope showing a positive gradient from top-heavy values in the center up to around $3$ in the outskirt, and a minor component with a much lower metallicity ([Z/H]$<-0.5$ dex) and a constant bottom-heavy IMF. The fraction of the secondary component increases with radius, reaching $50$\% at around R/R$_e\sim 0.2$. 

This picture is consistent with the two-phase formation scenario proposed by \citet{naab09} and \citet{oser10}, where the presence of a secondary accreted component is predicted. It is also consistent with previous studies focused on the history of mass assembly of the Fornax cluster that suggest that most of the interactions among the central galaxies happened in the early formation epoch. 

Our results also indicate that the positive IMF-metallicity relation does not hold locally within each galaxy. Instead, our results are consistent with a more metal-rich component that is associated with a top-heavy component, and a low-metallicity component that is associated with a bottom-heavy IMF. However, it is important to note that previous studies that found a positive correlation between the IMF and Z (e.g.: \citealt{martin15c}, \citealt{vandokkum17}, \citealt{parikh18}, \citealt{feldmeier-krause21}, \citealt{lonoce23}) were all based on a fixed SSP assumption.

This result could potentially demonstrate the validity of the two-stage galaxy formation model with a variation of the IMF proposed by \citet{weidner13}. This model infers the presence of a globally observed bottom-heavy IMF in metal-rich populations in massive ellipticals by assuming a first phase of high star formation rate with a top-heavy IMF, which enhances the presence of metals. This is followed by the formation of the bulk of stars with a bottom-heavy IMF. However, we have found that the majority of stars have top-heavy IMF. 

We conclude that, for the FSF codes studied here, the assumption made regarding the SFH allowed during the fit is crucial for the retrieval of all of the stellar properties, including the IMF. 
In particular, ignoring the composition of the underlying stellar population can affect the robustness of the properties that are retrieved; for example, the age.
Since the shape of the SFH is unpredictable \textit{a priori}, even in the case of old massive ellipticals, our suggestion for future studies is to always check for the presence of secondary components as the first step in the analysis. This can be done by exploiting codes like \textsc{Starlight} that, although providing less precise results, can indicate the most probable stellar population composition, including the presence of different components with different IMF. If the population is confirmed to be consistent with a SSP, then \textsc{ALF} and \textsc{PyStaff} can provide more precise results and add, for example, information about the elemental abundances. This would be the ideal recipe to treat a sample of galaxies that may have had a distribution of different SFH, allowing an unbiased comparison amongst them, including the exploration of correlations between stellar parameters.

A threshold on the secondary component mass fraction below which it would be safe to use \textsc{ALF} and \textsc{PyStaff} without incurring biases in age, metallicity and IMF slope, is unfortunately impossible to uniquely define. Indeed, the complexity of this degeneracy problem is very high. 

To improve the derivation of stellar properties and SFH in galaxies, a FSF code should be created that, in a reasonable amount of time, would be able to test for the presence of multiple components and simultaneously retrieve a set of stellar parameters for each component individually.
This would allow for robust retrieval of important quantities such as age, metallicity and IMF both for small and large samples of galaxies, as well as to facilitate the study of their evolution over cosmic time, including the newly-discovered, extremely distant galaxies observed by the James Webb Space Telescope.

\acknowledgments
We thank the University of Chicago for their support of this research, and the staff at Las Campanas Observatory for their help in getting these data. 
 AFK acknowledges funding from the Austrian Science Fund (FWF) [grant DOI 10.55776/ESP542].
%

\vspace{5mm}
\facility{Magellan: Baade (IMACS)}
\software{\textsc{ALF} \citep{conroy18}, EMCEE \citep{2013PASP..125..306F}, IDL, IRAF \citep{tody93},  MOLECFIT (\citealt{2015A&A...576A..77S,2015A&A...576A..78K}, \textsc{pPXF} \citep{2017MNRAS.466..798C}, \textsc{PyStaff} \citep{2018MNRAS.479.2443V}, \textsc{Starlight} \citep{cidfernandes05}.}





\appendix

\section{Elemental abundances results}
\label{app:elem}

\begin{figure*}[ht!]
\begin{centering}
\includegraphics[width=17.5cm]{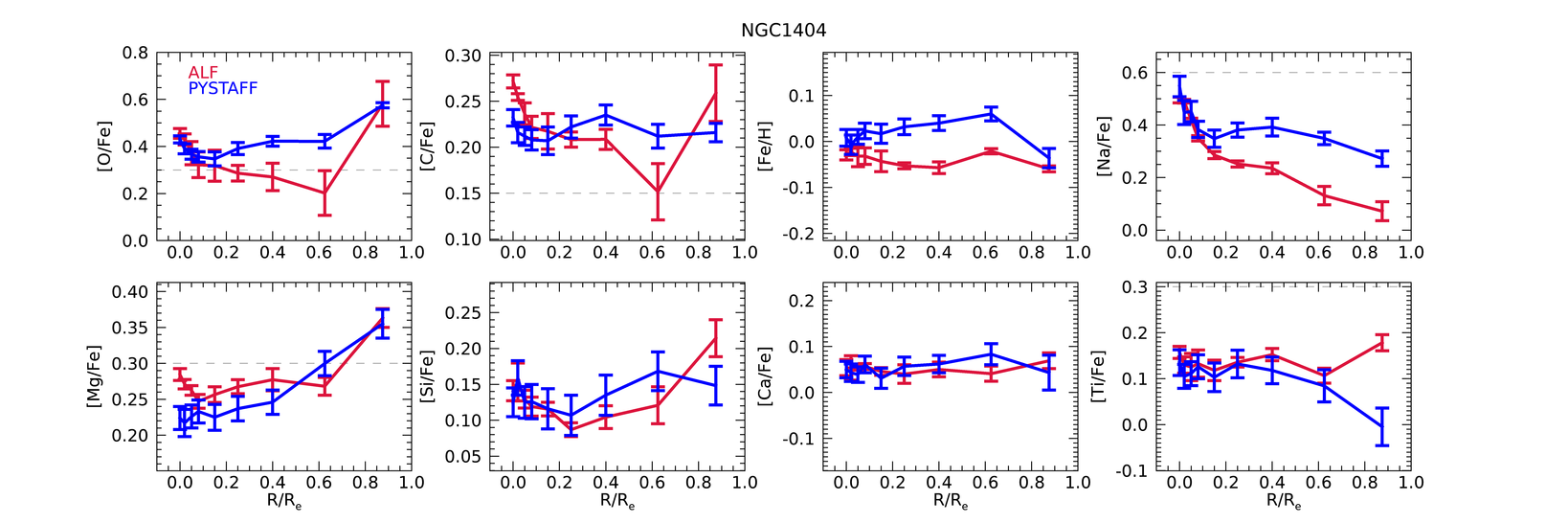}
\includegraphics[width=18cm]{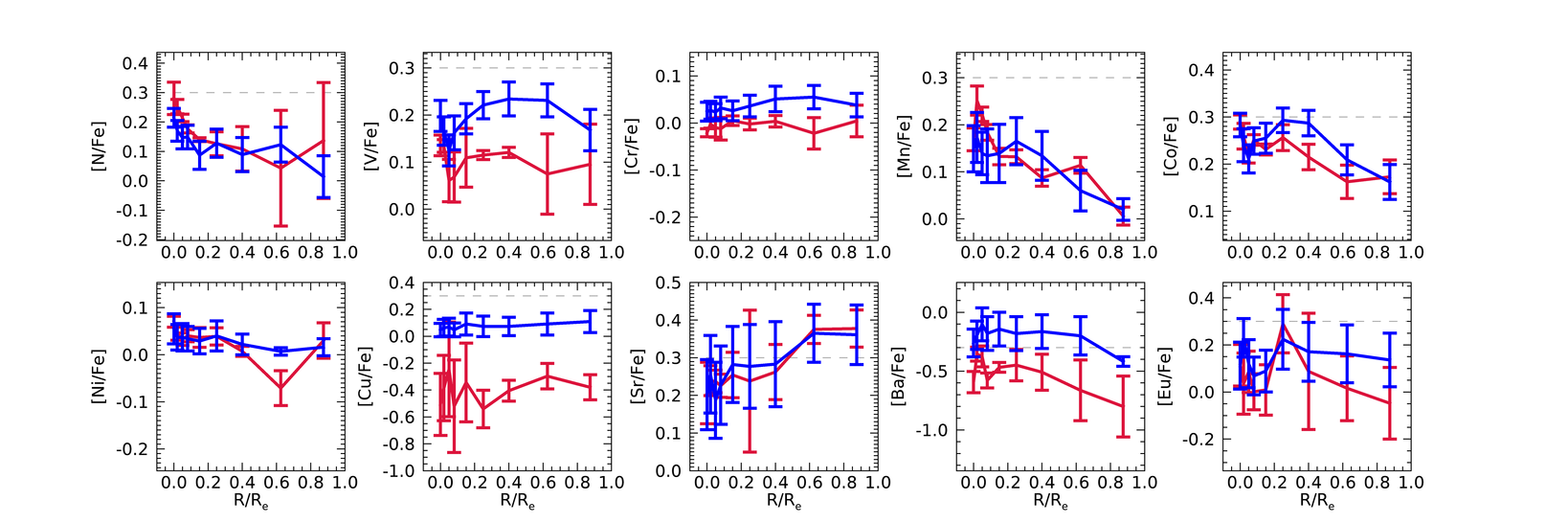}
\caption{\small{NGC1404 results. Comparison of elemental abundances values as retrieved by \textsc{ALF} (red) and \textsc{PyStaff} (blue). Horizontal dashed grey lines indicate the values of the response functions. All values of [O/Fe], [Mg/Fe], [Si/Fe], [Ca/Fe] and [Ti/Fe] have been corrected as suggested by the \textsc{ALF} documentation (see text for details).}}
\label{fig:res_elements14}
\end{centering}
\end{figure*}
\begin{figure*}[ht!]
\begin{centering}
\includegraphics[width=17.5cm]{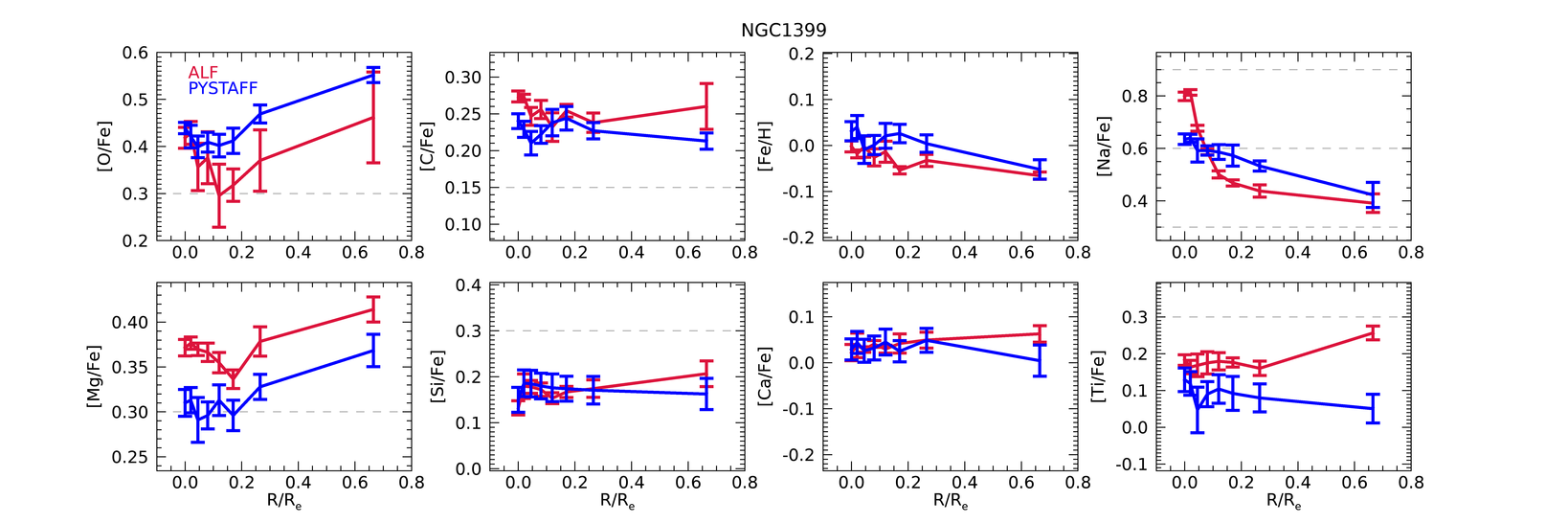}
\includegraphics[width=18cm]{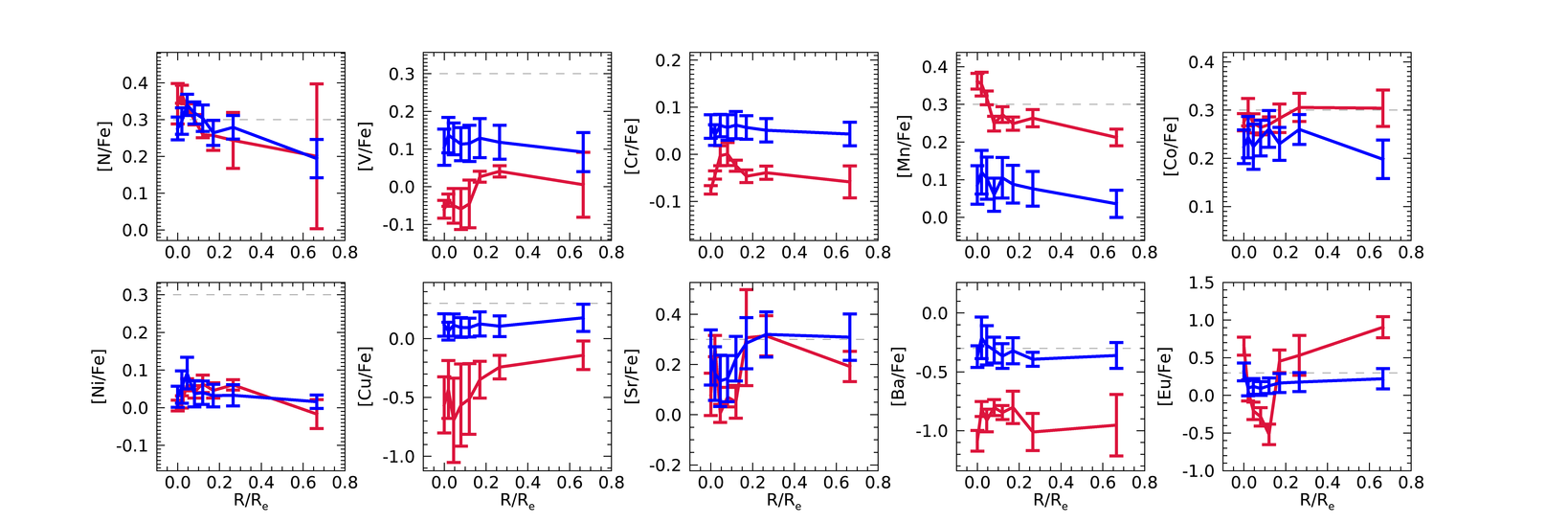}
\caption{\small{NGC1399 results. Same description as Figure \ref{fig:res_elements14}. }}
\label{fig:res_elements13}
\end{centering}
\end{figure*}

\textsc{ALF} and \textsc{PyStaff} are two of the few codes that allow the retrieval of non-solar values of numerous elemental abundances. Given the response functions provided by \citet{conroy18}, we could derive the radial gradients of $19$ elemental abundances for both our objects, NGC1404 and NGC1399. The retrieved values of each element as a function of galaxy radius are shown in Figures \ref{fig:res_elements14} and  \ref{fig:res_elements13}, with red lines for \textsc{ALF} and blue lines for \textsc{PyStaff}. As suggested in the \textsc{ALF} documentation, to correct for the fact that models with non-solar values of elemental abundances are built from stars located in the solar neighborhood, we applied further corrections to the following elemental abundances: O, Ca, Mg, Ti and Si, following the library corrective factors from \citet{schiavon07} and \citet{bensby14}. Such corrections increase for decreasing metallicity values, so these are more important for outer regions. For example, for our targets, assuming a SSP, the lowest metallicity values reach $\sim-0.3$ dex (Figures \ref{fig:results14}, \ref{fig:results13}): in this case the correspondent additive factors would be of the order of $0.1$ dex. 

Retrieving the values of elemental abundances is very important since in our wavelength range, all spectral features that change with the IMF slope also have dependencies on different elemental abundances with different proportions. Not constraining these parameters can lead to severe biases in the retrieved IMF, as discussed in \citet{lonoce21}. For this reason, although not all FSF codes allowed for their retrieval, we put all the possible elemental abundances as free parameters when fitting with \textsc{ALF} and \textsc{PyStaff}. 

We also tested if our results for age, metallicity, and IMF would change when fitting with \textsc{Starlight} and \textsc{pPXF} with the same set of templates as the normal fits, but with (fixed) non-solar elemental abundances. We applied the response functions to the basic templates with values of all $19$ elemental abundances taken from \textsc{Alf}'s results (Figures \ref{fig:res_elements14} and \ref{fig:res_elements13}), and ran the fit again on a subset of radial bins of each galaxy.

\begin{figure*}[ht!]
\begin{centering}
\includegraphics[width=12cm]{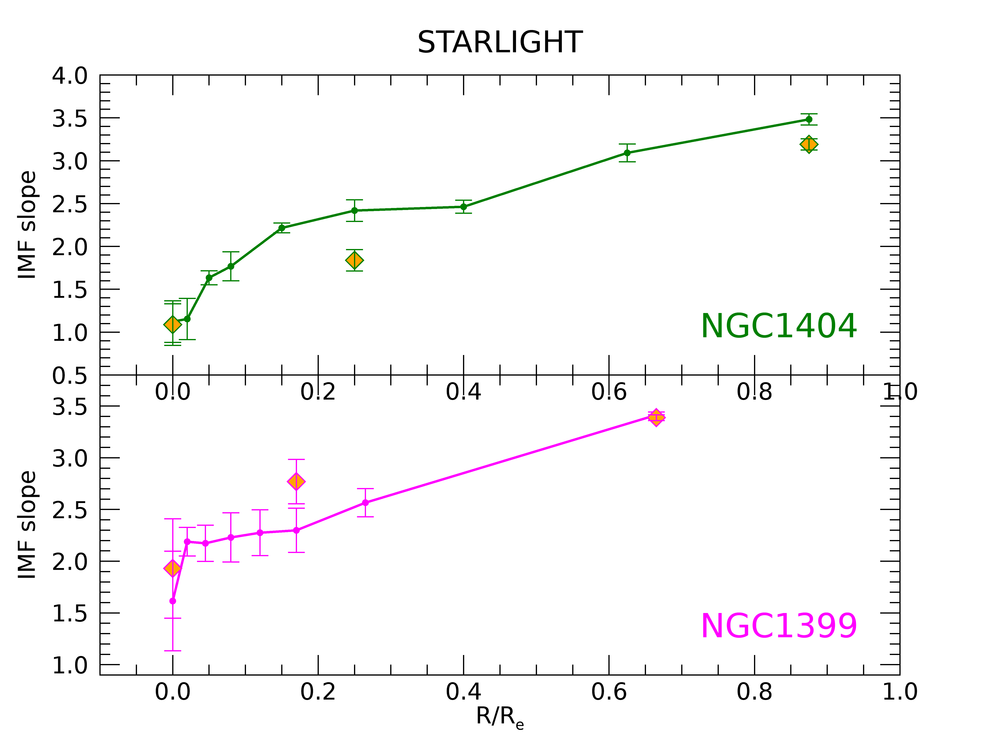}
\includegraphics[width=7cm]{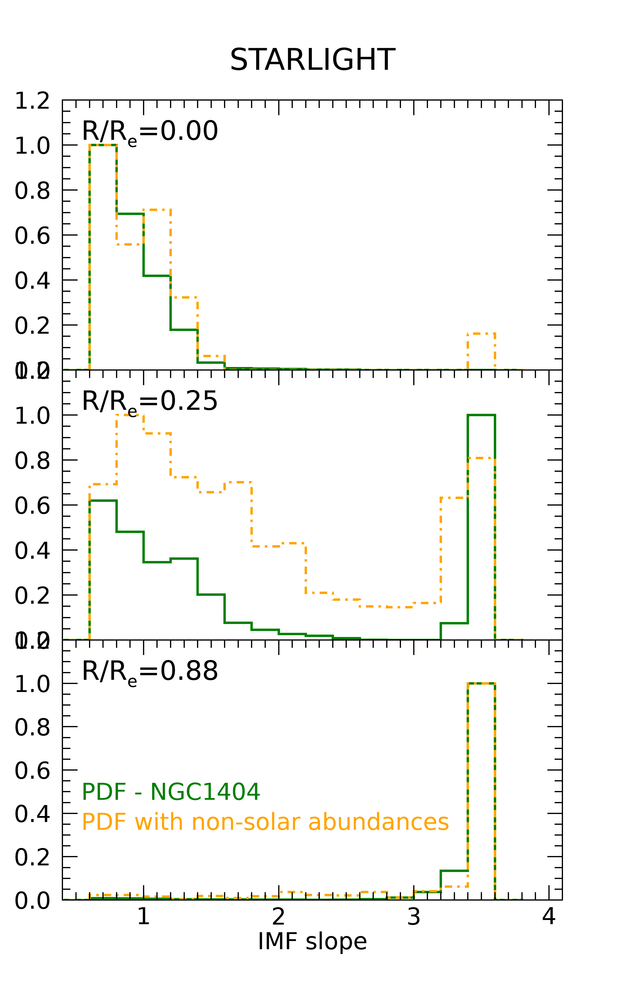}
\includegraphics[width=7cm]{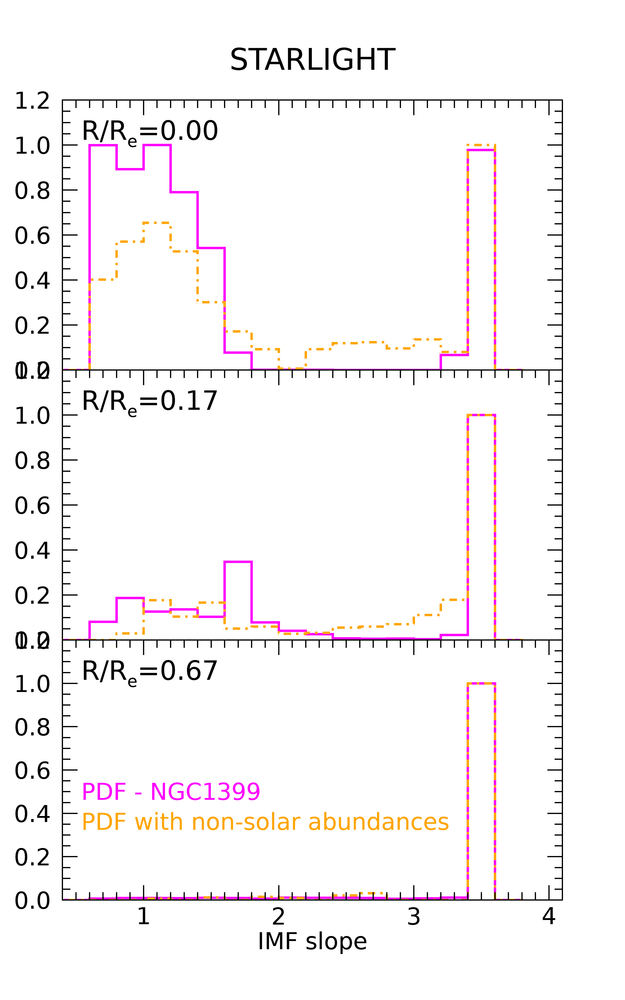}
\caption{\small{Comparison of IMF trends when fitting with \textsc{Starlight} with solar and non-solar elemental abundances. Non-solar abundances here are not meant to be fitted as free parameters as done for \textsc{ALF} and \textsc{PyStaff}, but fixed in the library models before fitting. Upper panel: IMF slope trends as weighted means as a function of radius for NGC1404 (green line) and NGC1399 (magenta line). Filled orange symbols are the retrieved values when non-solar elemental abundances are fixed in the models. Lower panel: PDFs distributions for the IMF slope values for NGC1404 (green) on the left and NGC1399 (magenta) on the right. Solid lines are the solar values results while dashed-dotted orange lines are the non-solar values analogues.}}
\label{fig:elem-SLimf}
\end{centering}
\end{figure*}

\begin{figure*}[ht!]
\begin{centering}
\includegraphics[width=12cm]{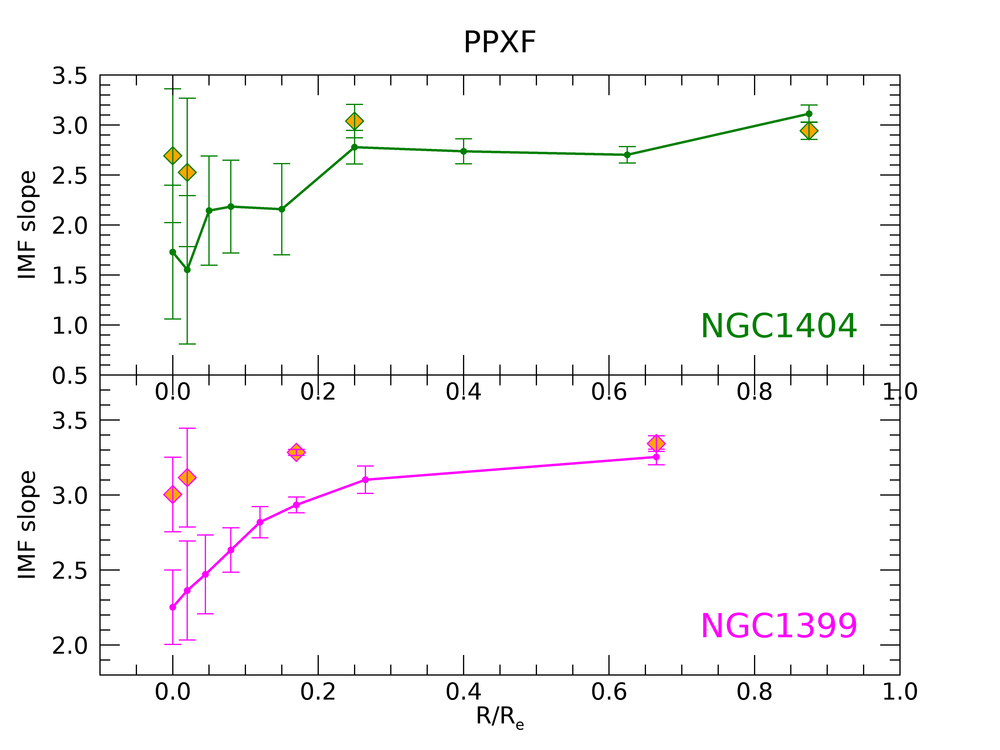}
\includegraphics[width=7cm]{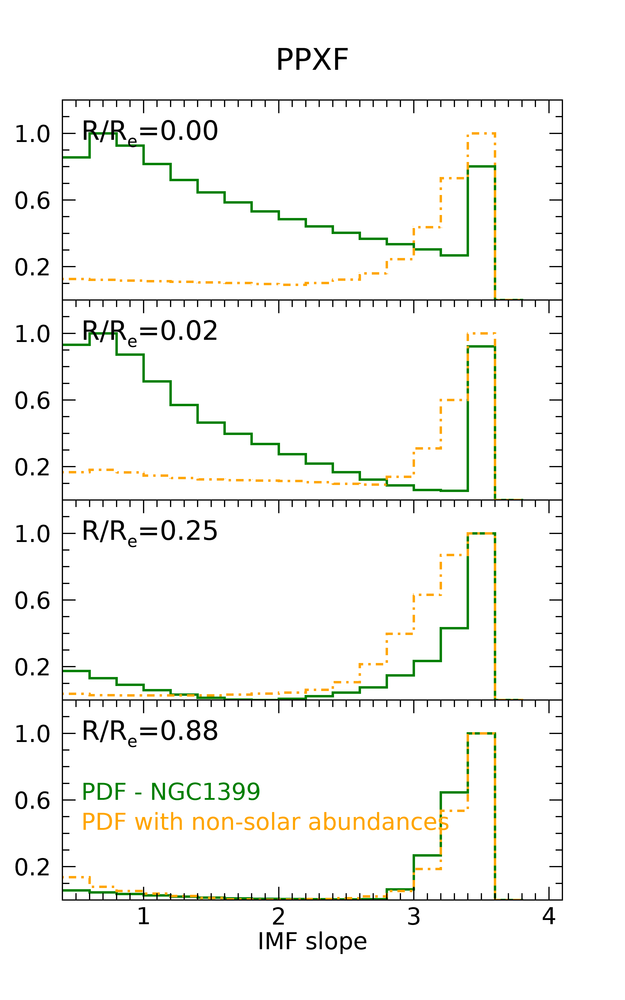}
\includegraphics[width=7cm]{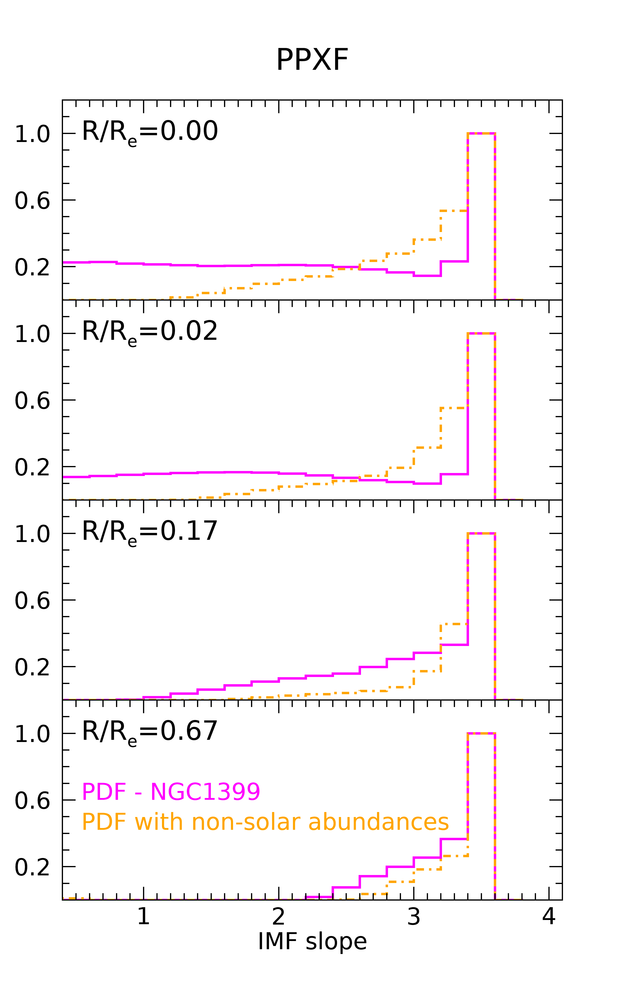}
\caption{\small{Same as Figure \ref{fig:elem-SLimf} but for \textsc{pPXF}.}}
\label{fig:elem-PPXFimf}
\end{centering}
\end{figure*}

The results are shown in Figures \ref{fig:elem-SLimf} and \ref{fig:elem-PPXFimf} for \textsc{Starlight} and \textsc{pPXF} respectively. In the upper panels we show the IMF weighted mean trends for NGC1404 (green) and NGC1399 (magenta) as they appeared in Figures \ref{fig:results14} and \ref{fig:results13}, superimposed with a few radial bins for which we tested the non-solar abundance models (orange filled symbols). In the lower panels we show the corresponding PDF distributions again in the case of fitting solar (solid green/magenta lines) and non-solar (dotted-dashed orange lines) abundance models. The comparison in the upper panels presents some inconsistent points as the middle (orange) point in Figure \ref{fig:elem-SLimf}, which deviates by $\sim2\sigma$ from the original value. However, we caution that the error bars on the weighted means, which were computed  with standard statistical techniques, do not reflect the actual broadening of the PDF, particularly when the PDF shows multiple peaks. Indeed, for the example of the middle point, by looking at its corresponding PDF in the lower panel, we see that the results are similarly pointing to the presence of two components peaked at consistent IMF slope values, only with slightly different weight distributions. \\
In general, by looking at the PDFs of the results of \textsc{Starlight}, we see a good consistency between those fitted with solar and non-solar abundances. The consistency is even better for age and metallicity (not shown). For \textsc{pPXF} we instead notice a discrepancy in the IMF slopes distributions (Figure \ref{fig:elem-PPXFimf} lower panels) in the two central bins, pushing the main component toward bottom-heavier slopes. This discrepancy is reflected also in the metallicity distributions (not shown) that appear to peak more toward solar and super-solar values (but not in the age distributions). This is likely a sign of a degeneracy that appears to bias the results of Z, IMF and elements. However, in particular for NGC1399, the probability of the presence of the second component with top-heavy IMF slope is non zero, and increases from the very central radial bin (R/R$_e=0.0$) to its adjacent one at R/R$_e=0.02$.\\
With this exercise we have shown that our results are generally stable when accounting for non-solar elemental abundances, although the most reliable way to derive the IMF and other stellar parameters is to fit for elemental abundances as free parameters. But more importantly, we have shown once again how important is to look at the probability distribution functions when fitting with codes such as \textsc{Starlight} and \textsc{pPXF}, to avoid misinterpretations of the parameter trends.

Returning to the results of abundance ratios, NGC1404 (Figure \ref{fig:res_elements14}) shows mild radial gradients in the elemental abundances, with the exception of Na (negative gradient, typical of elliptical galaxies), Mg (positive gradient) and Mn (negative gradient). The two codes give generally consistent results, mostly within R/R$_e=0.2$. The largest discrepancies are seen at R/R$_e>0.2$ for O, Fe, Na and V. We note that the extrapolation of values beyond those provided by the response functions, i.e. typically $\pm0.3$ dex, are performed in a different way in the two codes (as discussed in \citealt{lonoce21}). For \textsc{PyStaff} we adopted boundaries generally included within $\pm0.45$ dex (see Table \ref{tab:bounds}) to avoid extrapolation issues, while for \textsc{ALF} we could generally enlarge the limits to $-0.5$ to $+0.6$ dex to guarantee that all chains converged well during the fit. This difference can explain the discrepancies we observe in Figures \ref{fig:res_elements14} and \ref{fig:res_elements13} for Cu and Ba in the results for both galaxies.\\
For NGC1399, we note similarly mild gradients, again excluding Na, Mg and Mn, as well as O (positive gradient) and N (negative gradient), though the only gradient with a change of more than $0.2$ dex is Na. On the other hand, for this galaxy the consistency between results obtained with different codes is poorer. Na profiles are in tension with different trends, while O, Mg, Ti, V, Cr, Mn, Cu and Ba appear to be shifted by $\sim0.1$ dex. These shifts could be partially due to the different Na profiles obtained by the two codes, since Na shows a moderate correlation with O, Mg, Ti, V, Cr, C and Ca during the fit (see Appendix \ref{app:cross}). 

Since in the literature we often find the trend of the $\alpha$-elements instead of the detailed chemical characteristics, we have also calculated the [$\alpha$/Fe] radial profile for NGC1404 and NGC1399 by averaging our results on C, O, Mg, Ca, Si and Ti. For both galaxies the trend is flat with a constant value of [$\alpha$/Fe]$=0.2\pm0.12$ dex. If we split this quantity in three parts, averaging only C-O, Mg-Si and Ca-Ti as in \citet{lonoce23}, we find that Mg-Si follows the total [$\alpha$/Fe] with values around $0.2$ dex, while C-O presents values around $0.3$ dex and Ca-Ti around $0.1$ dex. In this case, and differently from \citet{lonoce23}, the total [$\alpha$/Fe] is effectively a tracer of Mg-Si which are generally used to derive the star formation timescale (in our case estimated to be $\sim1$ Gyr, following the equation by \citealt{thomas05}). Nevertheless, we found a difference among the three quantities, indicating that the abundance patterns are complex and worth being studied in detail. 


\section{Simulations results}  
\label{app:sim}

In Section \ref{sec:sim} we described how we set up our four sets of simulations, with different input mock spectra whose characteristics are listed in Table \ref{tab:sim}. Here we show and comment on the results we obtained from all performed simulations.

\begin{figure*}[ht!]
\begin{centering}
\includegraphics[width=18.0cm]{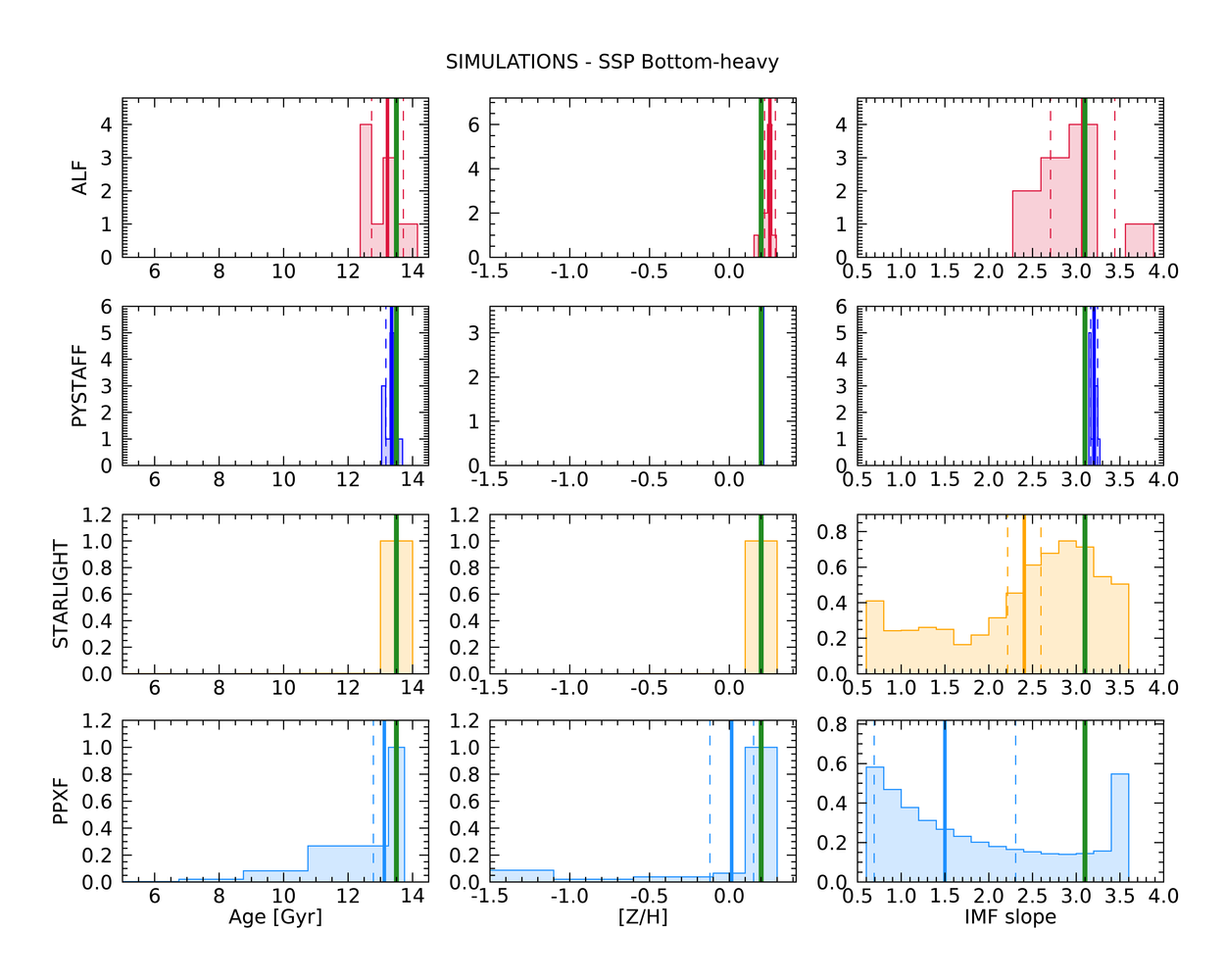}
\caption{\small{Simulation results SSP - BH. Each column shows the distribution of retrieved parameters from each code. From top to bottom: \textsc{ALF} (red), \textsc{PyStaff} (blue), \textsc{Starlight} (orange) and \textsc{pPXF} (light blue). Vertical colored lines (color-coded as the codes) indicate the weighted mean of the distribution of weighted means (solid) and their respective standard deviation (dashed). Vertical green lines show the input values. For \textsc{Starlight} and \textsc{pPXF}, whose solutions are a superposition of templates with different associated weights, we show the mean distribution of the 10 weights distributions. Input parameters of mock spectra can be found in Table \ref{tab:sim}.} }
\label{fig:sim_sspBH}
\end{centering}
\end{figure*}

\begin{figure*}[ht!]
\begin{centering}
\includegraphics[width=18.0cm]{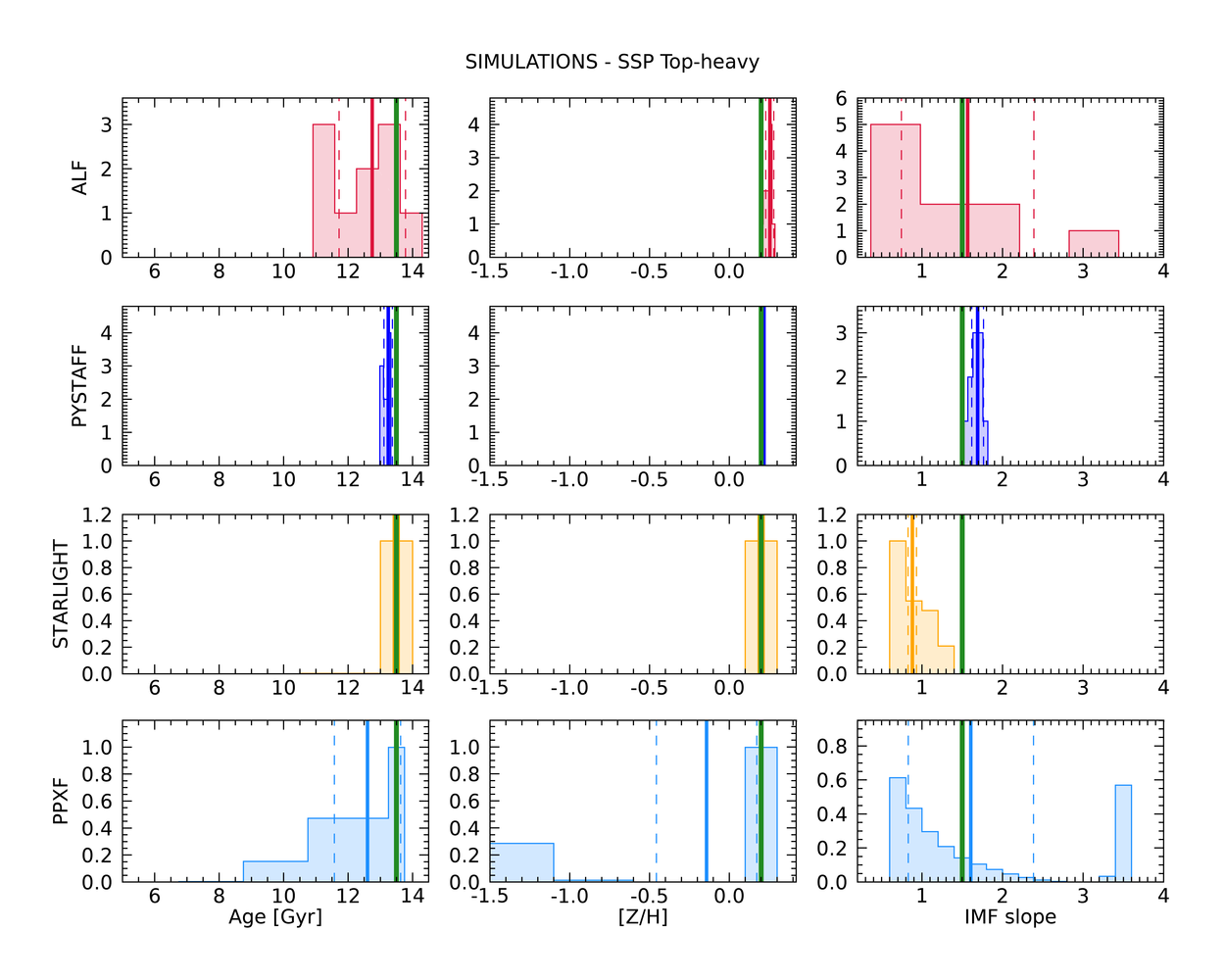}
\caption{\small{Simulation results SSP - TH. Same description as for Figure \ref{fig:sim_sspBH}. }}
\label{fig:sim_sspTH}
\end{centering}
\end{figure*}

\begin{figure*}[ht!]
\begin{centering}
\includegraphics[width=18.0cm]{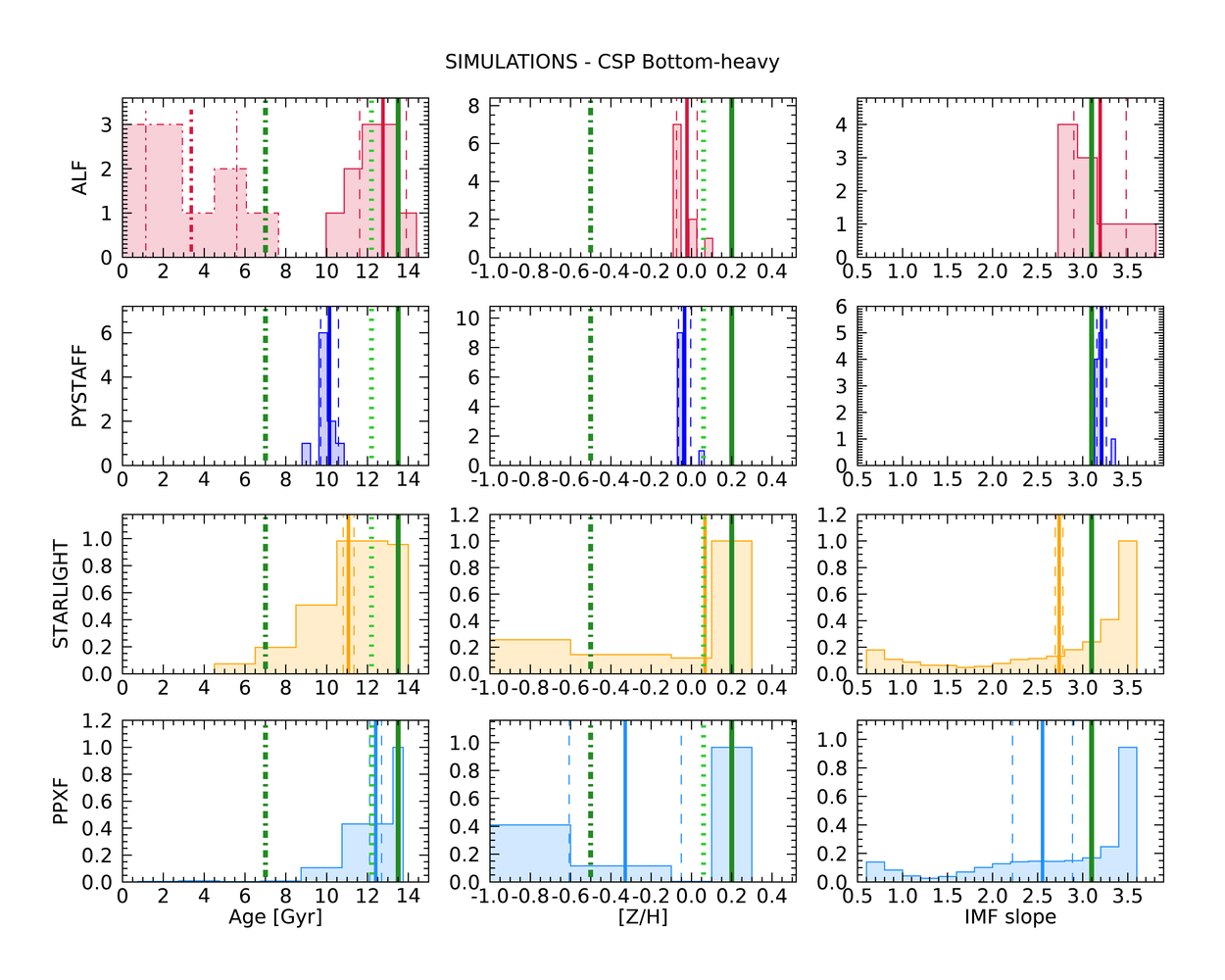}
\caption{\small{Simulation results CSP - BH. Same description as for  Figure \ref{fig:sim_sspBH}. Vertical dot-dashed dark green lines indicate the input values of the secondary stellar component. Vertical dotted light green lines indicate the light-weighted stellar parameter means for the composite population}. }
\label{fig:sim_sfhBH}
\end{centering}
\end{figure*}

\begin{figure*}[ht!]
\begin{centering}
\includegraphics[width=18.0cm]{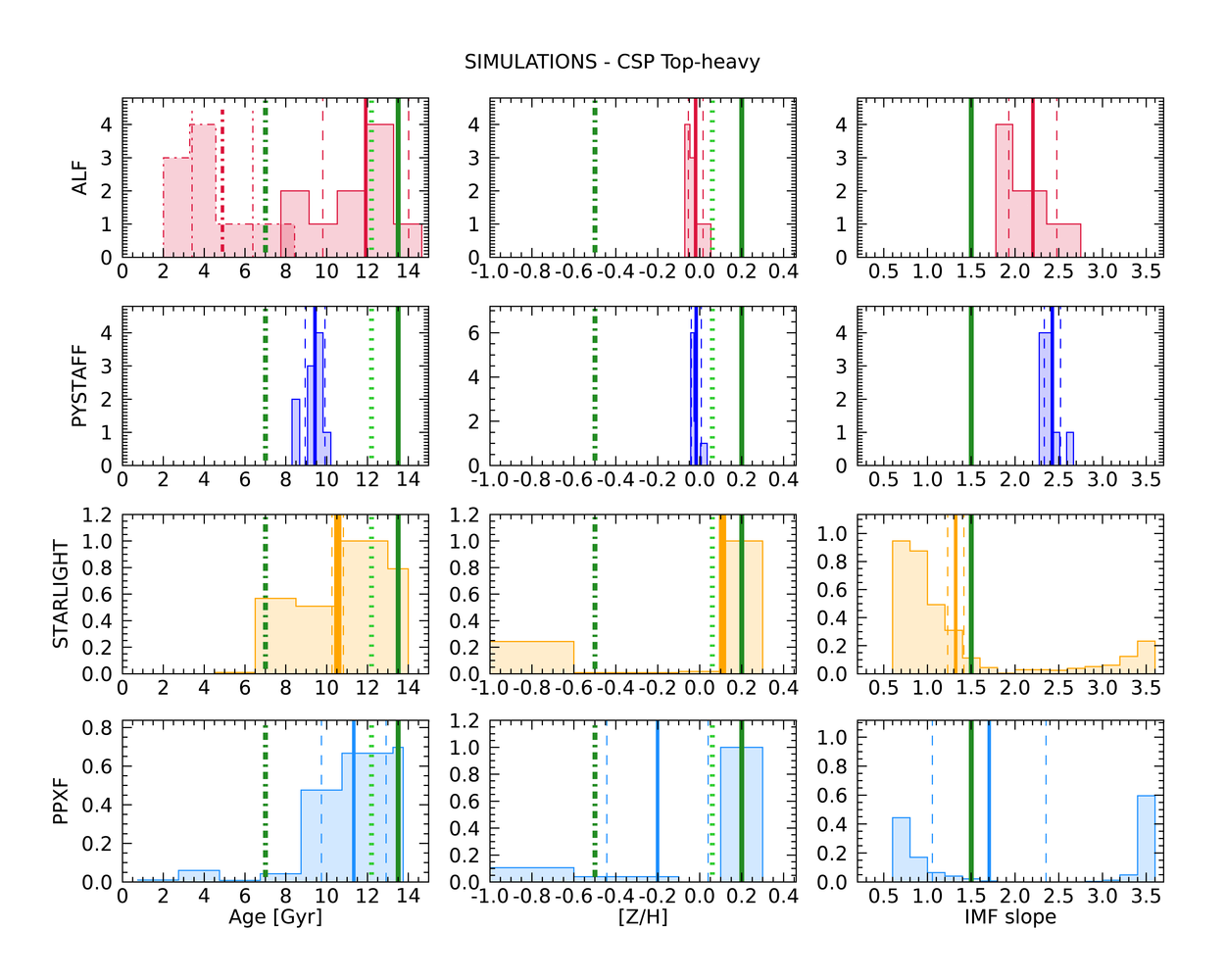}
\caption{\small{Simulation results CSP - TH. Same description as for Figure \ref{fig:sim_sfhBH}. }}
\label{fig:sim_sfhTH}
\end{centering}
\end{figure*}

\begin{figure*}[ht!]
\begin{centering}
\includegraphics[width=18.cm]{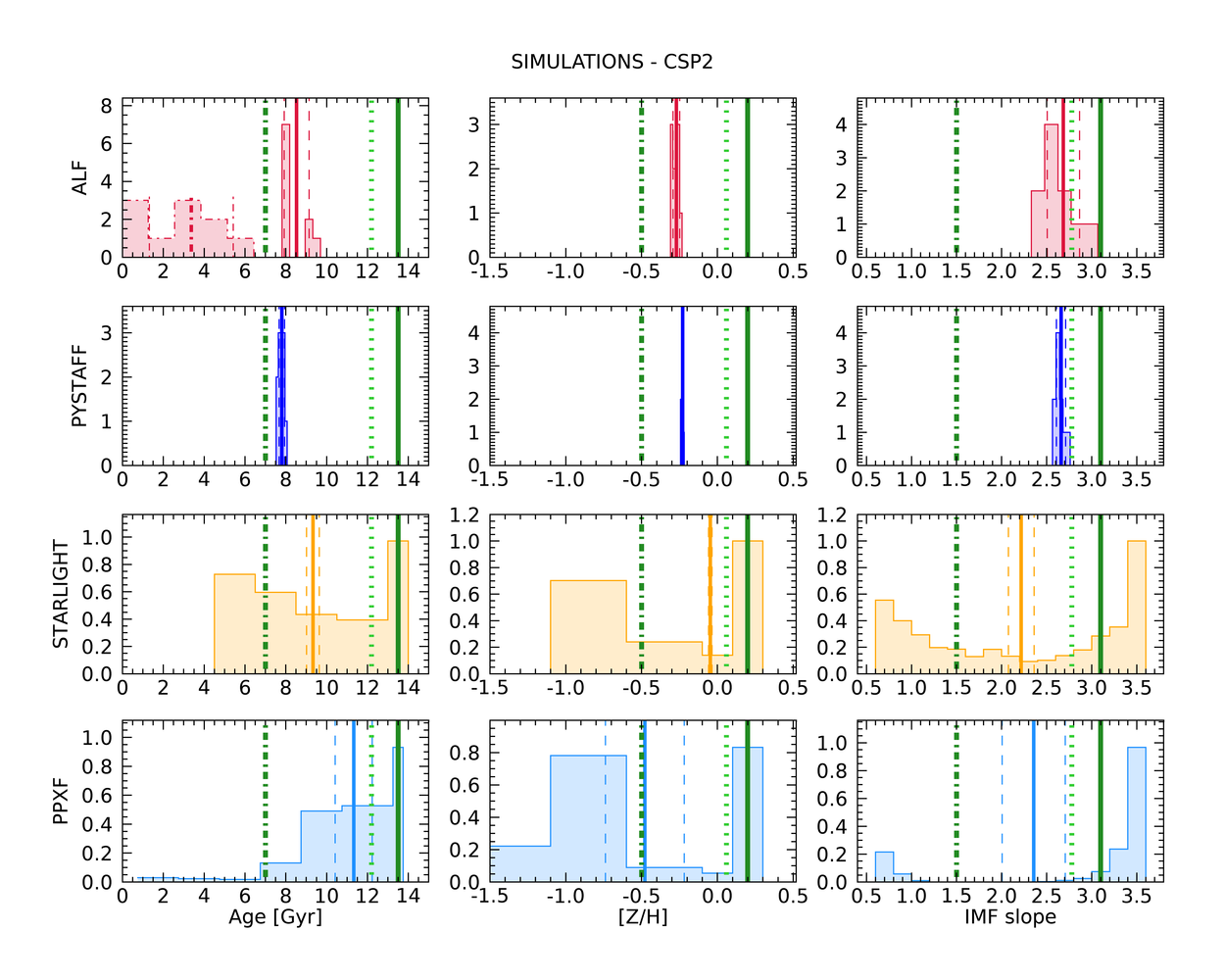}
\caption{\small{Simulation results CSP2: two ages ($13.5$ and $7.0$ Gyr), two [Z/H] values ($0.20$ and $-0.50$ dex), and two IMF slope values ($3.1$ and $1.5$). Same description as Figure \ref{fig:sim_sfhBH}. }}
\label{fig:sim_sfh2}
\end{centering}
\end{figure*}

\begin{figure*}[ht!]
\begin{centering}
\includegraphics[width=18.cm]{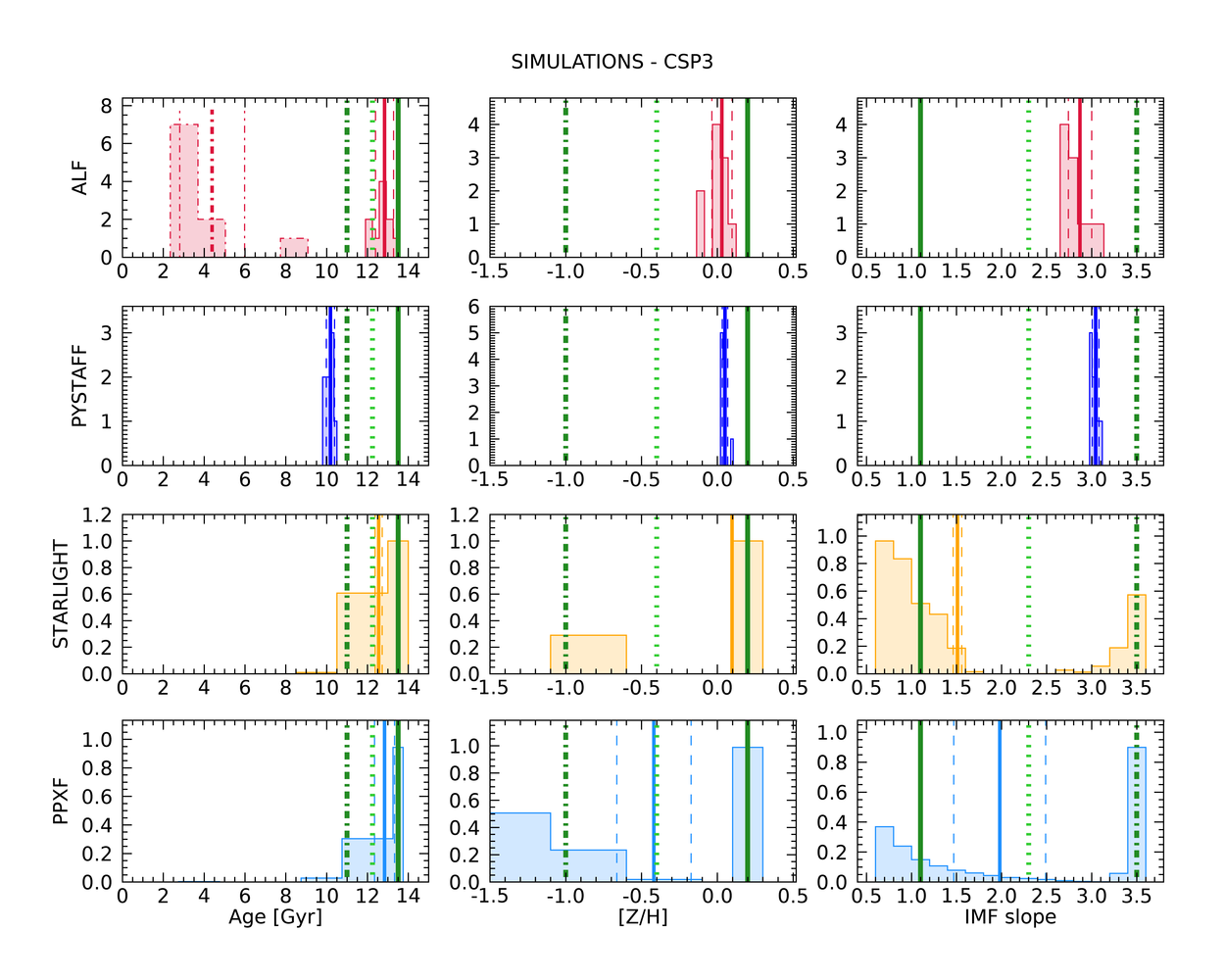}
\caption{\small{Simulation results CSP3: two ages ($13.5$ and $11.0$ Gyr), two [Z/H] values ($0.20$ and $-1.00$ dex), and two IMF slope values ($1.1$ and $3.5$). Same description as for Figure \ref{fig:sim_sfh2}.}}
\label{fig:sim_sfh3}
\end{centering}
\end{figure*}

In Figures \ref{fig:sim_sspBH} - \ref{fig:sim_sfh3}, we present the comparison among the four FSF codes on the retrieval of the stellar population properties (age, metallicity and IMF slope) for each set of simulations. In each figure, the top row refers to \textsc{ALF} results (red), followed by \textsc{PyStaff} (blue), \textsc{Starlight} (orange) and \textsc{pPXF} (light blue). In detail, for \textsc{ALF} and \textsc{PyStaff} the distributions of the mean values obtained from each of the 10 runs are shown; vertical solid lines (red and blue respectively) show the average of the distributions and dashed lines their $1\sigma$ limits. For \textsc{Starlight} and \textsc{pPXF}, whose weighted means can be biased by the presence of multiple peaks in the parameter distributions, we show the average weights distributions computed on the 10 simulated fits. However, solid and dashed lines (orange and light blue respectively) indicate the mean and standard deviation obtained from the weighted mean distributions as for the other codes: this is to highlight in each case the possible bias that can arise when blindly adopting the weighted means instead of inspecting the weights distributions. In all plots, green vertical lines show the input values from the mock spectra  and the light green dotted lines indicate the light-weighted means of each input quantity. 

In the following list we enumerate the results of stellar population properties for each set of simulations.
\begin{enumerate}

    \item \textbf{SSP - TH and BH:} (Figures \ref{fig:sim_sspBH}, \ref{fig:sim_sspTH}) Age and metallicity are very well determined for all FSF codes. Histogram bins vary since each code interpolates (or not) parameters of the base templates in a different way. \textsc{Starlight}, in particular, allows for up to $300$ base models; when also including IMF variations, this is inevitably at the expense of the age and metallicity values grid. 
    The IMF slope is retrieved well by \textsc{PyStaff} and \textsc{ALF}. \textsc{Starlight} and \textsc{pPXF} both show a broad distribution when the input is BH and a biased one for TH with a shift of $\sim0.7$ for \textsc{Starlight} and a double peak for \textsc{pPXF}. \textsc{pPXF} also shows a fraction of sub-solar metallicity values for both TH and BH. 
    
    \item \textbf{CSP - TH and BH:} (Figures \ref{fig:sim_sfhBH}, \ref{fig:sim_sfhTH}) In both cases, the presence of $20\%$ of a younger and lower metallicity component is significantly affecting the retrieval of all parameters. The main component's age is still well-determined by \textsc{ALF}, \textsc{Starlight} and \textsc{pPXF}, while \textsc{PyStaff} presents age values around $10$ Gyr, i.e. in between the ages of the two populations. Its metallicity value is correctly retrieved only by \textsc{Starlight} and \textsc{pPXF}. \textsc{ALF} and \textsc{PyStaff}, instead, show solar values, deviating from the main component value of $0.2$ dex. 
    The younger component's age is never well defined, with \textsc{ALF} detecting a younger component but with a $\sim2-3$ Gyr younger age, and \textsc{Starlight} just giving a broader age distribution on the side of younger ages.
    Alternately, the metallicity of the second component is detected by \textsc{Starlight} and \textsc{pPXF}, but the secondary peak is not very well shaped and constrained.
    Regarding the IMF, BH inputs generally work better: \textsc{ALF} and \textsc{PyStaff}, despite their difficulties in retrieving the age and metallicity, are able to recover unbiased IMF values, and both \textsc{Starlight} and \textsc{pPXF} show distributions centered on the input values, although with broad distributions. For TH IMF, we record more biases: both \textsc{ALF} and \textsc{PyStaff} are off by about $0.7-1.0$; \textsc{Starlight} shows a peak on an offset value ($\sim1$) and a smaller peak over the very BH range. \textsc{pPXF}, in a similar way, shows two almost equally peaked distributions at IMF slope $< 1.0$ and $\sim3.5$.
    
    \item \textbf{CSP2:} (Figure \ref{fig:sim_sfh2}) In this set, the second component not only has a different age and metallicity, but also a different IMF. In particular, the older and more metal-rich component is associated with a BH IMF slope (based on the assumption that a mass versus IMF-slope and/or metallicity versus IMF-slope relation holds within galaxies). In this case, the difference in IMF of the two components created a large bias in both the age and metallicity retrieval when fitting with \textsc{ALF} and \textsc{PyStaff}. This means that the presence of a secondary component, which in real galaxies 
    may have a different IMF in addition to a different age and metallicity, can severely affect not only the IMF retrieval but also the retrieval of the age and metallicity of the overall population. However, this only holds for codes that assume a fixed SSP during the fit. Indeed, \textsc{Starlight} and \textsc{pPXF}, created to allow for a complex SFH, succeed in detecting the differences in all parameters, including the IMF, though with broad distributions and generally biased double peaks.
    
    \item \textbf{CSP3:} (Figure \ref{fig:sim_sfh3}) This particular set was created to most closely mimic the composition of our target galaxies NGC1404 and NGC1399: two stellar populations with identical mass fractions, slightly different ages, and very different metallicity and IMF values. Similarly to the previous set, we found biased results for age and [Z/H] when fitting with \textsc{ALF} and \textsc{PyStaff}, with even more severe biases on the age of the younger component detected by \textsc{ALF} (off by $\sim7$ Gyr). On the other hand, \textsc{Starlight} and \textsc{pPXF} retrieve correct values, although the retrieved percentage of each component is different from the input one.   
\end{enumerate}



\begin{figure*}[ht!]
\begin{centering}
\includegraphics[width=18.0cm]{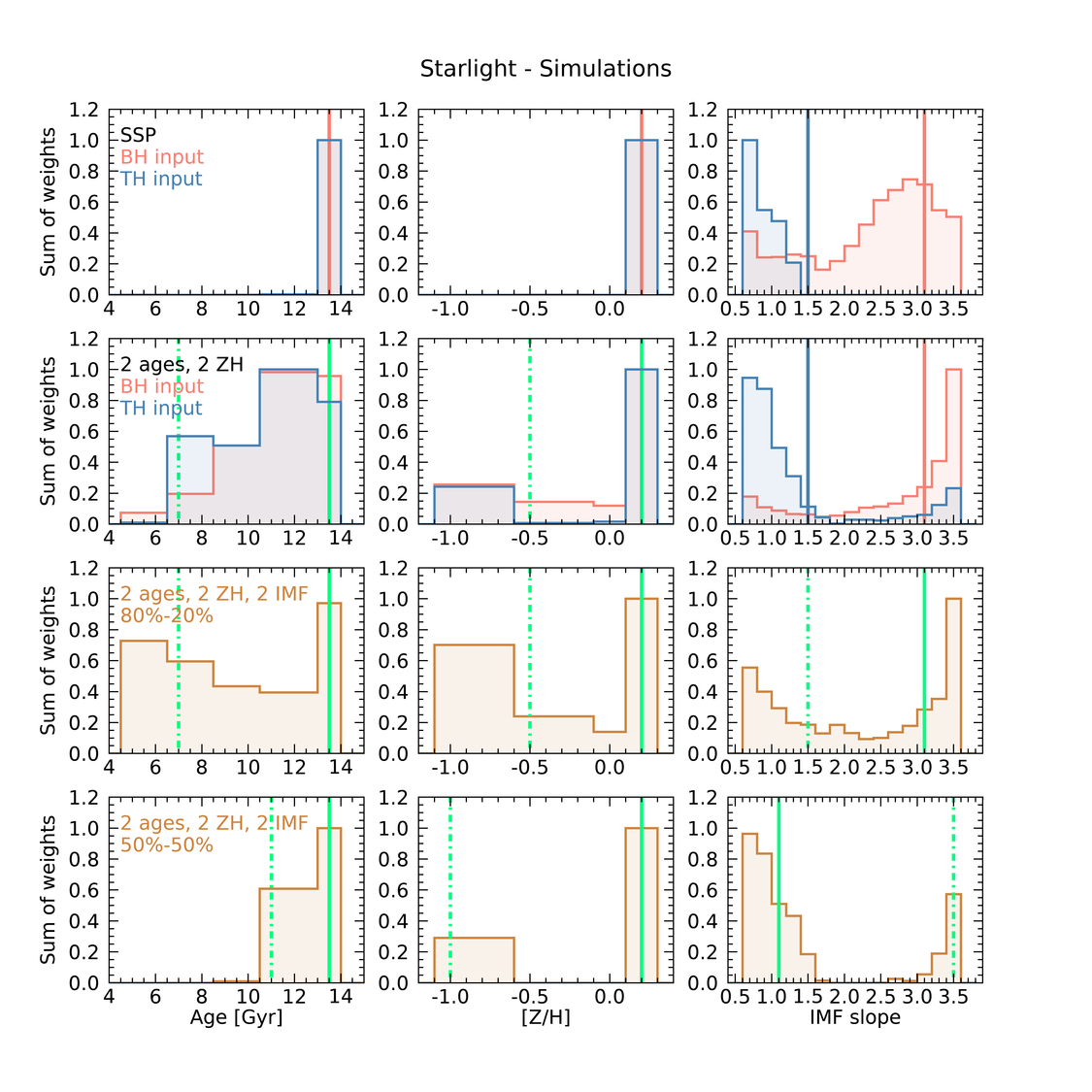}
\caption{\small{Simulation results - \textsc{Starlight}. \textit{Upper panels:} Distribution of weights summed in each parameter bin for age (left), metallicity (middle) and IMF slope (right), when the input mock spectrum is a SSP with BH IMF (blue) or TH (pink). \textit{Second row panels:} Same as upper panels but in this case the input mock spectra are built as a superposition of $80$\% of mass from an old and super-solar metallicity population and the remaining $20$\% of mass from a younger and metal-poor component (see text for more details). \textit{Lower panels:} Same as above, but in this case the two stellar components not only differ in their age and metallicity, but also in their IMF; in particular, on the third row from above, the older and more metallic component has been associated with a BH IMF, and the younger component has been associated with a TH IMF. On the bottom row, more similar to our real data results, the super-solar component has a TH IMF slope. Vertical lines indicate the input values for each parameter; in particular, green lines show the values of the two components: solid lines the main older component, dot-dashed lines the secondary younger one.}}
\label{fig:sim_SLall}
\end{centering}
\end{figure*}

\begin{figure*}[ht!]
\begin{centering}
\includegraphics[width=18.0cm]{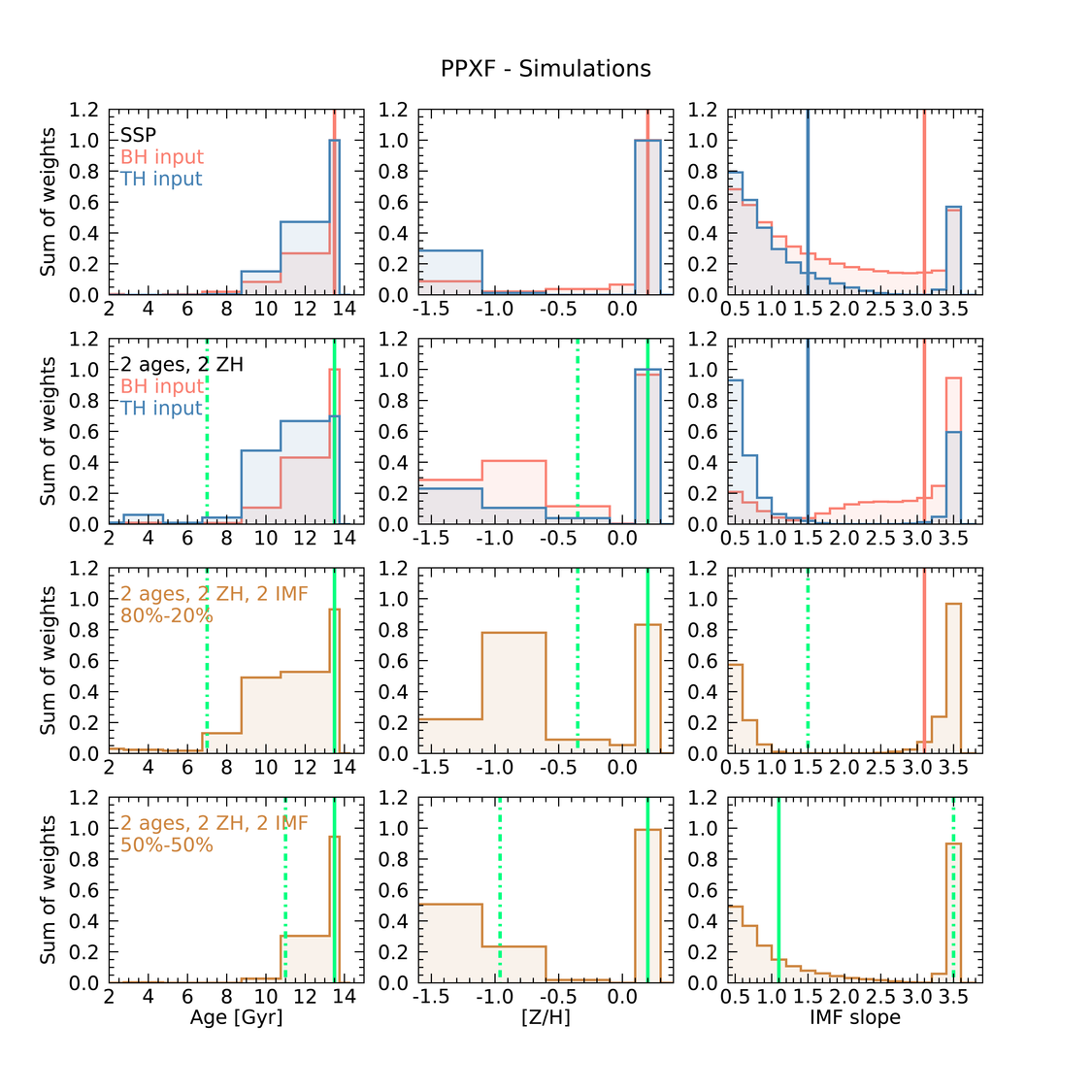}
\caption{\small{Simulation results - \textsc{pPXF}. Same as in Figure \ref{fig:sim_SLall}. }}
\label{fig:sim_PPall}
\end{centering}
\end{figure*}

\begin{figure*}[ht!]
\begin{centering}
\includegraphics[width=18.0cm]{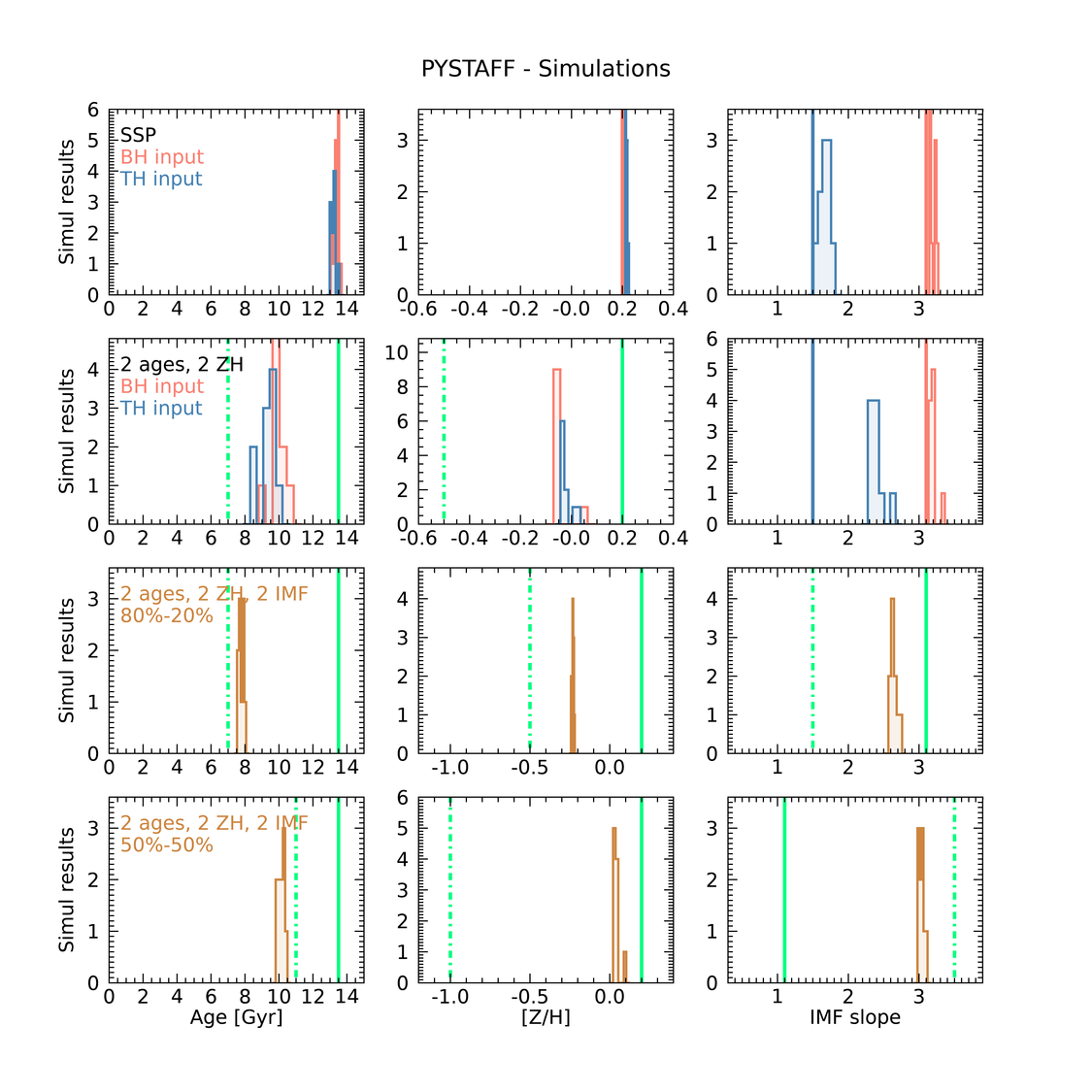}
\caption{\small{Simulation results - \textsc{PyStaff}. Same as in Figure \ref{fig:sim_SLall}.}}
\label{fig:sim_PYall}
\end{centering}
\end{figure*}

\begin{figure*}[ht!]
\begin{centering}
\includegraphics[width=18.0cm]{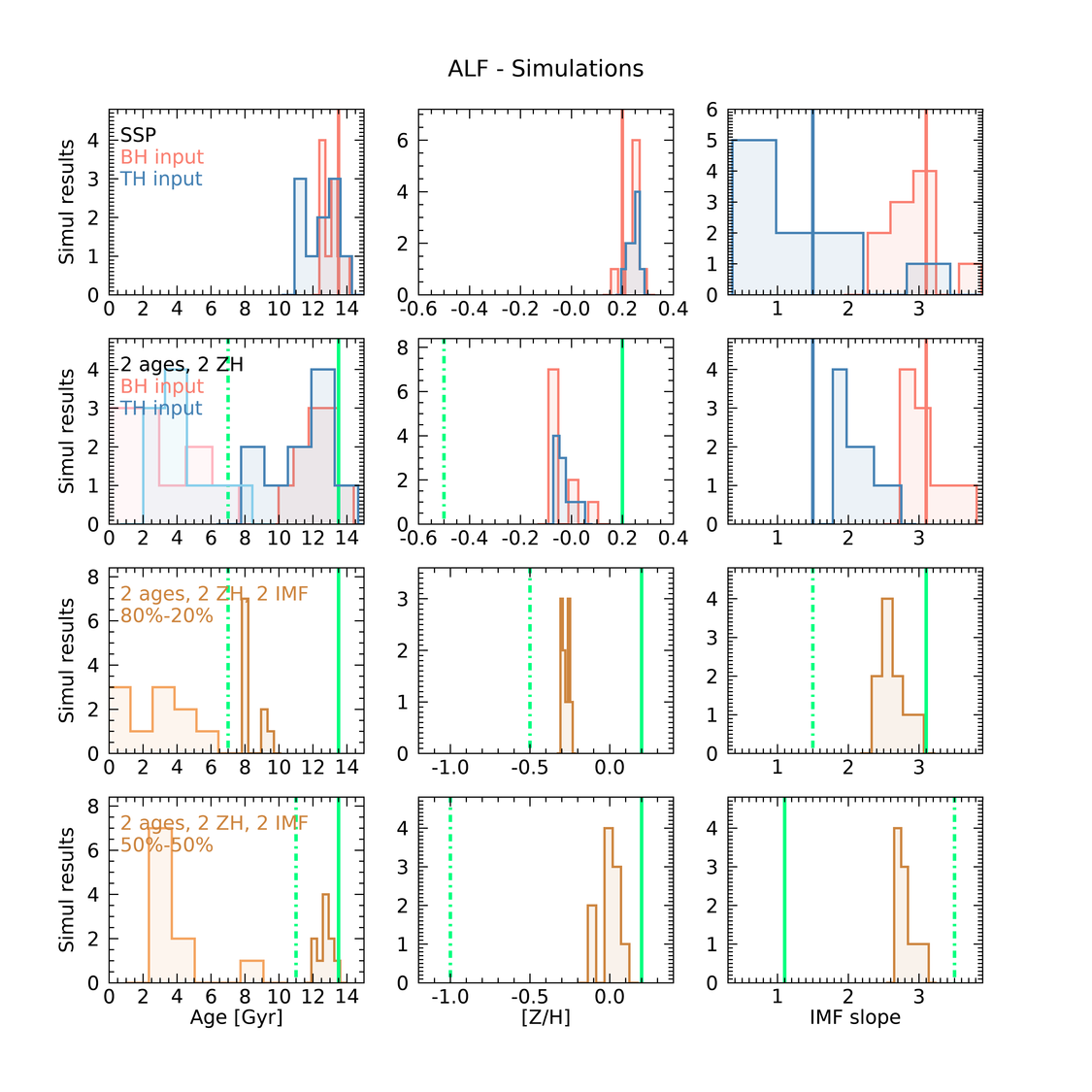}
\caption{\small{Simulation results - \textsc{ALF}. Same as in Figure \ref{fig:sim_SLall}. Light shaded age histograms refer to the younger stellar component as retrieved by \textsc{ALF}.}}
\label{fig:sim_Alfall}
\end{centering}
\end{figure*}

In order to directly visualize the behavior of a specific code in retrieving parameters with increasing SFH complexity of the input mock spectra, in Figures \ref{fig:sim_SLall} - \ref{fig:sim_Alfall} we again show the results of our simulations but this time grouped for each single code. In all figures, each row of panels shows in order from the top: SSP, CSP, CSP2 and CSP3. Vertical lines indicate the true input values. In all plots the decreasing quality of the retrieval when the complexity is increased is clearly visible: more biased results for \textsc{ALF} and \textsc{PyStaff} and less precision for \textsc{Starlight} and \textsc{pPXF}.

\vspace{0.5cm}

Of particular importance is how \textsc{PyStaff} (and similarly but less evidently for \textsc{ALF}) always offers very precise results, with distributions close to $\delta$-functions. This is true not only for accurate results obtained for SSP inputs, but also for completely non-accurate results found for complex SFHs. Such codes are thus not giving any hints as to the possibility of the presence of multiple components in the stellar populations of the objects they are fitting, not even through the error bars. However, we recall that \textsc{PyStaff}'s input spectra were re-sampled to $1.25$ \AA\space pixel$^{-1}$, decreasing the noise level with respect to the other codes.\\
These results place a great responsibility on the choice of the FSF code, either a SSP-based one or one without an a-priori assumed SFH. That is, the results depend on the FSF code adopted. In the former case, for example, the chosen code is going to produce \textit{precise} (but potentially not \textit{accurate}) results regardless of the true SFH of the studied object. This will prevent a fair comparison among galaxies from samples of objects with different SFHs. 

\subsection{Elemental abundances in simulations}
\label{app:elem_sim}

Although we created mock input spectra with the inclusion of non-solar elemental abundances (see Section \ref{sec:sim}), only \textsc{ALF} and \textsc{PyStaff} allow for their retrieval. In this section we show and discuss the comparison of the two codes in the retrieval of chemical elements when fitting the same simulation sets discussed above. 
In general we found good consistency of the results with the true input values, and \textsc{PyStaff} distributions are generally sharper than \textsc{ALF} ones. The results are shown in Figures \ref{fig:sim_elemSSP} - \ref{fig:sim_elemSFH3}. Differently from the results of the real data (see Appendix \ref{app:elem}, Figures \ref{fig:res_elements14} and \ref{fig:res_elements13}), in these plots we did not apply any correction to O, Mg, Ca, Ti and Si, as here we are just testing the ability of the codes in retrieving the input values.
\\
For SSP inputs, for both TH and BH IMF, both codes successfully retrieved the input values, with only [Na/Fe] offset by $<0.3$ dex with \textsc{ALF}. We thus confirm the strength of such codes in determining the overall stellar population properties of galaxies when assuming a SSP. 
For the simplest CSP, with same IMF, we notice a few more elemental abundances with results deviating by $>1\sigma$ from the input values, i.e.: O, C, Si, Mg, Ca, Ni and V.  
When also changing the IMF in the inputs, i.e. CSP2 and CSP3, we do not see any significant additional bias on the elemental abundances, attesting to an overall robustness of their measurement. However, we caution that the two stellar populations in our simulations have the same elemental abundance values, which is unlikely to be the case in a real galaxy.

\begin{figure*}[ht!]
\begin{centering}
\includegraphics[width=13.0cm]{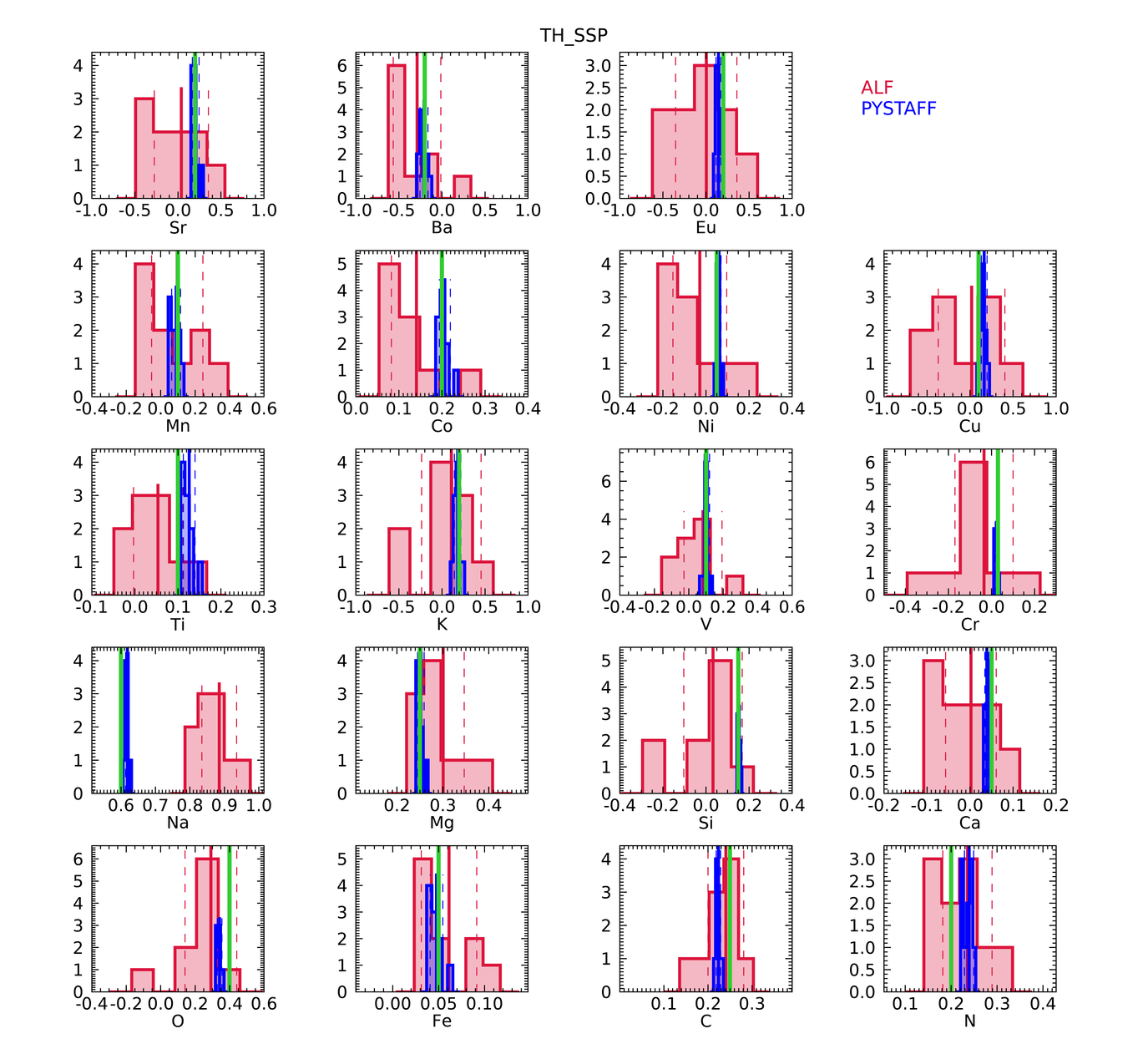}
\includegraphics[width=13.0cm]{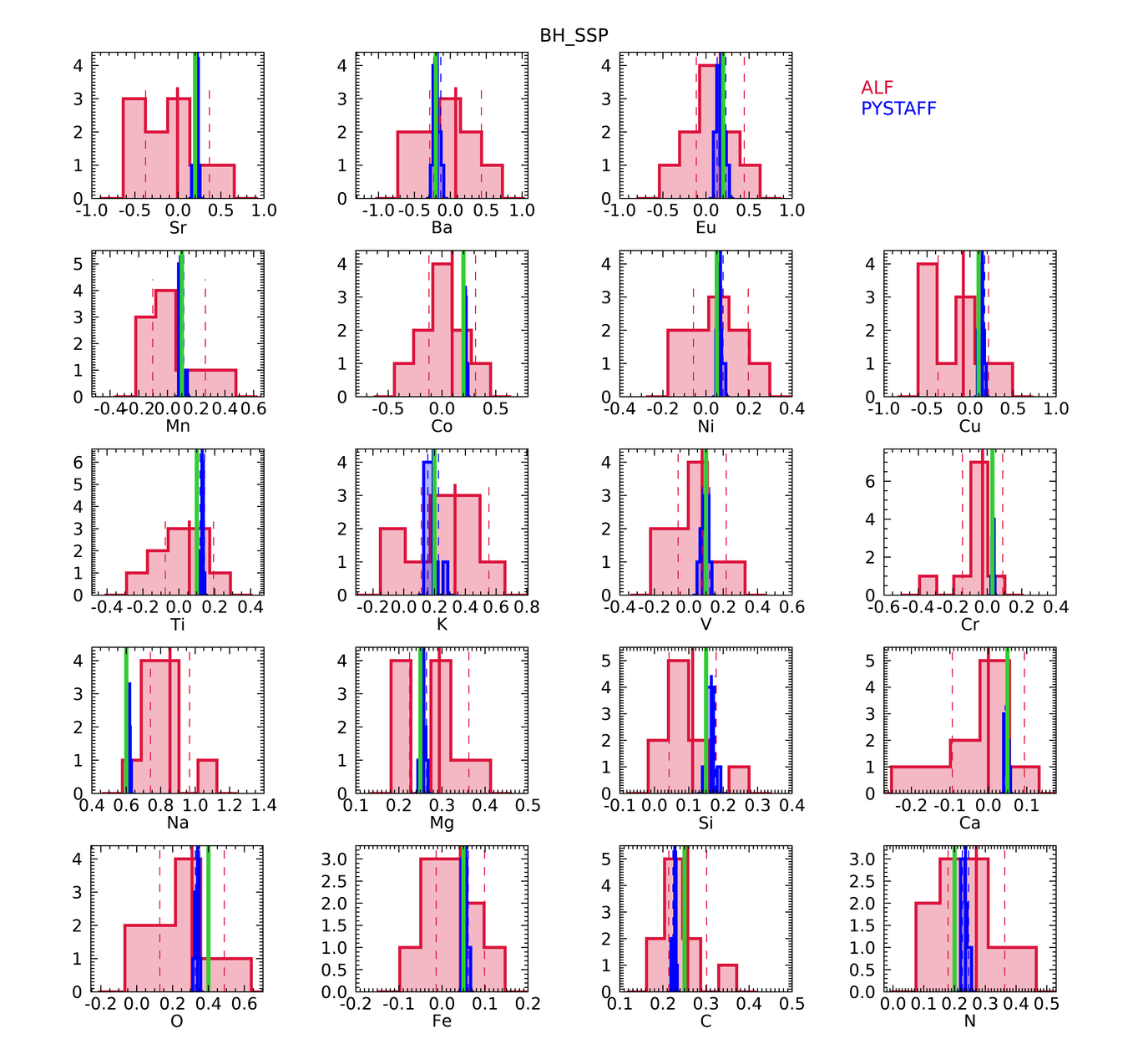}
\caption{\small{Simulation results - Elemental abundances - SSP. Comparison of \textsc{ALF} results (red) and \textsc{PyStaff} results (blue) in retrieving the elemental abundances values from simulated spectra. Vertical lines indicate the true input values. Vertical solid and dashed lines indicate respectively the mean and standard deviation of each distribution for \textsc{ALF} (red) and \textsc{PyStaff} (blue).}}
\label{fig:sim_elemSSP}
\end{centering}
\end{figure*}

\begin{figure*}[ht!]
\begin{centering}
\includegraphics[width=13.0cm]{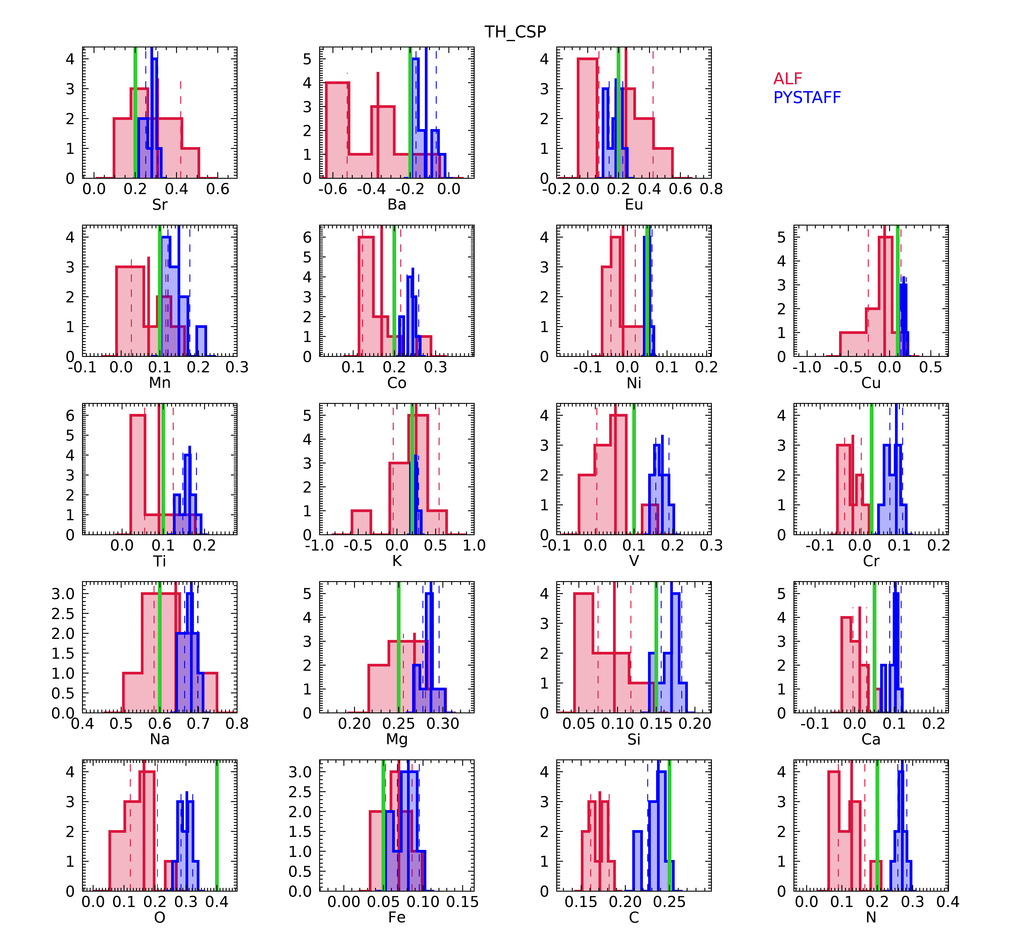}
\includegraphics[width=13.0cm]{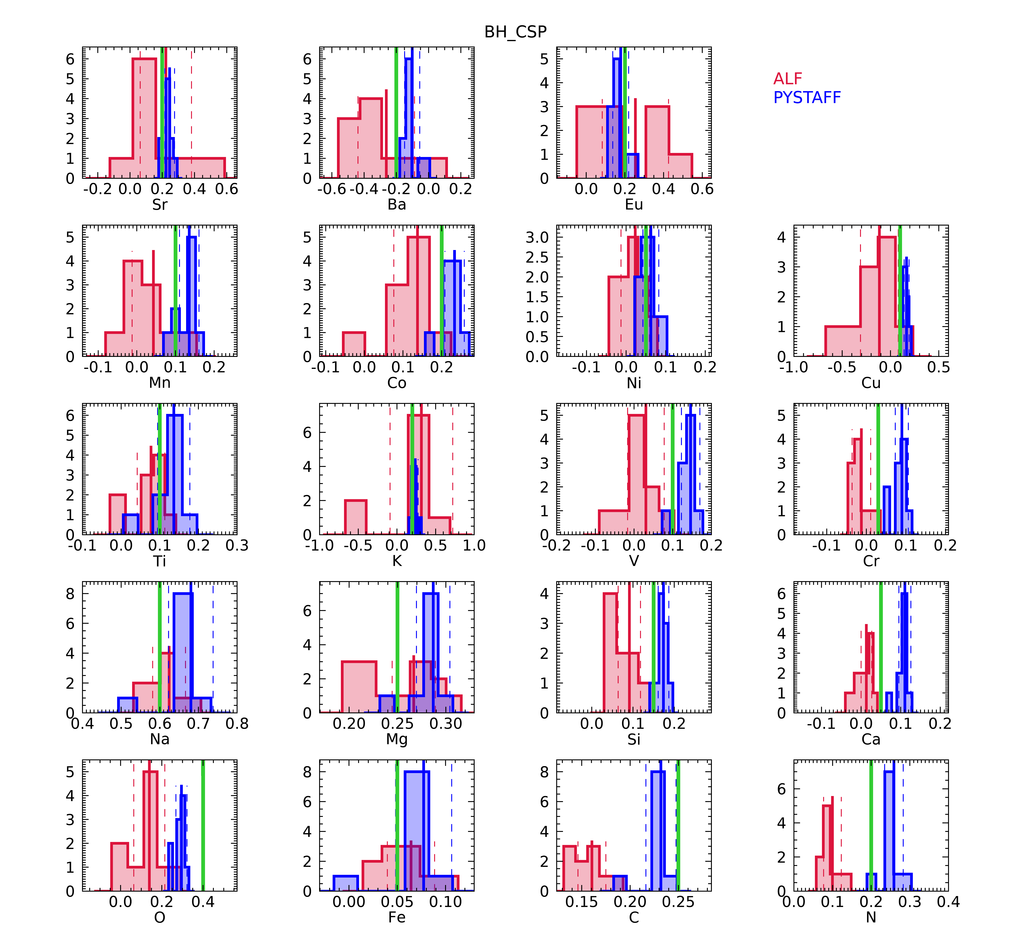}
\caption{\small{Simulation results - Elemental abundances - CSP. Same description as in Figure \ref{fig:sim_elemSSP}.}}
\label{fig:sim_elemSFH}
\end{centering}
\end{figure*}

\begin{figure*}[ht!]
\begin{centering}
\includegraphics[width=17.0cm]{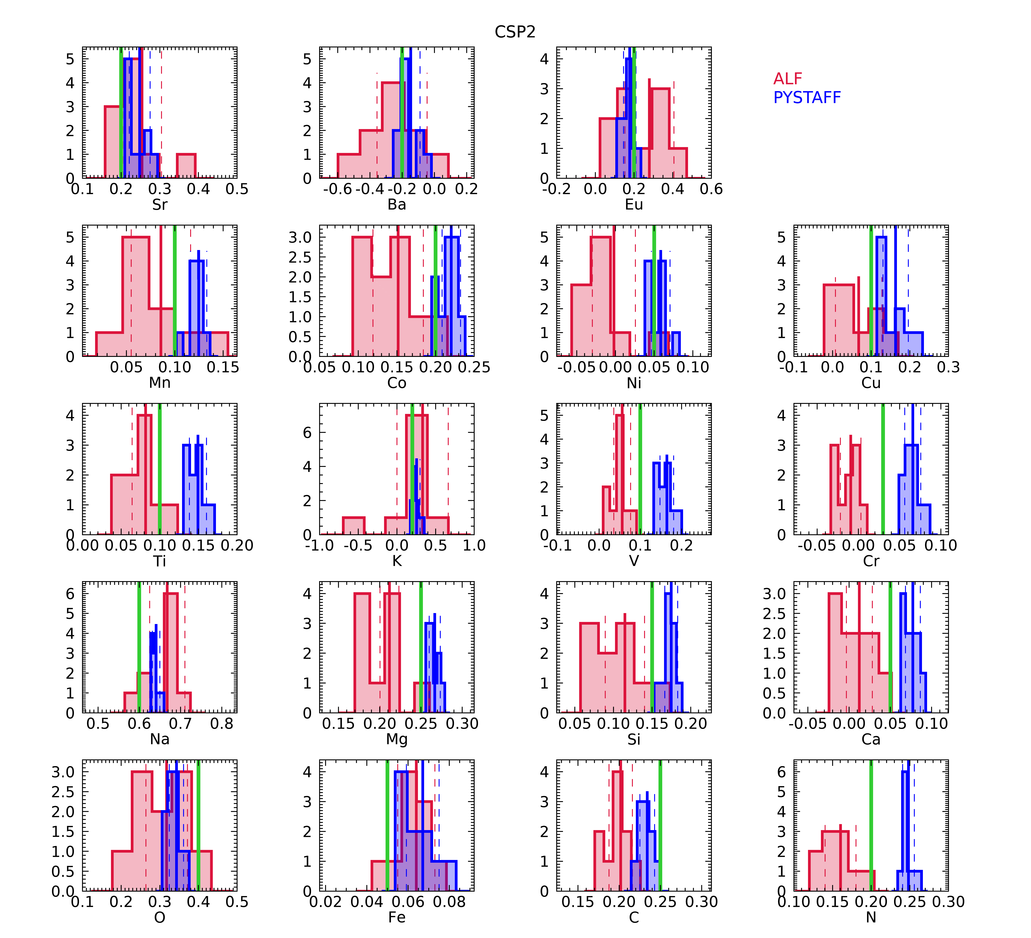}
\caption{\small{Simulation results - Elemental abundances - CSP2. Same description as in Figure \ref{fig:sim_elemSSP}.}}
\label{fig:sim_elemSFH2}
\end{centering}
\end{figure*}

\begin{figure*}[ht!]
\begin{centering}
\includegraphics[width=17.0cm]{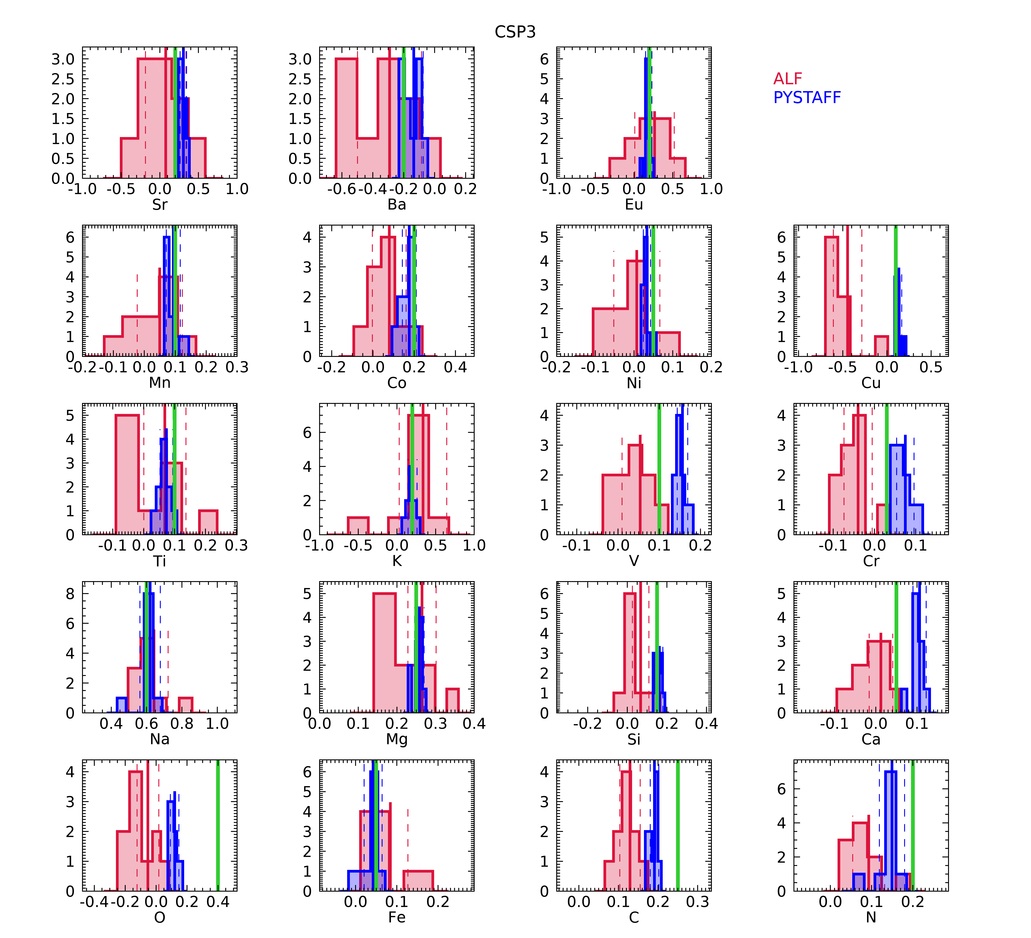}
\caption{\small{Simulation results - Elemental abundances - CSP3. Same description as in Figure \ref{fig:sim_elemSSP}.}}
\label{fig:sim_elemSFH3}
\end{centering}
\end{figure*}

\section{Residuals analysis}
\label{app:residuals}

As briefly discussed in Section \ref{sec:analysis}, to compare the goodness of each fit produced by the four FSF codes, one is naturally tempted to quantify and discuss the level of the residuals and find the code whose best fit is closest to the input spectrum. In this Appendix we want to show some examples of the  residuals spectra obtained for each code and demonstrate that the comparison is actually misleading, as hinted in the main text. 

\begin{figure*}[ht!]
\begin{centering}
\includegraphics[width=18.0cm]{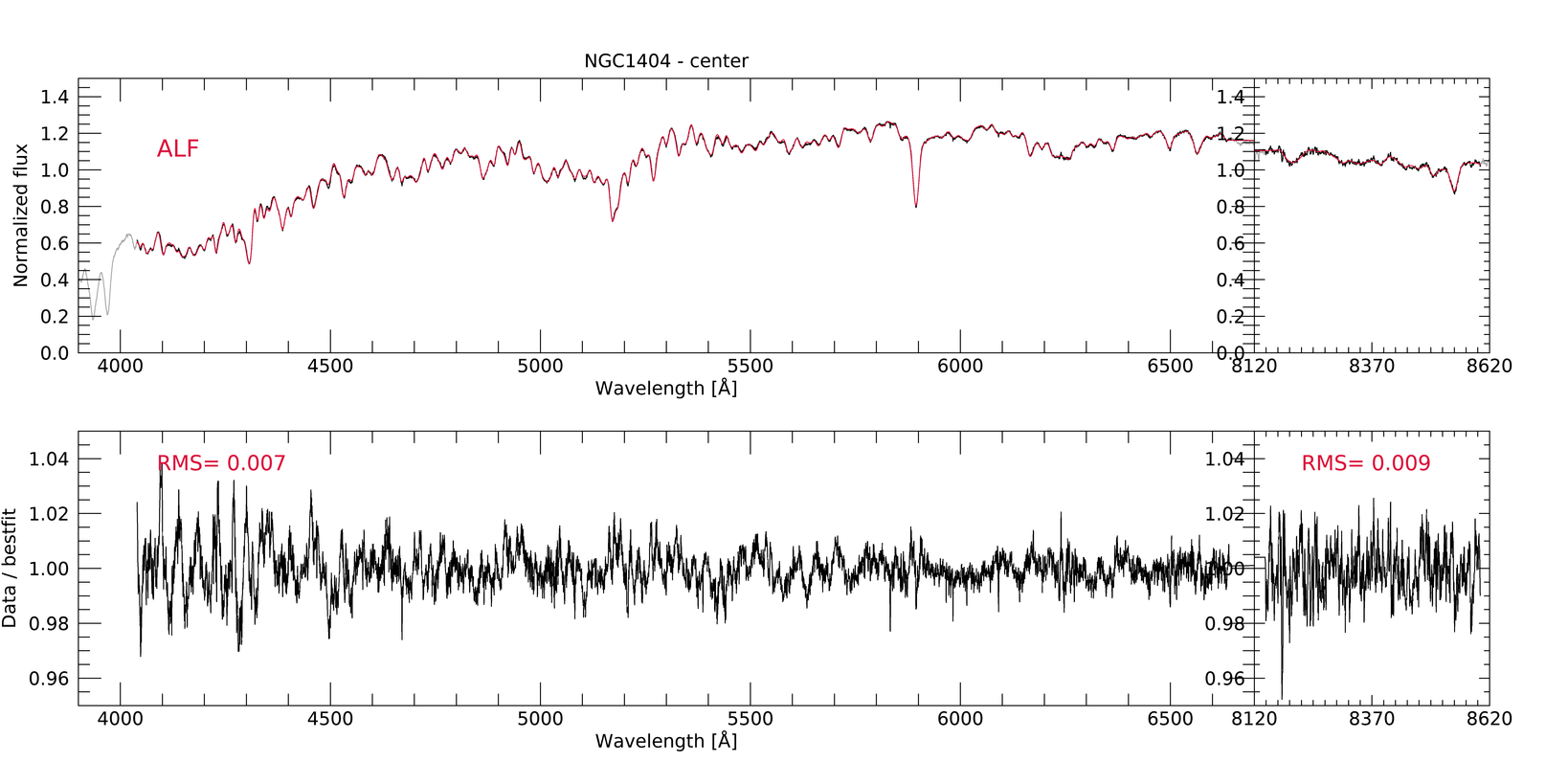}
\caption{\small{Example of best-fit match with the input spectrum (top panel) and residuals spectrum (bottom panel) for the central radial bin spectrum of NGC1404 fitted with \textsc{ALF}. Top panel: grey line is the original spectrum, black line is the adapted spectrum as required by each code (with changed spectral resolution and/or sampling, see Section \ref{sec:analysis}), and the red line is the best-fit spectrum. Bottom panel: values indicate the average value of residuals in the optical and NIR region.}}
\label{fig:residualsALF1}
\end{centering}
\end{figure*}

\begin{figure*}[ht!]
\begin{centering}
\includegraphics[width=18.0cm]{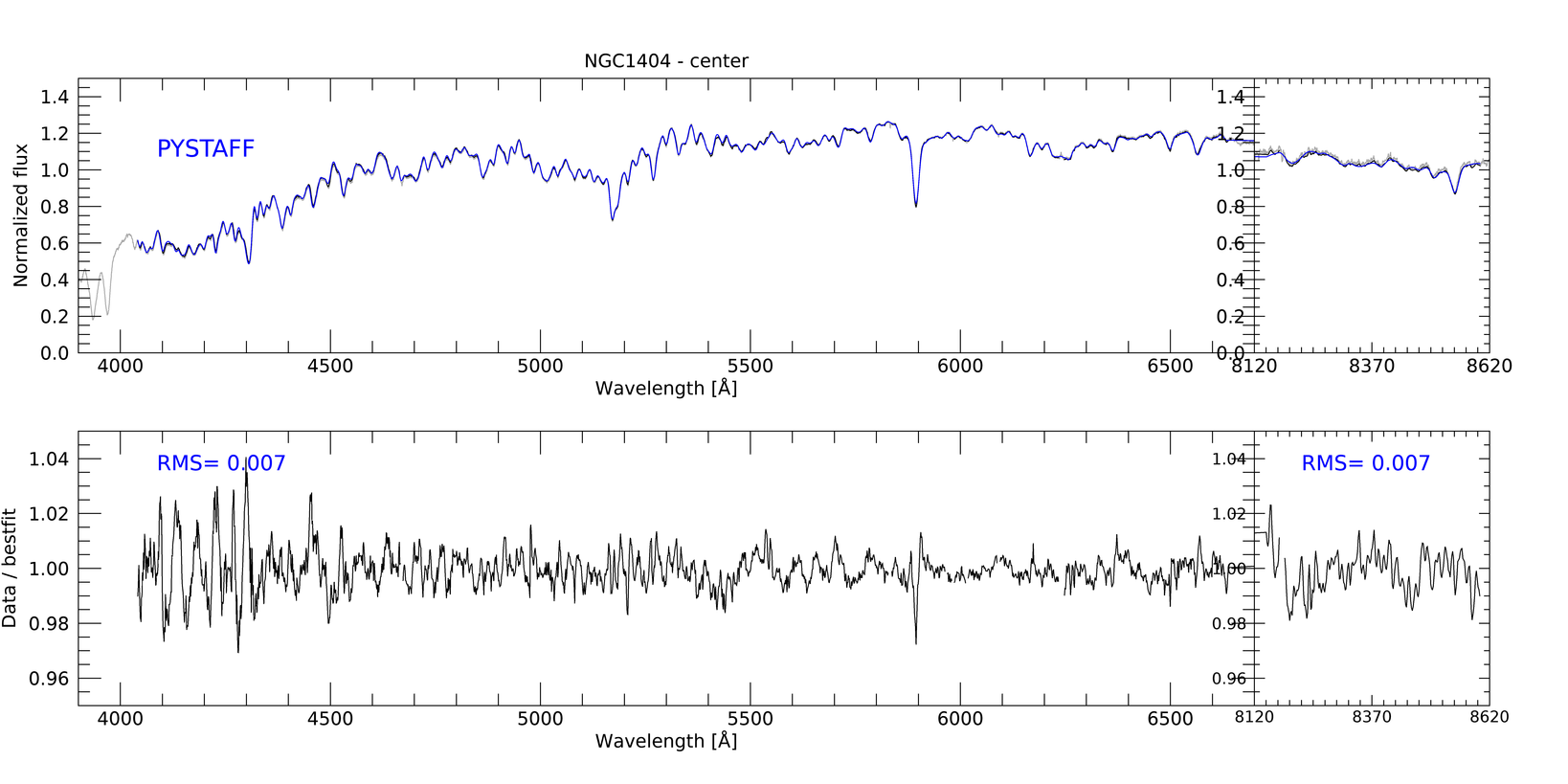}
\caption{\small{Example of best-fit match with the input spectrum (top panel) and residuals spectrum (bottom panel) for the central radial bin spectrum of NGC1404 fitted with \textsc{PyStaff}. Top panel: grey line is the original spectrum, black line is the adapted spectrum as required by each code (with changed spectral resolution and/or sampling, see Section \ref{sec:analysis}), and the blue line is the best-fit spectrum. Bottom panel: values indicate the average value of residuals in the optical and NIR region.}}
\label{fig:residualsPYSTAFF1}
\end{centering}
\end{figure*}

\begin{figure*}[ht!]
\begin{centering}
\includegraphics[width=18.0cm]{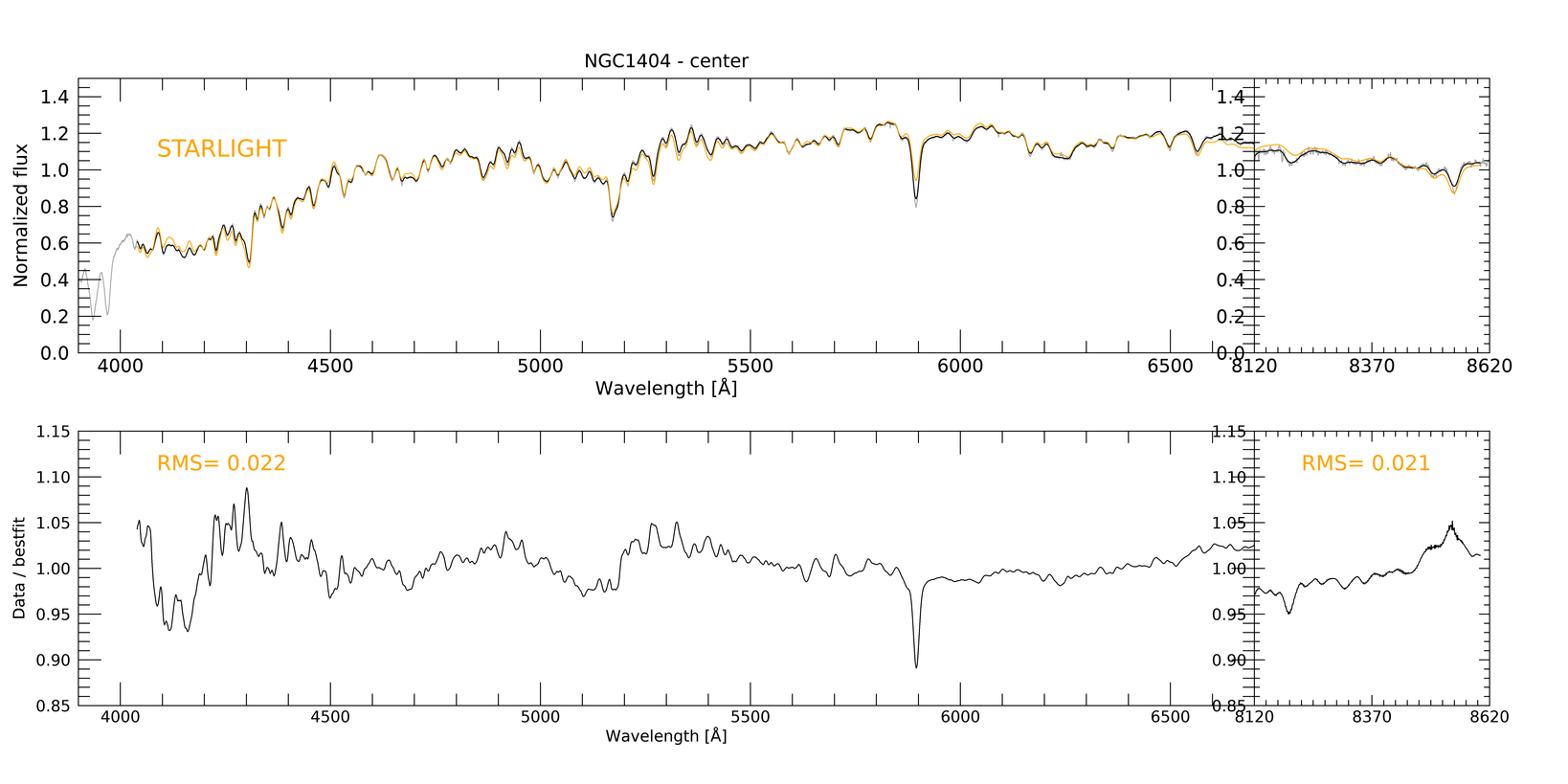}
\caption{\small{Example of best-fit match with the input spectrum (top panel) and residuals spectrum (bottom panel) for the central radial bin spectrum of NGC1404 fitted with \textsc{Starlight}. Top panel: grey line is the original spectrum, black line is the adapted spectrum as required by each code (with changed spectral resolution and/or sampling, see Section \ref{sec:analysis}), and the orange line is the best-fit spectrum. Bottom panel: values indicate the average value of residuals in the optical and NIR region.}}
\label{fig:residualsSL1}
\end{centering}
\end{figure*}

\begin{figure*}[ht!]
\begin{centering}
\includegraphics[width=18.0cm]{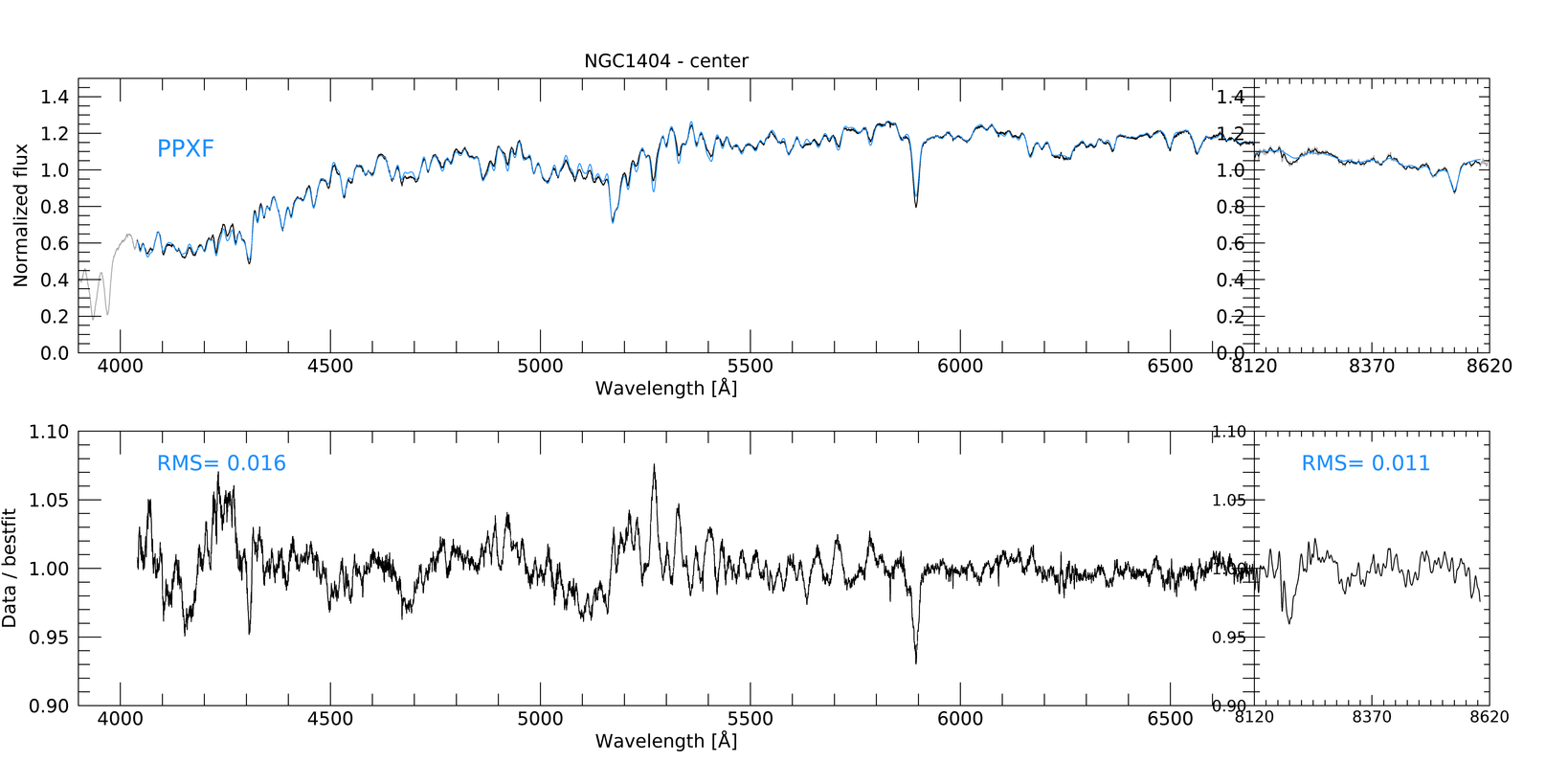}
\caption{\small{Example of best-fit match with the input spectrum (top panel) and residuals spectrum (bottom panel) for the central radial bin spectrum of NGC1404 fitted with \textsc{pPXF}. Top panel: grey line is the original spectrum, black line is the adapted spectrum as required by each code (with changed spectral resolution and/or sampling, see Section \ref{sec:analysis}), and the light blue line is the best-fit spectrum. Bottom panel: values indicate the average value of residuals in the optical and NIR region.}}
\label{fig:residualsPPXF1}
\end{centering}
\end{figure*}

In Figures \ref{fig:residualsALF1} - \ref{fig:residualsPPXF1}, we show examples of the comparison of the fitted input spectrum with the best fit model retrieved by each code. In particular, for these examples we choose the innermost radial bin spectrum of NGC1404. In the upper panels of these figures we plotted the original spectrum (as it appears also in Figure \ref{fig:spectra}) in grey and superimposed a black spectrum which is the adapted version as required by each code, i.e. the spectrum that has been actually fitted. We recollect that the differences are changes in the spectral resolution and sampling. As can be noticed from the plots, this has an effect of broadening the spectral features and an effect on the effective signal-to-noise. The colored lines are the best-fit spectra. Although interesting, the differences among codes in the actually fitted spectra prevent a direct comparison of their fitted spectral features. \\
In the lower panels we show the ratio of the fitted spectrum with the best-fit spectrum, and  the computed average RMS in the optical and NIR regions. As expected, we see that the fitting codes that use a polynomial function to adjust the best-fit spectrum to the input one, produce much lower RMS.\\
Moreover, since \textsc{Starlight} and \textsc{pPXF} could not fit for non-solar abundance ratios, their fitting procedure had fewer degrees of freedom to take into account all of the observed spectral features. This fact is reflected again in worse residuals spectra.\\
To have a more complete picture of the level of residuals in our fitted spectra, we collected the average RMS in the optical region of both the central and outermost radial bin spectra of both galaxies NGC1399 and NGC1404. The results are shown in Table \ref{tab:residuals}. For each spectrum we also inspected the residuals of the fits obtained with \textsc{Starlight} and \textsc{pPXF} when forced to fit models at non-solar abundance ratios. We warn that elemental abundances were not free parameters (since this is not possible with these two codes), but all models have been transformed to a set of fixed non-solar abundances values. In particular, the fixed values were taken from the output values extracted by \textsc{ALF}. This test was done to check if part of the reason why \textsc{Starlight} and \textsc{pPXF} retrieved worse residuals was because they missed the information on the elemental abundance ratios. Indeed, \textsc{pPXF} significantly improved the RMS, while for \textsc{Starlight} there was only mild improvement.

\begin{table*}
\begin{centering}
 \caption{Residuals average values in the optical region.}
 \label{tab:residuals}
 \begin{tabular}{lcccc}
 \hline
 \hline
                 & \textsc{ALF} & \textsc{PyStaff}  & \textsc{Starlight} & \textsc{pPXF} \\ 
 \hline
 NGC1399 CENTER     &  0.6 \%    &  0.6 \%  &  4.1 \%  &  2.1 \%  \\
 NGC1399 CENTER*    &  -         &  -       &  3.0 \%  &  1.1 \%  \\
 NGC1399 OUTER      &  0.8 \%    &  0.8 \%  &  3.7 \%  &  1.4 \%  \\
 NGC1399 OUTER*     &  -         &  -       &  3.4 \%  &  1.1 \%  \\
 NGC1404 CENTER     &  0.7 \%    &  0.7 \%  &  2.2 \%  &  1.6 \%  \\
 NGC1404 CENTER*    &  -         &  -       &  1.3 \%  &  0.9 \%  \\
 NGC1404 OUTER      &  0.8 \%    &  0.8 \%  &  1.6 \%  &  1.1 \%  \\
 NGC1404 OUTER*     &  -         &  -       &  2.0 \%  &  0.9 \%  \\
 \hline
  *Fitted from models with elemental abundance \\
  values fixed to those retrieved by \textsc{ALF}.\\
 \hline
 SIMULATIONS: \\
 \hline
 SSP-BH &  0.5 \% &  0.5 \%  & 2.5 \% &  0.9 \% \\
 SSP-TH &  0.4 \% &  0.4 \%  & 2.5 \% &  0.4 \% \\
 CSP-BH &  0.4 \% &  0.4 \%  & 1.5 \% &  0.5 \% \\
 CSP-TH &  0.4 \% &  0.4 \%  & 1.5 \% &  0.5 \% \\
 CSP2   &  0.4 \% &  0.4 \%  & 1.0 \% &  0.5 \% \\
 CSP3   &  0.4 \% &  0.4 \%  & 1.7 \% &  0.4 \% \\
 \hline
\end{tabular}
\end{centering}
\end{table*}

\begin{figure*}[ht!]
\begin{centering}
\includegraphics[width=19.0cm]{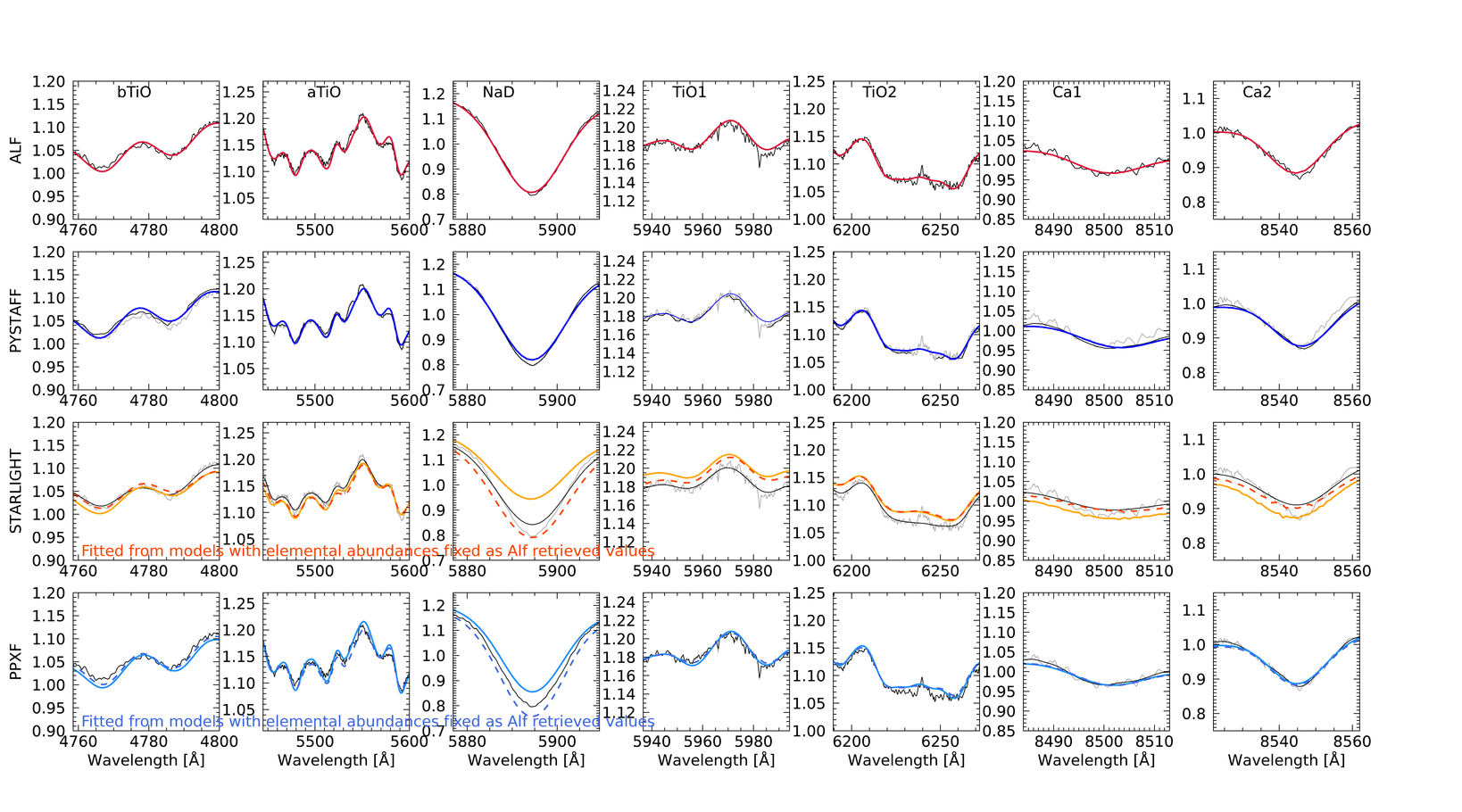}
\caption{\small{Example of best-fit match with the input spectrum for the central spectrum of NGC1404 with a closer look at some significant spectral features in the optical and NIR region. As in Figures \ref{fig:residualsALF1}-\ref{fig:residualsPPXF1}, gray lines are the original spectrum, black lines the actually fitted spectrum, and the color-coded lines the best-fit of each fitting code. For \textsc{Starlight} and \textsc{pPXF}, we superimposed with darker color and dashed lines the best-fits found with the adding of fixed non-solar elemental abundances, as explained in the text.}}
\label{fig:residualsZOOM}
\end{centering}
\end{figure*}

A further insight into this discussion can be found in Figure \ref{fig:residualsZOOM} where we show a zoom of some significant spectral features in the optical and NIR spectrum of the center of NGC1404. Here it is even more clear how much the different spectral resolution and sampling can affect the fitting of absorption lines. We have also superimposed the best fit obtained with \textsc{Starlight} and \textsc{pPXF} when forced to fit non-solar abundance ratio models (as discussed above). It is clear how some features, like the NaD at $\sim5900$\AA, do require a non-solar abundance ratio to be fully described. 

Finally, we have done the same exercise with simulated data. 
Following the results obtained in Appendix \ref{app:sim}, we then calculated the residuals for each simulation and for each code, and summarized the results in Table \ref{tab:residuals}. Without the limits that naturally occur with real data, the best-fit RMS of simulations are lower than those of observed data. In particular, for \textsc{ALF},  \textsc{PyStaff}  and \textsc{pPXF}, those that use polynomial functions, the average RMS is so low as to be of the same order of the stellar population model uncertainties, i.e. $\sim0.5\%$, as discussed in \cite{vandokkum17}. While \textsc{Starlight} presents higher RMS, around $1$-$2\%$, predominantly due to the lack of continuum fitting and adjustment.
Notice that we did not see any significant change in the residuals level when changing the complexity of the input spectrum. Indeed, we would have expected, for example, worse residuals when fitting with \textsc{ALF} or \textsc{PyStaff} a more complex SFH input like CSP3 than a SSP, since they cannot retrieve more than two stellar components. We attribute this finding to the fact that the polynomial function adjustment is correcting the best-fit spectrum at the limit of models uncertainties, regardless of the underlying complexity. We warn that this conclusion does not mean that the analysis of the stellar population properties is affected by the polynomial function adjustment. Only \textsc{Starlight} shows a mild improvement in RMS with increasing input SFH complexity, probably due to its ability to combine multiple stellar populations to give the optimal output best fit.

\section{Cross-correlations}
\label{app:cross}

\begin{figure*}[ht!]
\begin{centering}
\includegraphics[width=17.5cm]{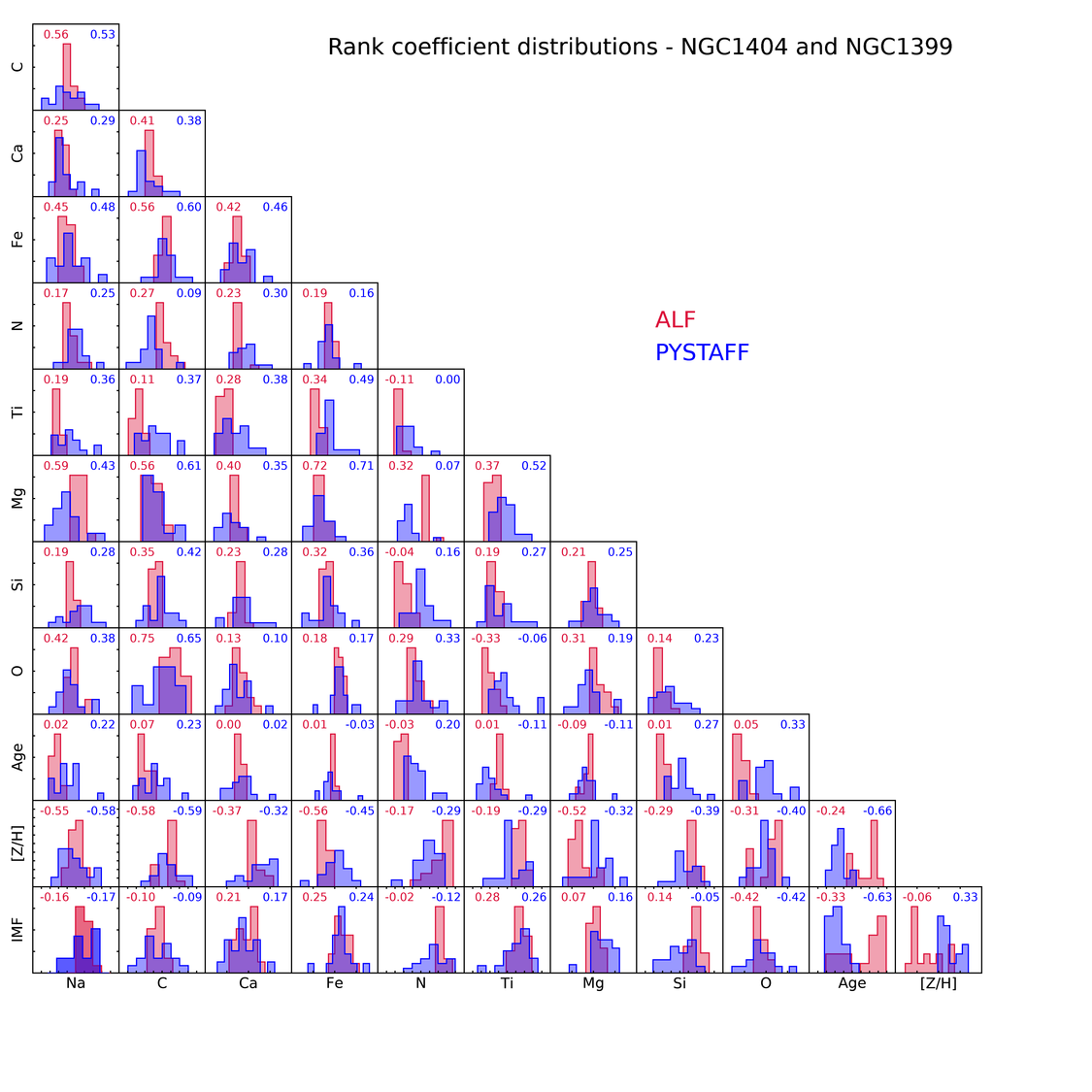}
\caption{\small{Distributions of rank coefficients as derived from \textsc{ALF} (red) and \textsc{PyStaff} (blue) fitting. Numbers in each box represent the mean value of $\rho$ for each distribution. Data refer to all fits performed on NGC1404 and NGC1399 spectra.}}
\label{fig:cross}
\end{centering}
\end{figure*}

During spectral fitting, degeneracy among parameters can occur for three main reasons: i) spectral features that can be explained by, for example, increasing the value of one parameter and decreasing that of another parameter or vice-versa (causing, e.g., the well-known age-metallicity degeneracy), or ii) internal fitting processes that can assign the same weight equally to one solution or another one, without true physical meaning, or iii) measurement uncertainties of the observed spectrum. Such degeneracies cause biases in the retrieved parameters and can lead to fake correlations among physical quantities. It is thus generally recommended to check for such degeneracies and quantify them before measuring rank coefficients between the retrieved stellar parameters. 

In this work, we are studying the differences among FSF codes and their ability in retrieving stellar parameters. In this section, we further focus on characterizing the cross-correlations among the retrieved parameters that characterize different codes.
In particular, we undertook such a comparison for \textsc{ALF} and \textsc{PyStaff} since they both operate with MCMC and provide the value of the $\chi^2$ for all steps of each chain. For both codes, we derived the rank coefficients for all pairs of parameters retrieved in each fit. Specifically, we adopted the Spearman rank coefficient $\rho$, associated with P, i.e. the significance of its deviation from zero. \\

We combined all data coming from all of the fits to NGC1404 and NGC1399 spectra in order to have a statistically significant number of data points. The distributions of rank coefficients for each pair of parameters are shown in Figure \ref{fig:cross}, for \textsc{ALF} (red) and \textsc{PyStaff} (blue), and for the following parameters: age, [Z/H], IMF slope, and abundances of O, Si, Mg, Ti, N, Fe, Ca, Ca and Na. In each panel of Figure \ref{fig:cross} the numbers show the mean value of $\rho$, color-coded in the same way as the histograms.

In general, the two codes show rather similar results, with matching positive-positive and negative-negative correlations. The well-known age-metallicity correlation is confirmed, with a stronger impact for \textsc{PyStaff} ($\rho=-0.66$) than for \textsc{ALF} ($\rho=-0.24$). Similarly, we see a correlation between age and IMF-slope and metallicity and IMF-slope, with \textsc{PyStaff} again showing larger rank coefficients. Among elements, the strongest degeneracies occur between C-O, Mg-Fe, C-Fe and Mg-C for both codes.

To estimate the strength of correlations among age, metallicity and IMF slope when fitting with \textsc{Starlight} and \textsc{pPXF}, we followed the procedure suggested by \citet{cidfernandes05}. In that work, the authors quantified the level of the well-known age-metallicity degeneracy by computing the residuals of each simulation output value from its input one. In a similar way, we computed the residuals by subtracting the weighted mean value of each parameter retrieved from each of the $100$ Monte-Carlo realizations fitted for each galaxy radius, with the corresponding mean value obtained from the final distribution. To derive the rank coefficient, we combined all residuals from all radii of both galaxies. The two codes give different results: while they both show the anti-correlation between age and metallicity (with different strength, \textsc{Starlight} being more biased), the IMF-age degeneracy is positive for \textsc{pPXF} but negative for \textsc{Starlight}, although less than $0.3$ in absolute value. We summarize the results in Table \ref{tab:cross} for the three common parameters: age, metallicity and IMF slope. 

We encourage the use of the results shown in Figure \ref{fig:cross} and Table \ref{tab:cross} in evaluating which code is best to use when deriving correlations between two specific stellar parameters. For example, if a study of the IMF-metallicity relation is performed and the assumption of SSP is robust, then \textsc{ALF} is the most preferable code since its fitting degeneracy is lower and it causes a lower level of bias in the analysis results.

\begin{table*}
\begin{centering}
 \caption{Cross-correlations results}
 \label{tab:cross}
 \begin{tabular}{cccccc}
 \hline
 \hline
par1 & par2 & $\rho$ - \textsc{PyStaff}  & $\rho$ - \textsc{ALF}  & $\rho$ - \textsc{Starlight} & $\rho$ - \textsc{pPXF} \\ 
 \hline
 age    &   Z    &  -0.24  &  -0.66  &  -0.53  &  -0.19  \\
 age    &   IMF  &  -0.63  &  -0.33  &  -0.05  &  -0.48  \\
 Z      &   IMF  &  +0.33  &  -0.06  &  -0.15  &  +0.27  \\
 \hline
\end{tabular}
\end{centering}
\end{table*}



\bibliography{Fornax_biblio}{}
\bibliographystyle{aasjournal}



\end{document}